\documentclass[usenatbib]{mnras}
\usepackage{graphicx}
\usepackage[fleqn]{amsmath}
\usepackage[T1]{fontenc}
\usepackage{ae,aecompl}
\usepackage[dvipsnames]{xcolor}
\usepackage{amssymb,amsfonts,hyperref,color,subfig}
\hypersetup{linkcolor=red,citecolor=blue,filecolor=blue,urlcolor=magenta}
\usepackage{etoolbox}
\makeatletter
\patchcmd\@combinedblfloats{\box\@outputbox}{\unvbox\@outputbox}{}{%
   \errmessage{\noexpand\@combinedblfloats could not be patched}%
}%
 \makeatother
\title[Two vortices in the MWC 758 disc]{Dust traps in the protoplanetary disc MWC 758: \\
  two vortices produced by two giant planets?}

\author[Baruteau et al.]{Cl{\'e}ment Baruteau,$^{1}$\thanks{E-mail: clement.baruteau@irap.omp.eu}
Marcelo Barraza,$^{2,3,7}$
Sebasti{\'a}n P{\'e}rez$,^{2,3}$
Simon Casassus,$^{2,3}$
\newauthor
Ruobing Dong,$^{4}$
Wladimir Lyra,$^{5,6}$
Sebasti{\'a}n Marino,$^{7}$
Valentin Christiaens,$^{2,3,8}$
\newauthor
Zhaohuan Zhu,$^{9}$
Andr{\'e}s Carmona,$^{1}$
Florian Debras$^{1}$
and
Felipe Alarcon$^{2,3}$
\\
$^{1}$IRAP, Universit{\'e} de Toulouse, CNRS, UPS, Toulouse, France\\
$^{2}$Departamento de Astronom{\'i}a, Universidad de Chile, Casilla 36-D, Santiago, Chile\\
$^{3}$Millennium Nucleus 'Protoplanetary Disks', Chile\\
$^{4}$Steward Observatory, University of Arizona, Tucson, AZ, 85719, USA\\
$^{5}$Department of Physics and Astronomy, California State University Northridge, 1811 Nordhoff St, Northridge CA 91130, USA\\
$^{6}$Jet Propulsion Laboratory, California Institute of Technology, 4800 Oak Grove Drive, Pasadena, CA, 91109, USA\\
$^{7}$Max Planck Institute for Astronomy, Königstuhl 17, D-69117 Heidelberg, Germany\\
$^{8}$Monash Centre for Astrophysics (MoCA) and School of Physics and Astronomy, Monash University, Clayton Vic 3800, Australia\\
$^{9}$Department of Physics and Astronomy, University of Nevada, Las Vegas, 4505 South Maryland Pkwy, Las Vegas, NV 89154, USA\\
}

\date{Accepted 2019 March 13. Received 2019 March 12; in original form 2018 July 17}

\pubyear{2019}
\begin{document}
\label{firstpage}
\pagerange{\pageref{firstpage}--\pageref{lastpage}}
\maketitle

\begin{abstract}
  Resolved ALMA and VLA observations indicate the existence of two
  dust traps in the protoplanetary disc MWC 758. By means of 2D
  gas+dust hydrodynamical simulations post-processed with 3D dust
  radiative transfer calculations, we show that the spirals in
  scattered light, the eccentric, asymmetric ring and the
  crescent-shaped structure in the (sub)millimetre can all be caused
  by two giant planets: a 1.5-Jupiter mass planet at 35~au (inside the
  spirals) and a 5-Jupiter mass planet at 140~au (outside the
  spirals). The outer planet forms a dust-trapping vortex at the inner
  edge of its gap (at $\sim$85~au), and the continuum emission of this
  dust trap reproduces the ALMA and VLA observations well.  The outer
  planet triggers several spiral arms which are similar to those
  observed in polarised scattered light. The inner planet also forms a
  vortex at the outer edge of its gap (at $\sim$50~au), but it decays
  faster than the vortex induced by the outer planet, as a result of
  the disc's turbulent viscosity. The vortex decay can explain the
  eccentric inner ring seen with ALMA as well as the low signal and
  larger azimuthal spread of this dust trap in VLA
  observations. Finding the thermal and kinematic signatures of both
  giant planets could verify the proposed scenario.
\end{abstract}


\section{Introduction}
MWC 758 is a 3.5 $\pm$ 2.0 Myr Herbig A5 star \citep{Meeus12} located
at a distance of 160.2 $\pm$ 1.7 pc \citep{Gaia2018}. Its mass has
been estimated as 1.5 $\pm$ 0.2 $M_{\odot}$ \citep{Isella2010,
  Reggiani2018}. The disc around MWC 758 is a transition disc with a
nearly 50 au (0\farcs30) cavity in the submillimetre
\citep{Andrews2011}. Recent high angular resolution observations have
revealed stunning non-axisymmetric emission features in the MWC 758
disc.  These asymmetries so far consist of multiple spiral arms and
arcs in near-infrared scattered light \citep{Grady2013, Benisty2015,
  Reggiani2018, Ren2018}, an asymmetric ring of emission at
$\sim$0\farcs32 as well as a compact crescent-shaped structure at
$\sim$0\farcs53 in the (sub)millimetre emission
\citep{Marino2015mwc,Boehler2018,Dong18,Casassus2019}. The asymmetries
in the (sub)millimetre emission, which we will refer to as Clump 1
(crescent at $\sim$0\farcs53) and Clump 2 (asymmetric ring at
$\sim$0\farcs32; see the right-hand panels in Fig.~\ref{fig:mapsALMA})
have been interpreted as dust traps at local pressure maxima arising
from two large-scale vortices \citep{Marino2015mwc}. The cm-wavelength
VLA observations presented in \citet{Casassus2019} support the dust
trapping scenario for Clump 1, and suggest marginal trapping for Clump
2.  Still, the mechanism behind the formation of the possible
dust-trapping vortices in the MWC 758 disc remains elusive and is the
subject of this paper.

One way to form a large-scale vortex in a protoplanetary disc is
through the Rossby wave instability \citep[RWI;][] {Lovelace1999,
  Li2000, Li2001}. This instability can set in when there is a radial
minimum in the gas vortensity\footnote{\label{def_vortensity}In 2D,
  the gas vortensity (or potential vorticity), which we denote by
  $\omega$, is the ratio of the z-component of the curl of the (2D)
  velocity to the surface density. We denote by $\omega_0$ the initial
  radial profile of the gas vortensity. Vortensity tends to be
  conserved along streamlines, diffused by the action of turbulent
  viscosity, and created at shocks or at locations where surfaces of
  constant density and pressure are not aligned (baroclinic source
  term).}, which in practice may occur where there is a radial
pressure bump.  A radial pressure bump can form for instance at the
transition between magnetically active and inactive regions in
protoplanetary discs, where a sharp transition in the effective
turbulent viscosity occurs \citep{Varniere06, Regaly12, Lyra15}, or at
the edges of the gap that a massive planet carves in its disc
\citep[e.g.,][]{Lyra2009, Lin2012planetvortex}. More often, it is the
gap's outer edge that develops a pressure maximum, but a very massive
planet of typically a few Jupiter masses may also form and maintain a
pressure maximum at the inner edge of its gap \citep{Bae2016}.

Whatever its trigger, the RWI leads to the formation of one or several
vortices, which tend to merge and form a single large-scale
anticyclonic vortex. An anticyclonic vortex forms a patch of closed
elliptical streamlines about a local pressure maximum. The vortex flow
tends to maintain dust on the same elliptical streamlines, gas drag
tends to drive dust towards the vortex centre, while dust turbulent
diffusion tends to spread it out
\citep{Chavanis2000,Youdin2010,LyraLin2013}. In addition, the vortex's
self-gravity causes dust particles to describe horseshoe U-turns
relative to the vortex centre, much like in the circular restricted
three-body problem, despite the vortex not being a point mass
\citep{Baruteau2016}.  The competition between the aforementioned
effects implies that dust particles of increasing size get trapped
farther ahead of the vortex centre in the azimuthal direction
\citep{Baruteau2016}. Vortices triggered by the RWI could play a key
role in planet formation by slowing down or stalling the dust's inward
drift due to gas drag, while potentially allowing dust to grow to
planetesimal sizes or even planetary sizes
\citep{Lyra2009,Sandor2011}.

A planetary origin for the two possible vortices in the MWC 758 disc
is appealing as it could also account for the detection of a
point-like source inside the submillimetre cavity in the $L'$-band
high-contrast imaging observations of \citet{Reggiani2018}, and for
the spirals in near-infrared scattered light.  The aim of this paper
is to present theoretical support for the scenario where the
asymmetric structures in the (sub)millimetre and the spirals in
scattered light could be due to the presence of two massive planets in
the MWC 758 disc. For this purpose, we have carried out
two-dimensional (2D) gas+dust hydrodynamical simulations of the
protoplanetary disc around MWC 758, and used 3D dust radiative
transfer calculations to compare synthetic maps of continuum and
scattered light with observations.  In Section~\ref{sec:methods}, we
describe the physical model and numerical setup of the hydrodynamical
simulations and the radiative transfer calculations. Their results are
then presented in Section~\ref{sec:results}.  Discussion and summary
follow in Section~\ref{sec:conclusion}.

\section{Physical model and numerical methods}
\label{sec:methods}

\subsection{Hydrodynamical simulations}
We carried out 2D gas+dust hydrodynamical simulations using the code
Dusty FARGO-ADSG. It is an extended version of the grid-based code
\href{http://fargo.in2p3.fr/-FARGO-ADSG-}{FARGO-ADSG}
\citep{Masset2000,BaruteauMasset2008a,BaruteauMasset2008b} with dust
modelled as Lagrangian test particles \citep{Baruteau2016,Fuente2017}.

\subsubsection{Planets}
We assume that MWC 758, which we take as a $M_{\star} = 1.5 M_\odot$
star \citep{Isella2010,Reggiani2018}, has two planetary companions: a
1.5 Jupiter-mass planet at 35 au, and a 5 Jupiter-mass planet at 140
au.  The mass of the inner planet is chosen such that it opens a mild
gap in the gas around its orbit (see first paragraph in
Section~\ref{sec:hydro}), to be consistent with the non-detection of a
gap in scattered light around this location ($\sim$0\farcs22,
\citealp{Benisty2015}). The mass of the outer planet is taken as the
upper mass limit for a companion at this location ($\sim$0\farcs87),
as estimated by \citet{Reggiani2018} based on their $L'$-band
observations and the use of the BT-Settl hot start evolutionary model
for planetary luminosities at near-infrared wavelengths
\citep{Allard14}.  Section~\ref{sec:planets_mwc758} contains a
discussion on the location and mass of the possible planets in the MWC
758 disc. In our simulations, the planets do not migrate in the disc
(they remain on quasi-circular orbits). To avoid a violent relaxation
of the disc due to the sudden introduction of the planets, their mass
is gradually increased over 20 orbits of the inner planet. In the
following, whenever time is expressed in orbits, it refers to the
orbital period at the inner planet's location, which is about 170 yr.

\subsubsection{Gas}
\label{sec:methods_gas}
A locally isothermal equation of state is assumed for the gas, where
its temperature remains fixed in time. Based on \citet{Boehler2018}'s
thermal Monte Carlo simulation of the MWC 758 disc (see their figure
9), we take the midplane temperature to decrease as $r^{-1}$ and equal
to 85 K at 35 au ($r$ denotes the radial cylindrical coordinate
measured from the central star). The disc's aspect ratio $h$, which is
the ratio of the midplane isothermal sound speed to the Keplerian
velocity, is therefore uniform and equal to 0.088 (we assume a mean
molecular weight of 2.4).

The initial surface density of the disc gas, which we denote by
$\Sigma_0$, is assumed proportional to $r^{-1}$ and equal to $\sim$1.7
g cm$^{-2}$ at 35 au.  In a quasi steady state, when the planets have
carved a gap around their orbit, the azimuthally-averaged gas surface
density varies from 1 to 2 g cm$^{-2}$ between 40 and 90 au, which is
overall consistent with the values of the gas surface density obtained
by \citet{Boehler2018} based on their ALMA band 7 observations of
$^{13}$CO and C$^{18}$O in the MWC 758 disc (see their Figure
8). Despite the Toomre $Q$-parameter being rather large (it is
$\sim$30 at 85 au, the radial location of Clump 1), gas self-gravity
is included, as it is found to impact both the vortex lifetime
\citep{Zhu2016,Regaly2017} and the dust dynamics \citep{Baruteau2016}
given our range of disc parameters.  The importance of gas
self-gravity will be further emphasised in
Section~\ref{sec:impact_sg}. To mimic the effect of a finite vertical
thickness, a softening length of $0.3H(r)$ is used in the calculation
of the self-gravitating acceleration, with $H(r) = h\times r$ the
disc's pressure scale height. Likewise, a softening length of
$0.6H(r)$ is used in the calculation of the planets acceleration on
the gas.

Turbulent transport of angular momentum is modelled by a constant
alpha turbulent viscosity, $\alpha = 10^{-4}$. This rather low level
of turbulence is representative of the discs midplane from a few tens
to a hundred au, as suggested by observations of the dust continuum in
the submillimetre \citep[see, e.g.,][for the modelling of the HL Tau
disc]{Pinte16}, and according to 3D, non-ideal, local
magnetohydrodynamic (MHD) simulations (although larger $\alpha$ values
in the midplane can be obtained depending on the disc model, in
particular the amplitude of the vertical magnetic field that threads
the disc, see, e.g., \citealp{Simon15, Simon18}).

The continuity and momentum equations for the gas are solved on a
polar grid centred on the star, and the indirect terms due to the
acceleration of the star by the disc and the planets are taken into
account. The grid extends from 10.5 au to 350 au in the radial
direction, and from 0 to $2\pi$ in the azimuthal direction. We use 900
cells in the radial direction with a logarithmic spacing (required for
the gas self-gravitating acceleration to be computed by Fast-Fourier
Transforms; see \citealp{BaruteauMasset2008b}). We use 1200 cells
evenly spaced in azimuth.  Given our initial surface density profile,
the initial mass of the disc gas amounts to $\sim$0.01 $M_{\star}$.
To minimise the reflection of the planets wakes, we use wave-killing
zones as boundary conditions at the inner and outer radial edges of
the grid, where the disc fields are damped towards their initial
value.  We have checked that a different choice of boundary condition
does not affect our results.

\subsubsection{Dust}
\label{sec:methods_dust}
Dust is modelled as Lagrangian test particles that feel the gravity of
the star, of the planets, of the gaseous disc (since gas self-gravity
is included), and gas drag. Dust turbulent diffusion is also included
as stochastic kicks on the particles position following the method in
\citet{Charnoz2011} (see \citealp{Ataiee18} for more details).
However, the dust self-gravity, dust drag (or feedback), growth and
fragmentation are not taken into account in the simulations.

We use 10$^5$ particles with a size distribution $n_{\rm simu}(s)
\propto s^{-1}$ for the particles size $s$ ranging from 10 $\mu$m to
10 cm (the quantity $n_{\rm simu}(s)ds$ represents the number of
super-particles in the size interval $[s,s+ds]$ in the simulation).
This particular scaling of the dust's size distribution is chosen for
computational reasons, as it implies that there is approximately the
same number of particles per decade of size. An important note is that
the radiative transfer calculations do need a realistic size
distribution for the dust, but the only input that they need from the
hydrodynamical simulations is the spatial distribution of the dust
particles. This is the reason why we can choose any size distribution
in the simulations, as long as there is enough particles per bin size
to properly resolve their dynamics.

The dust particles are introduced in the disc gas at $300$ orbits
after the beginning of the simulation, when the planets have already
started to open a gap around their orbit. The particles are uniformly
distributed between $52$ au and $102$ au, so that they approximately
all remain between the gaps over the duration of the simulation.  This
is meant to maximise the particles resolution at the two dust traps
from which Clump 1 and Clump 2 originate in our scenario. Furthermore,
inspired by the dust trapping predictions of \citet[][see their
Section~3.2.2]{Casassus2019}, we assume that the dust particles have
an internal density $\rho_{\rm int} = 0.1$ g cm$^{-3}$, independent of
particles size, instead of a more conventional internal density of a
few g cm$^{-3}$. Our dust particles can therefore be considered as
moderately porous particles. This rather low density is overall
consistent with the collection by Rosetta of large (>10 $\mu$m) porous
aggregate particles\footnote{More specifically, internal densities
  between 0.1 and 1 g cm$^{-3}$ are required to explain the dust's
  observed accelerations via radiation pressure or by a {\it rocket
    force} due to sublimation of surface ice on the day side of
  ejected grains \citep[see, e.g.,][for a
  review]{GuttlerReviewRosettaDust}.}  in the near coma of comet
67P/Churyumov-Gerasimenko \citep{Bentley16_67P,Langevin16_67P}.  A
brief discussion on how the particles internal density impacts our
results is given in Section~\ref{sec:impact_internal_density}.

The dust particles that we simulate are much smaller than the mean
free path in our disc model, and the Epstein regime of drag is
therefore relevant. In this regime, the particle's Stokes number (St),
which is the ratio of the particle's stopping time to the dynamical
time, can be expressed as
\begin{equation}
{\rm St} \approx 0.15 \times 
\left(\frac{s}{1\;{\rm cm}}\right)
\left(\frac{\rho_{\rm int}}{0.1\;{\rm g\;cm}^{-3}}\right)
\left(\frac{1\;{\rm g\;cm}^{-2}}{\Sigma}\right),
\label{eq:St}
\end{equation}
where $\Sigma$ denotes here the gas surface density interpolated at
the particle's location. When our simulations reach a quasi steady
state, St varies from $\sim$3$\times10^{-5}$ to $\sim$2.  Note that
the so-called short-friction time approximation is used in the
simulations for the smallest dust particles for which the local
stopping time is shorter than the hydrodynamical timestep.  A summary
of the parameters used in the simulations can be found in
Table~\ref{table:parameters}.
\begin{table}
\centering
\caption{\label{table:parameters} Simulations parameters}
\begin{tabular}{lr}
\hline
\hline
Parameter                           & Value           \\
\hline
Star mass                        	        & 1.5 M$_{\odot}$       \\
Inner planet's mass                   & 1.5 M$_{\rm Jup}$         \\
Inner planet's location               & 35 au         \\
Outer planet's mass                  & 5 M$_{\rm Jup}$         \\
Outer planet's location              & 140 au           \\
Disc's aspect ratio                       & 0.088       \\
Alpha turbulent viscosity         & $10^{-4}$   \\
Dust's initial location  & $\in [52-102]$ au\\
Dust's size range & $\in [10\,\mu{\rm m} - 10\,{\rm cm}]$\\
Dust's internal density & 0.1 g cm$^{-3}$\\
\hline
\end{tabular}
\end{table}

\subsection{Radiative transfer calculations}
\label{sec:RTsetup}
Our results of 2D hydrodynamical simulations are post-processed with
the 3D radiative transfer code
\href{http://www.ita.uni-heidelberg.de/~dullemond/software/radmc-3d}{RADMC3D}
(version 0.41, \citealp{Dullemond2015}).  The dust's spatial
distribution obtained in our simulations is used as input to compute
continuum emission maps at ALMA band 7 (0.9 mm) and VLA (9.0 mm)
wavelengths (Section~\ref{sec:TEsetup}). The gas surface density is
used as input to compute a polarised scattered light image in the
$Y$-band (1.04 $\mu$m, Section~\ref{sec:PIsetup}).  The spatial grid
used in RADMC3D is a 3D extension of the simulations grid in spherical
coordinates (the grid's vertical extent and the number of cells in
colatitude are specified in Sections~\ref{sec:TEsetup}
and~\ref{sec:PIsetup}).  We use $10^9$ photon packages for the thermal
Monte Carlo calculation of the dust temperature and for the
ray-tracing computation of the (sub)millimetre continuum and
near-infrared polarised scattered light images. We assume that the
star has a radius of 2.0 $R_{\odot}$ and an effective temperature of
7340 K \citep[][or see Gaia archive
\href{http://gea.esac.esa.int/archive/}{online}]{Gaia2018}.  We assume
the disc to be located at 160 pc \citep{Gaia2018}, with an inclination
of $21^{\circ}$ and a position angle of $62^{\circ}$
\citep{Isella2010, Boehler2018}.  Since the disc is rotating clockwise
in the observations, but counter-clockwise in the simulation, the
effective inclination adopted in the radiative transfer calculations
is $201^{\circ}$. The python program used to compute the
(sub)millimetre continuum and near-infrared polarised scattered light
images with RADMC3D from the results of Dusty FARGO-ADSG simulations,
\texttt{fargo2radmc3d}, is publicly available at
\href{https://github.com/charango/fargo2radmc3d}{https://github.com/charango/fargo2radmc3d}.

\subsubsection{(Sub)millimetre continuum emission}
\label{sec:TEsetup}
The calculation of the dust's continuum emission requires to specify
(i) the size distribution $n(s)$ and the total mass that the dust
which we simulate would have in the MWC 758 disc, (ii) the vertical
distribution of the dust's mass volume density, and (iii) the
absorption and scattering opacities:
\begin{enumerate}
\item[(i)] The dust's size distribution and its total mass are free
  parameters which we have varied to best reproduce the current ALMA
  and VLA observations of the MWC 758 disc. As will be shown in
  Section~\ref{sec:results}, good results are obtained for
  $n(s)\propto s^{-3}$, a minimum particle size of 10 $\mu$m (as in
  the simulations), a maximum particle size of 1 cm, and a dust-to-gas
  mass ratio of 2\%. This corresponds to a total dust mass
  $\sim$$1.6\times 10^{-4} M_{\star}$ (or $\sim$80 $M_{\oplus}$)
  between $\sim$40 au and $\sim$100 au, which is about $5-7$ times
  larger than the value reported in \citet{Boehler2018}.  The mass
  difference likely points to optical depth effects (see
  Section~\ref{sec:results}). The impact of the dust's size
  distribution and its total mass will be discussed in
  Sections~\ref{sec:impact_sizedistribution}
  and~\ref{sec:impact_totalmass}.  In practice, the aforementioned
  size range, [10 $\mu$m -- 1 cm], is decomposed into 30
  logarithmically spaced size bins, and from the spatial distribution
  of the dust particles in the 2D hydrodynamical simulation we compute
  the dust's surface density for each size bin $i$, which we denote by
  $\sigma_{i, \rm{dust}}$. The quantity $\sigma_{i, \rm{dust}}$ can be
  expressed as
\begin{equation}
\sigma_{i, \rm{dust}}(r,\varphi) = \frac{N_i (r,\varphi)}{ {\cal A}(r)}
\times \frac{M_{i, \rm{dust}}}{\sum\limits_{r,\varphi}{N_i(r,\varphi)}},
\label{eq:sigma_i_dust}
\end{equation}
where $N_i$ denotes the number of dust particles per bin size and in
each grid cell of the simulation, $\cal A$ is the surface area of each
grid cell, and $M_{i, \rm{dust}}$ is the dust mass per bin size, which
takes the form
\begin{equation}
  M_{i, \rm{dust}} = \xi M_{\rm gas} \times \frac{s_{i+1}^{4-p} - s_{i}^{4-p}}{s_{\rm max}^{4-p} - s_{\rm min}^{4-p}} = \xi M_{\rm gas} \times \frac{s_i^{4-p}}{\sum\limits_{i}{s_i^{4-p}}},
\label{eq:mass_i_dust}
\end{equation}
where [$s_i$,$s_{i+1}$] is the size range of the $i^{\rm th}$ size
bin, $s_{\rm min}$ and $s_{\rm max}$ are the minimum and maximum
particle sizes, $-p$ is the power-law exponent of the dust's size
distribution $n(s)$, $M_{\rm gas}$ the total mass of gas in the
simulation, and $\xi$ the dust-to-gas mass ratio. As stated above, the
synthetic maps of continuum emission shown in
Section~\ref{sec:results} are for $s_{\rm min}$ = 10 $\mu$m, $s_{\rm
  max}$ = 1 cm, $p = 3$, and $\xi = 2\%$.
\medskip
\item[(ii)] For the vertical distribution of the dust's mass volume
  density, hydrostatic equilibrium is assumed and for each size bin a
  Gaussian profile is adopted in which the dust's scale height $H_{i,
    {\rm dust}}$ of the $i^{\rm th}$ size bin is given by \citep[see,
  e.g.,][]{Riols18}:
\begin{equation}
H_{i, {\rm dust}} = H \times \left( \frac{D_z}{D_z + {\rm St}_i} \right)^{1/2},
\label{eq:Hd}
\end{equation}
where $H$ is the gas pressure scale height, St$_i$ the average Stokes
number of the dust particles in the $i^{\rm th}$ size bin, and $D_z$
is a dimensionless turbulent diffusion coefficient for the gas in the
vertical direction evaluated at the disc midplane, which for
simplicity we take equal to the alpha turbulent viscosity in our 2D
hydrodynamical simulations (note, however, that $D_z$ may appreciably
differ from the alpha turbulent viscosity in 3D MHD simulations,
depending on the level of turbulent activity across the disc's
vertical extent; see, e.g., \citealp{YMLJ18}).  The spatial grid used
to produce the continuum emission maps covers $2H$ on both sides of
the disc midplane with 36 cells logarithmically-spaced in colatitude.
\medskip
\item[(iii)] To compute opacities for dust particles with an internal
  density of 0.1 g cm$^{-3}$, we assume that the dust is a mixture of
  a silicate matrix (internal density of 3.2 g cm$^{-3}$), water ices
  (internal density of 1.0 g cm$^{-3}$) and a vacuum
  inclusion. Assuming that the mix aggregate has 30\% of its solids
  being silicates and 70\% water ices, the level of porosity (volume
  fraction of vacuum) needed to produce grains with a density of 0.1 g
  cm$^{-3}$ is $\sim$92$\%$. We apply the Bruggeman rules to compute
  the optical constants of the mix. The optical constants of water
  ices are obtained from the
  \href{http://www.astro.uni-jena.de/Laboratory/Database/databases.html}{Jena
    database}, those of astrosilicates are from \citet{Draine1984}. We
  use the Mie theory to compute the absorption and scattering
  opacities for anisotropic scattering, and the mean scattering angle
  \citep{Bohren1983}. We have checked that both our absorption and
  scattering opacities are in accordance with those calculated by
  \citet{Kataoka14}.
\end{enumerate}

\begin{figure}
\centering
\resizebox{\hsize}{!}
{
\includegraphics[width=0.99\hsize]{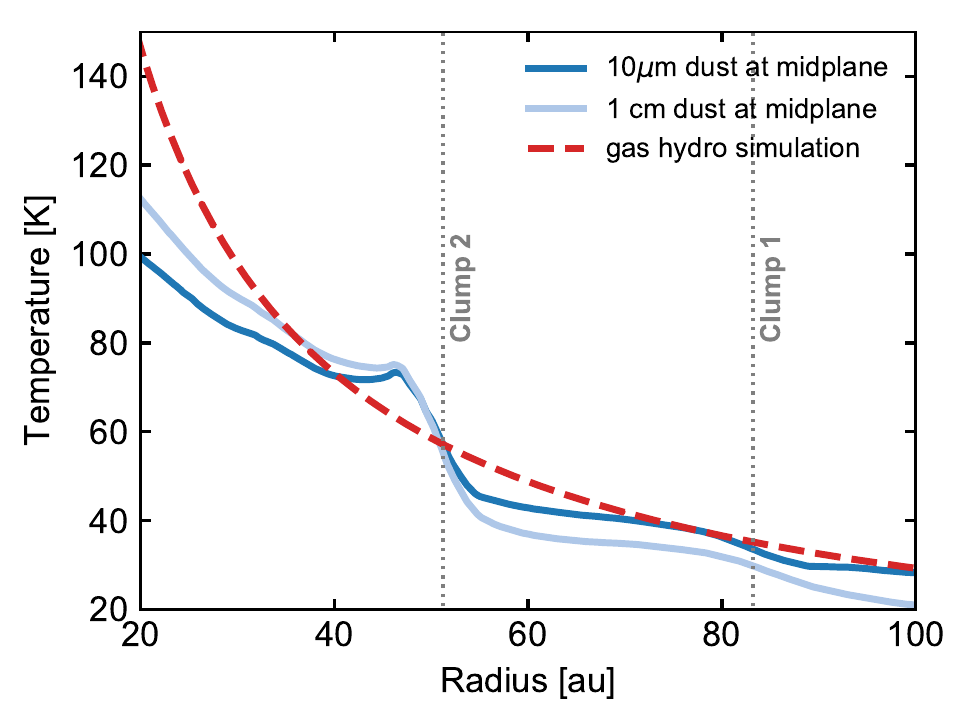}
}
\caption{Azimuthally-averaged radial profile of the dust's midplane
  temperature (solid curves) for the smallest (10 $\mu$m) and largest
  (1 cm) dust particles contributing to the (sub)millimetre continuum
  synthetic maps, and radial profile of the gas temperature in the
  hydrodynamical simulation (dashed curve). The dotted lines mark the
  location of Clumps 1 and 2.}
\label{fig:MCtemp}
\end{figure}

The raw maps of continuum emission (maps prior to beam convolution)
computed by RADMC3D include both absorption and scattering, assuming
Henyey-Greenstein anisotropic scattering. The dust temperatures are
computed first with a thermal Monte Carlo calculation. As can be seen
in Fig.~\ref{fig:MCtemp}, the azimuthally-averaged radial profile of
the dust's midplane temperature obtained with our thermal Monte Carlo
calculation for the smallest (10 $\mu$m) and largest (1 cm) dust
particles is similar to the gas temperature profile of our
hydrodynamical simulation near Clumps 1 and 2, where the vast majority
of the dust particles is concentrated (it is also consistent with
\citealp{Boehler2018}'s temperature profile, see
Section~\ref{sec:methods_gas}, despite different size distributions,
masses and opacities for the dust). The continuum image is then
computed by ray-tracing.

Some of the synthetic maps of continuum emission shown in
Section~\ref{sec:results} include noise. This is particularly helpful
for the maps at 9 mm, since the peak intensities at Clumps 1 and 2 in
the VLA image of \citet{Casassus2019}, which combines data sets in the
A, B and C array configurations, are only about 15 and 7 times higher
than the rms noise level, respectively. Instead of simulating the
exact same uv coverage and thermal noise as in the observed datasets,
we adopt a much simpler strategy, which is to add white noise to the
raw maps of continuum emission. This is done by adding at each pixel
of the raw maps a random number that follows a Gaussian probability
distribution with zero mean and standard deviation consistent with the
rms noise in the observations.  For the synthetic maps at 9 mm, the
standard deviation is set to 2$\mu$Jy/beam, which is the rms noise
level in \citet{Casassus2019}'s VLA image. Similarly, the noise
standard deviation in the synthetic maps at 0.9 mm is 20$\mu$Jy/beam,
which is the rms noise level in \citet{Dong18}'s ALMA image.

The raw flux maps at 0.9 mm are finally convolved with the same beam
as in the ALMA image of \citet{Dong18} ($0\farcs052\times0\farcs042$
PA -7.1$^\circ$), and the raw flux maps at 9 mm with the same beam as
in the VLA image of \citet{Casassus2019} ($0\farcs12\times0\farcs10$
PA 65$^\circ$). When included, the noise in the synthetic images thus
has a spatial scale that is similar to that of the beam.

\subsubsection{Near-infrared polarised scattered light}
\label{sec:PIsetup}
Near-infrared polarised scattered light traces (sub)micron-sized
grains at the surface of the disc, which are not included in our
hydrodynamical simulations. To produce scattered light predictions, we
include 12 bins of small grains ranging from 0.01 $\mu$m to 0.3
$\mu$m, and having the same density distribution and scale height as
the gas in the simulations (these small grains are expected to be well
coupled to the gas in our disc model). We artificially truncate the
dust density within 0\farcs15 ($\sim$24 au) to avoid a strong
contribution of the disc's inner parts to the polarised intensity,
which is not observed \citep{Benisty2015}.  We also reduce the dust
density beyond 0\farcs35 ($\sim$56 au, by a factor $\propto r^{-2}$)
to decrease the scattering off dust grains outside the inner edge of
the outer planet's gap. The outer reduction is meant to mimic
shadowing effects due to the gas spirals propagating between the two
planets.

We further assume that the small grains are compact monomers forming a
mixture of $60\%$ silicate and $40\%$ amorphous carbon, with an
equivalent internal density of 2.7 g cm$^{-3}$. The optical constants
for amorphous carbons are taken from \citet{Li1997}. The grains have a
size distribution $n(s)\propto s^{-3.5}$, and their total mass is
$\sim$$1.6\times10^{-5} \rm M_{\star}$ (or $\sim$8 $M_{\oplus}$) in
the results of Section~\ref{sec:results}. The grid used to compute the
polarised scattered light images covers 5 pressure scale heights on
both sides of the disc midplane with 30 cells evenly-spaced in
colatitude.

To obtain the synthetic images of polarised scattered light, the dust
temperatures are first computed with a thermal Monte Carlo
calculation.  We then compute the emergent Stokes maps $Q$ and $U$ at
1.04 $\mu$m ($Y$-band), which represent linear polarised
intensities. Scattering is assumed to be anisotropic, and a full
treatment of polarisation is adopted using the scattering matrix as
implemented in RADMC3D. When included, white noise is added to the
Stokes maps by adding at each pixel of the maps random numbers that
have Gaussian probability distributions with zero mean and a standard
deviation of 0.4\% the maximum value of each map. This value is found
to give a level of noise in the final convolved image of polarised
scattered light that is close to that in the $Y$-band polarised
intensity observation of \citet{Benisty2015}, which we will compare
our synthetic images to. A mask of $0\farcs2$ in radius is further
applied to the Stokes maps so as to enhance the brightness of the
spirals. The Stokes maps are then convolved with a circular beam of
FWHM $0\farcs026$, which corresponds to the angular resolution
achieved in \citet{Benisty2015}. Next, the convolved Stokes maps are
post-processed to obtain the local Stokes $Q_{\phi}$ following the
procedure described in \citet{Avenhaus2017}. Each pixel of the
$Q_{\phi}$ synthetic image is finally scaled with the square of the
deprojected distance from the central star in order to compare with
the SPHERE image of \citet{Benisty2015}.

\section{Results}
\label{sec:results}
We present in this section the results of our hydrodynamical
simulation, starting in Section~\ref{sec:hydro} with the time
evolution of the gas surface density and of the dust's spatial
distribution in response to the two planets. We show that the planets
form dust-trapping vortices at the edges of their gap as a result of
the RWI. Vortices may not be long-lived structures, however, and dust
particles can progressively lose the azimuthal trapping of their
vortex when the latter decays on account of the disc's turbulent
viscosity. Based on the ALMA and VLA continuum observations of
\citet{Dong18} and \citet{Casassus2019}, which are displayed in the
right-hand panels of Figs.~\ref{fig:mapsALMA} and~\ref{fig:mapsVLA},
we argue that Clump 1 is consistent with azimuthal trapping in a
vortex, while Clump 2 is more consistent with loss of azimuthal
trapping in a decaying vortex. This is what we show in
Section~\ref{sec:results_submm}, where we compare our (sub)millimetre
continuum synthetic maps with the observations.  We then show in
Section~\ref{sec:results_Yband} that the two main spirals in the
SPHERE image of \citet{Benisty2015}, which is displayed in the right
panel of Fig.~\ref{fig:mapsPI}, can be reproduced by two of the spiral
waves induced by the outer planet in our disc model. Our synthetic
maps have a fair number of free parameters, and our aim here is not to
find a set of parameters that would fit the observations, but rather
show that the two-vortex/two-planet scenario can indeed reproduce the
most salient features in the MWC 758 disc.

\subsection{Gas and dust evolutions}
\label{sec:hydro}

\begin{figure*}
\centering
\resizebox{\hsize}{!}
{
\includegraphics[width=0.33\hsize]{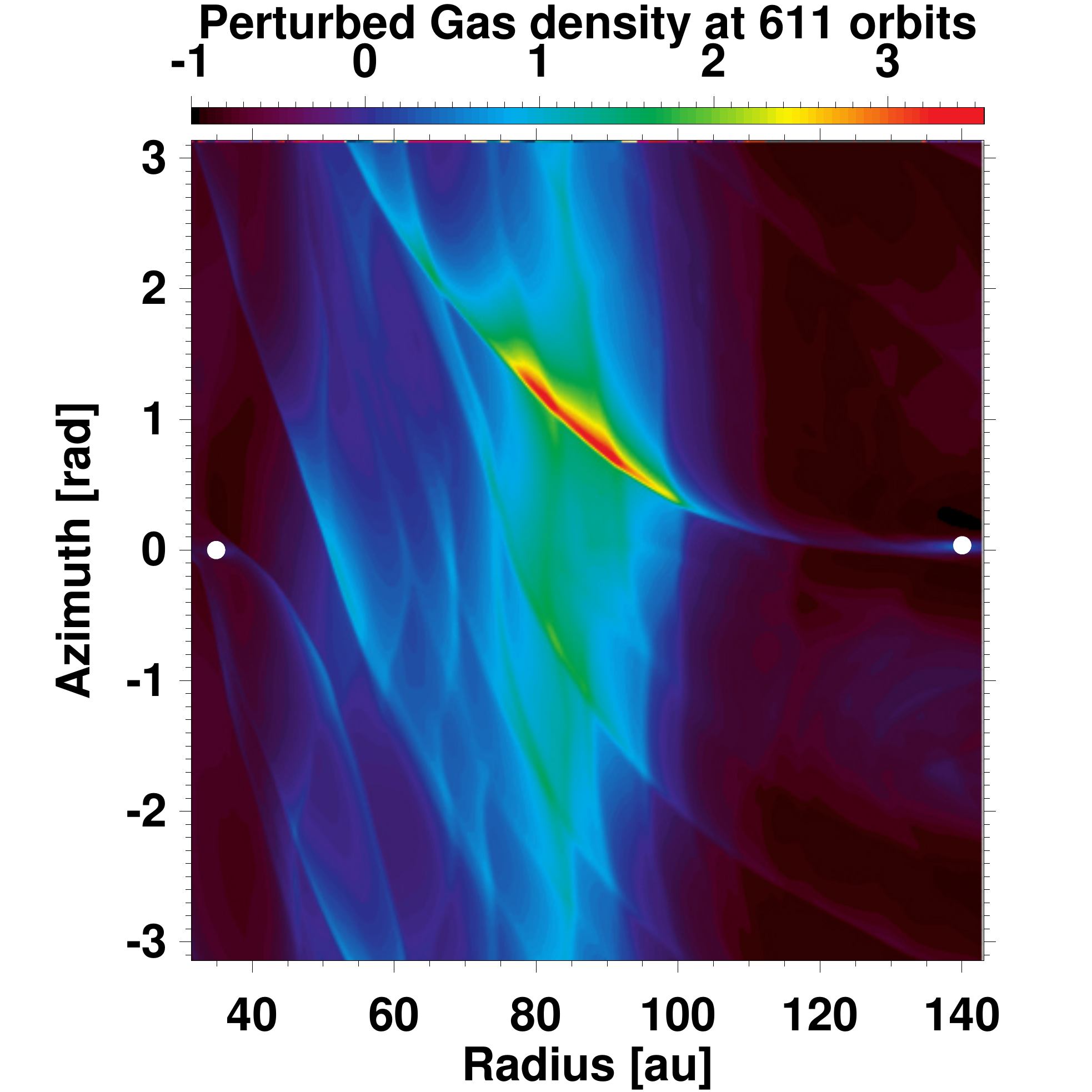}
\includegraphics[width=0.33\hsize]{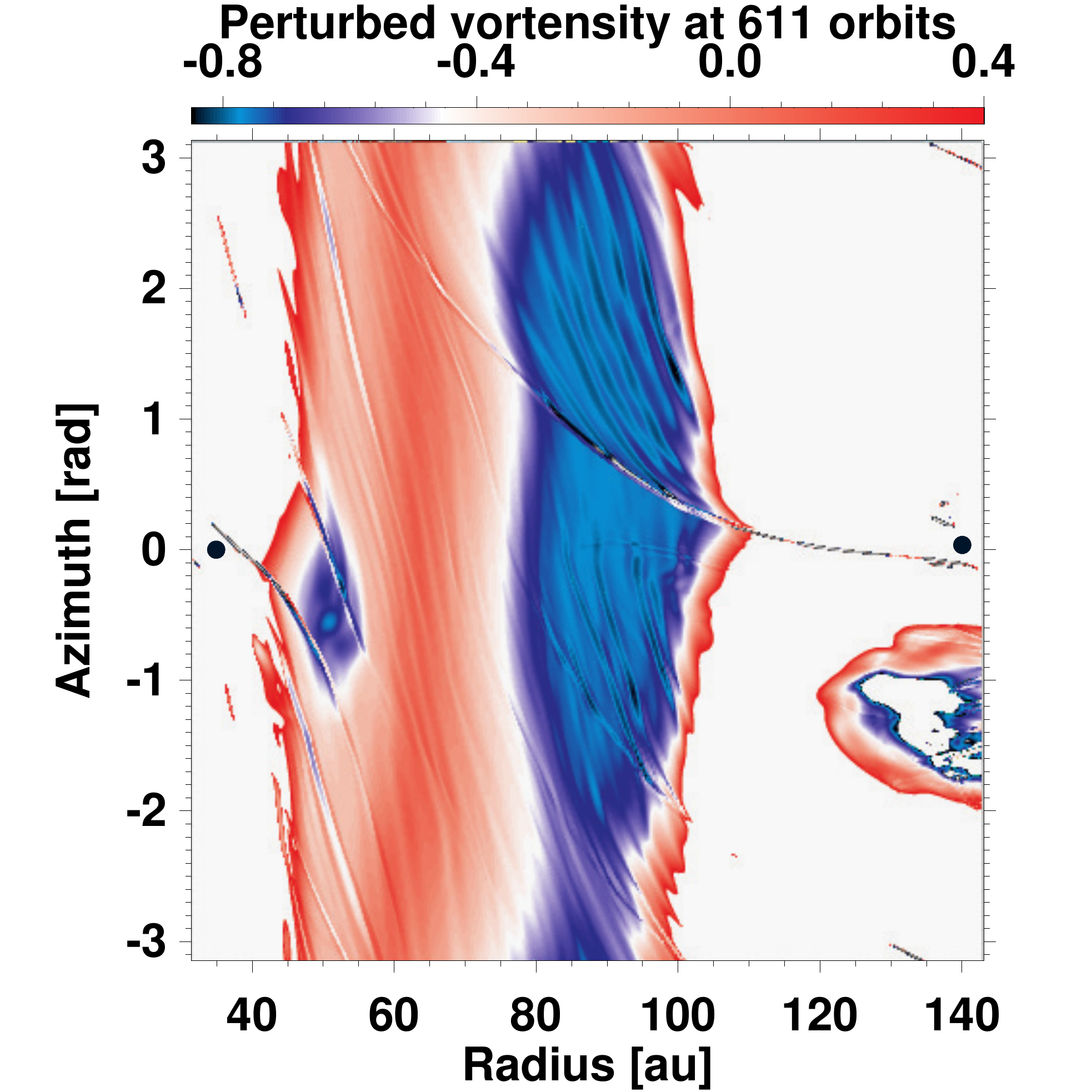}
\includegraphics[width=0.33\hsize]{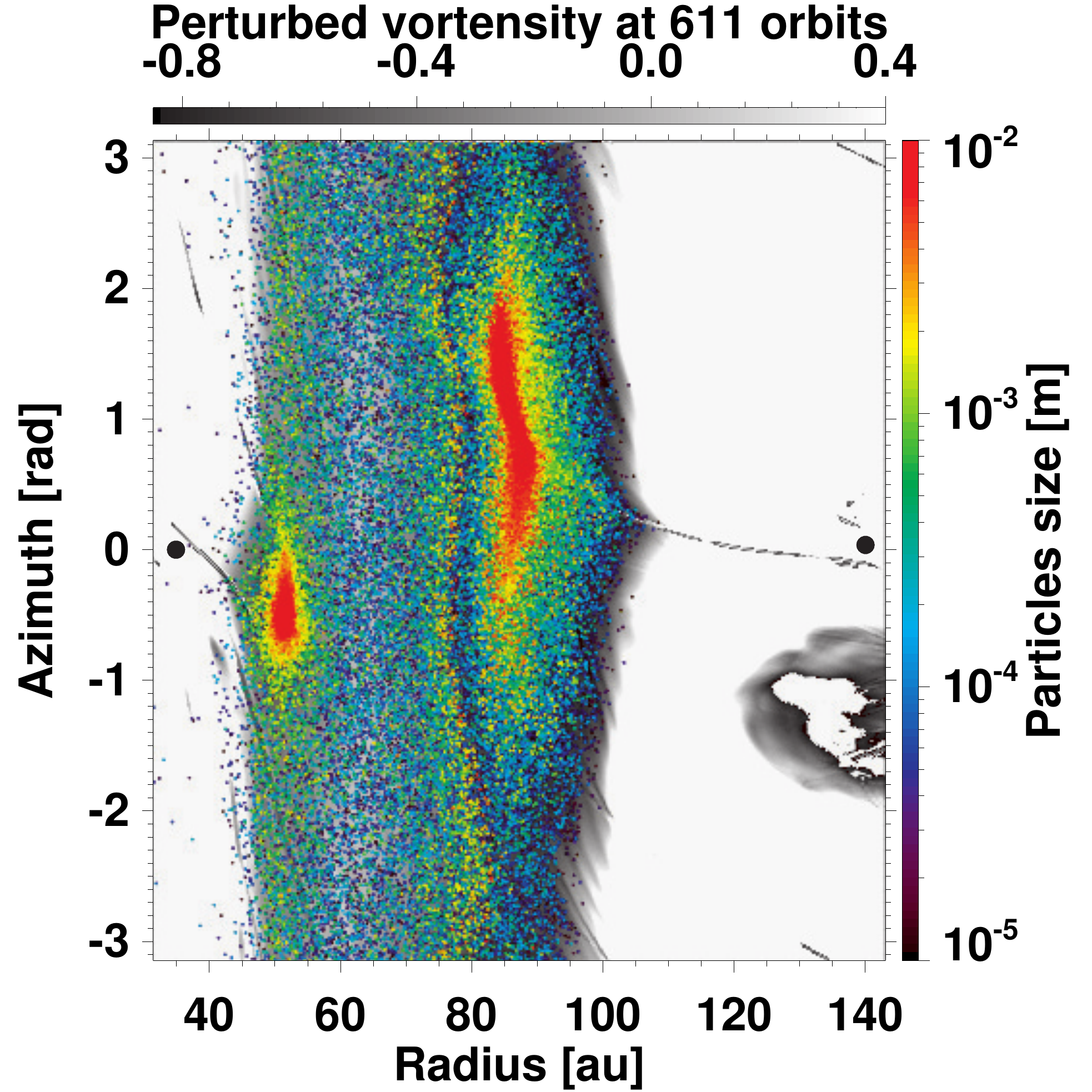}
}
\resizebox{\hsize}{!}
{
\includegraphics[width=0.33\hsize]{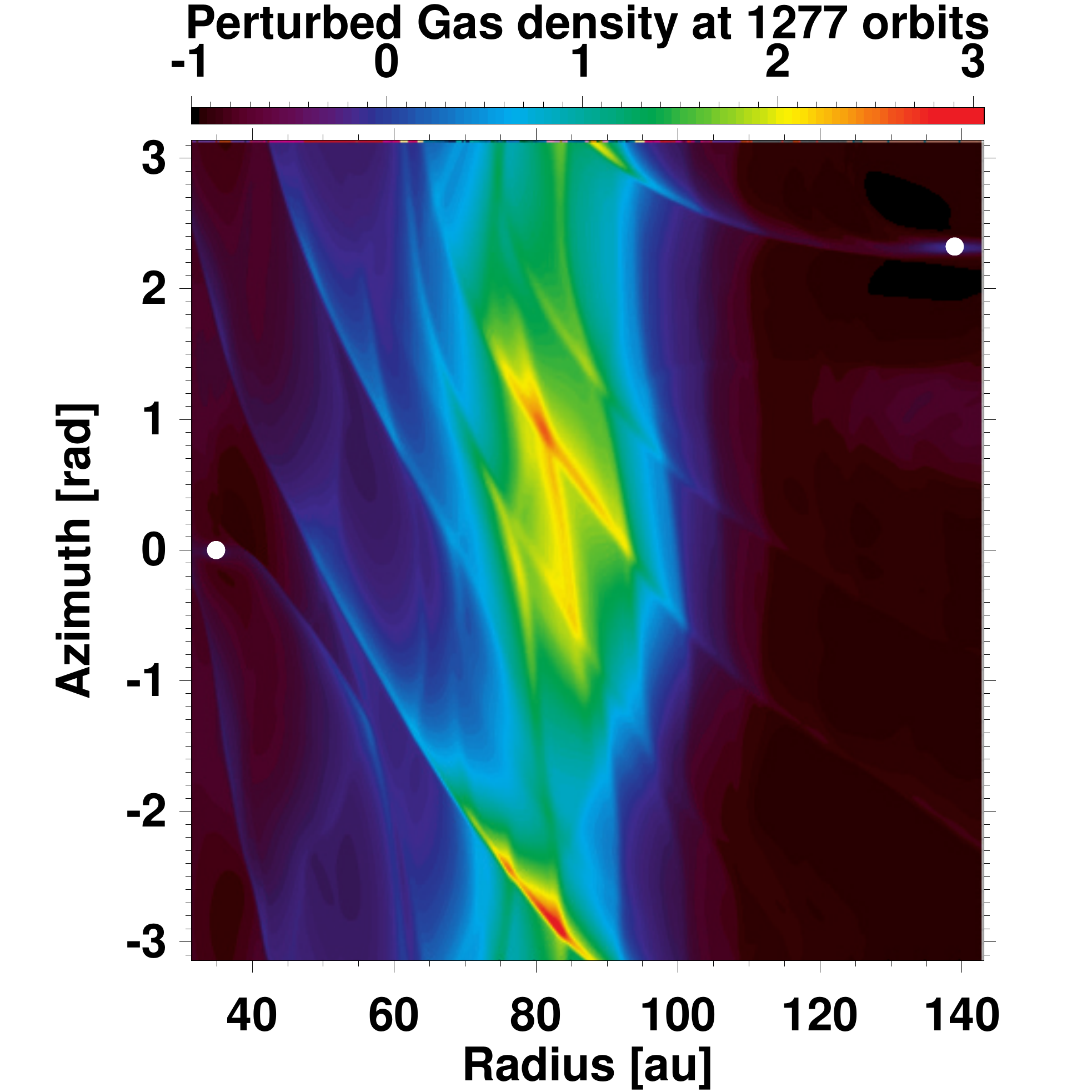}
\includegraphics[width=0.33\hsize]{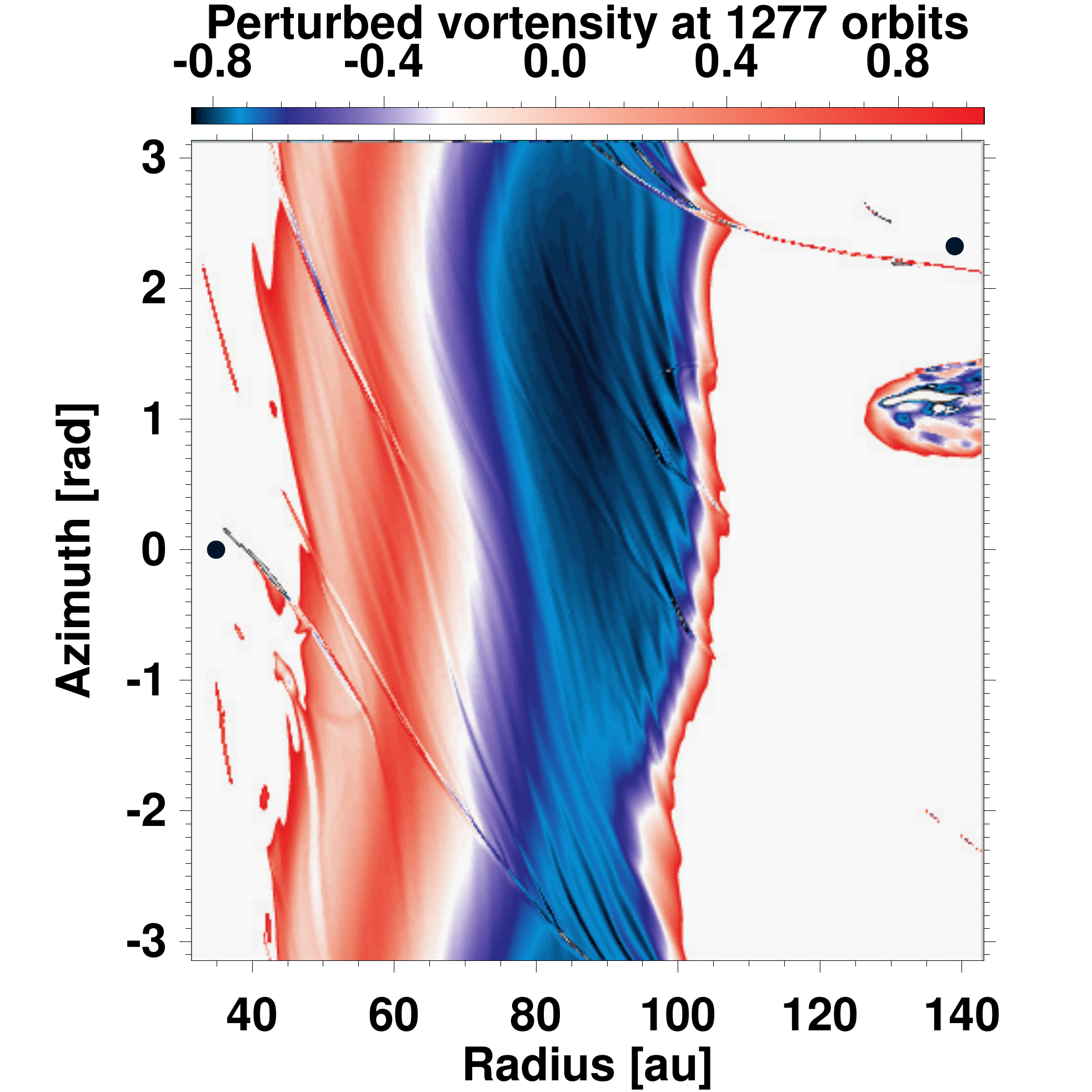}
\includegraphics[width=0.33\hsize]{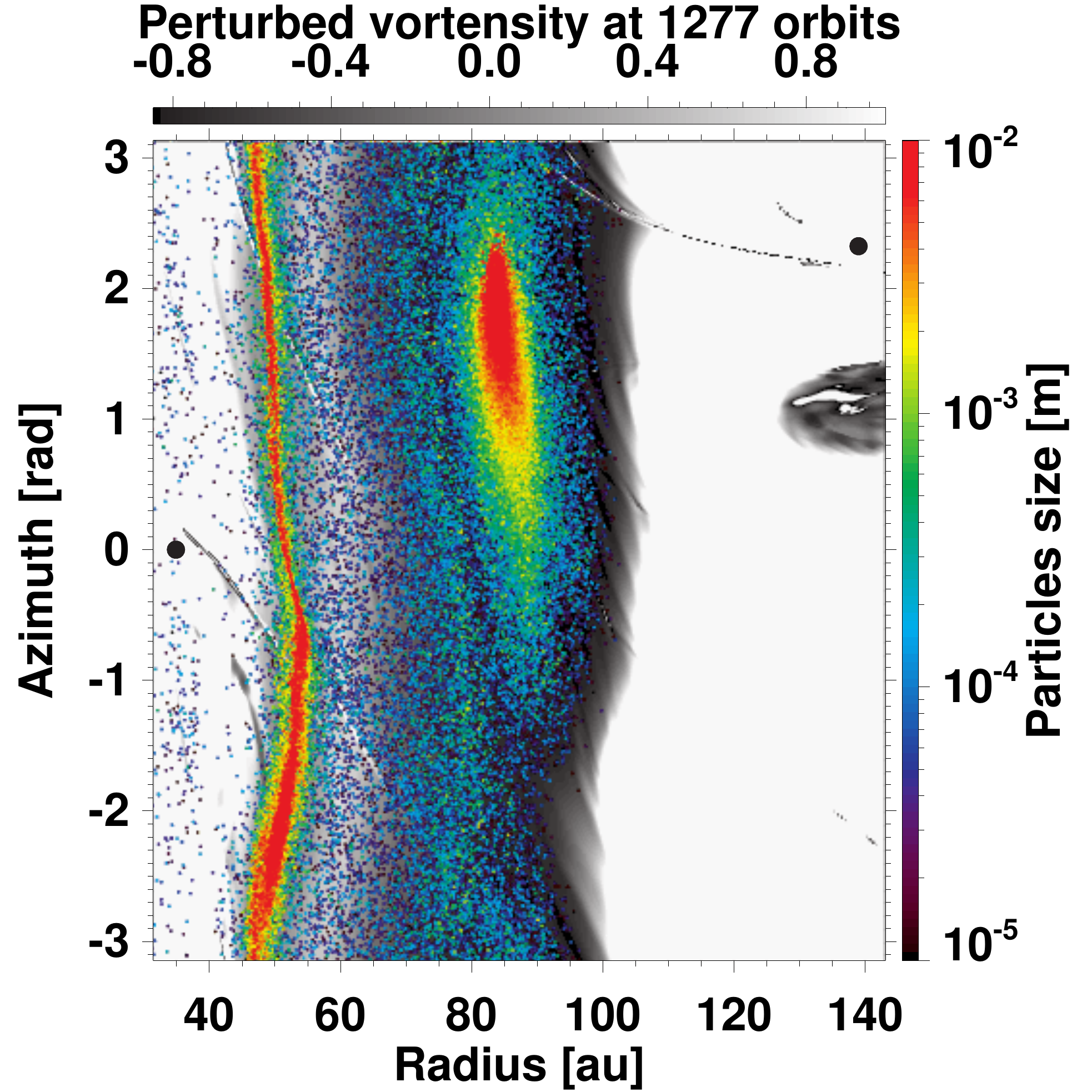}
}
\resizebox{\hsize}{!}
{
\includegraphics[width=0.33\hsize]{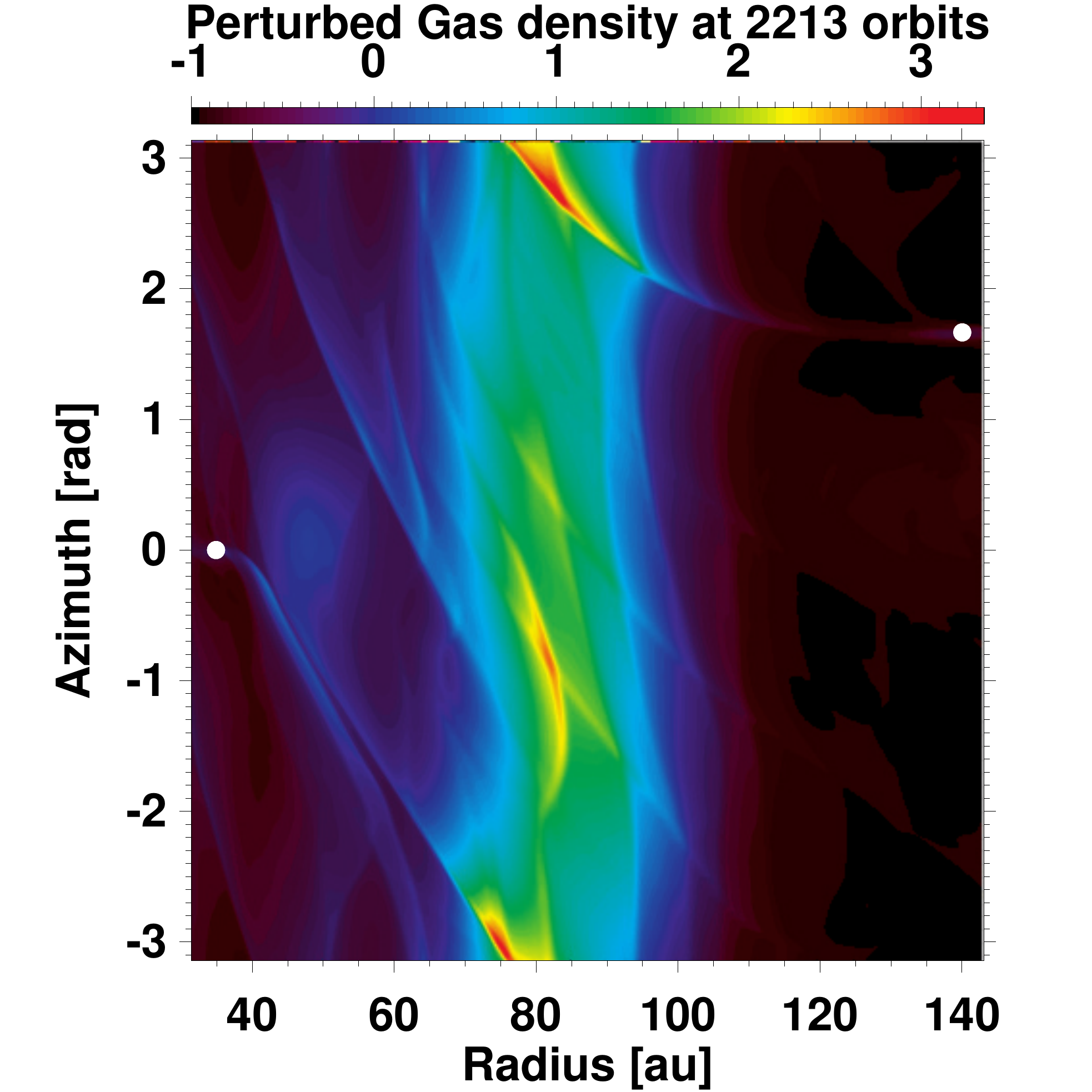}
\includegraphics[width=0.33\hsize]{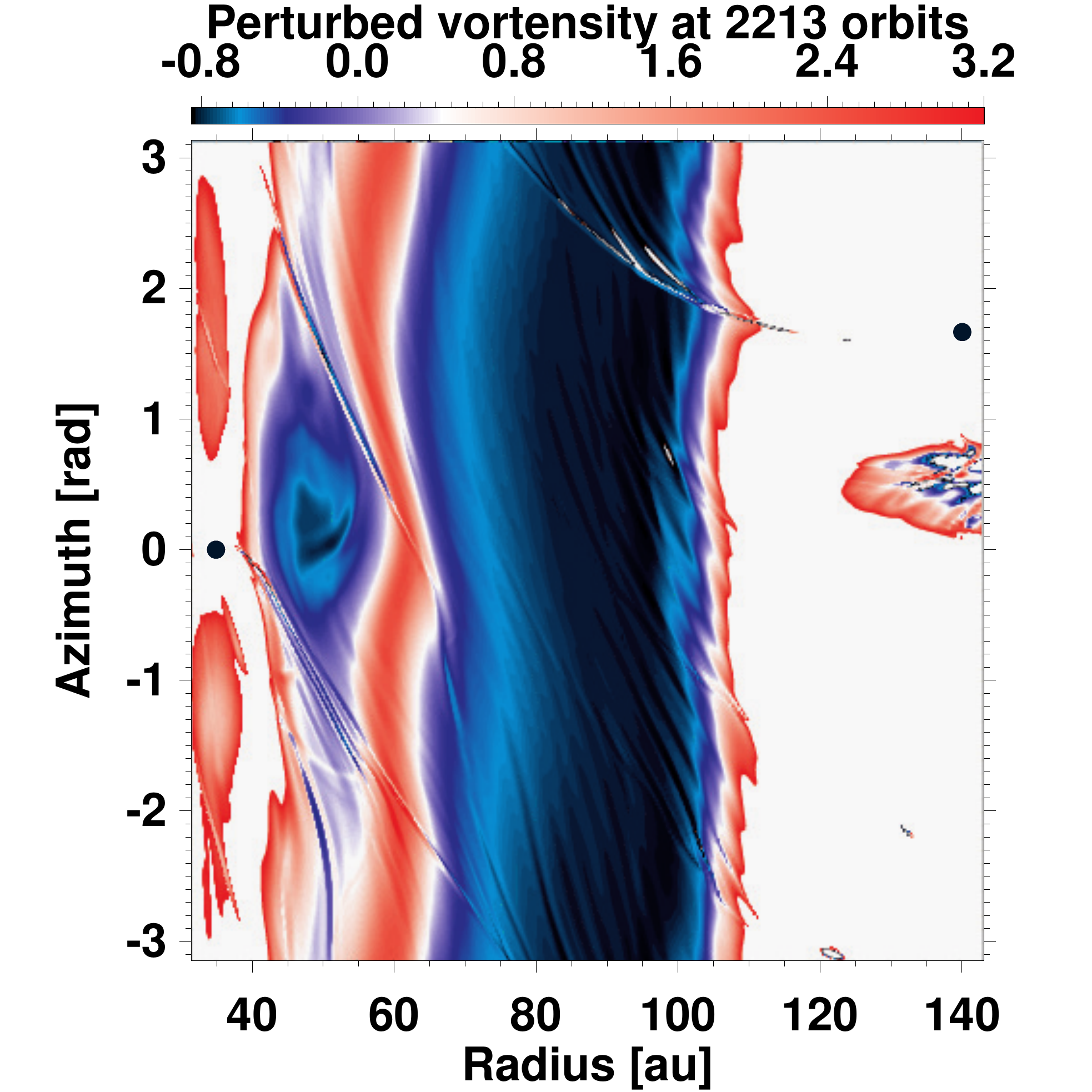}
\includegraphics[width=0.33\hsize]{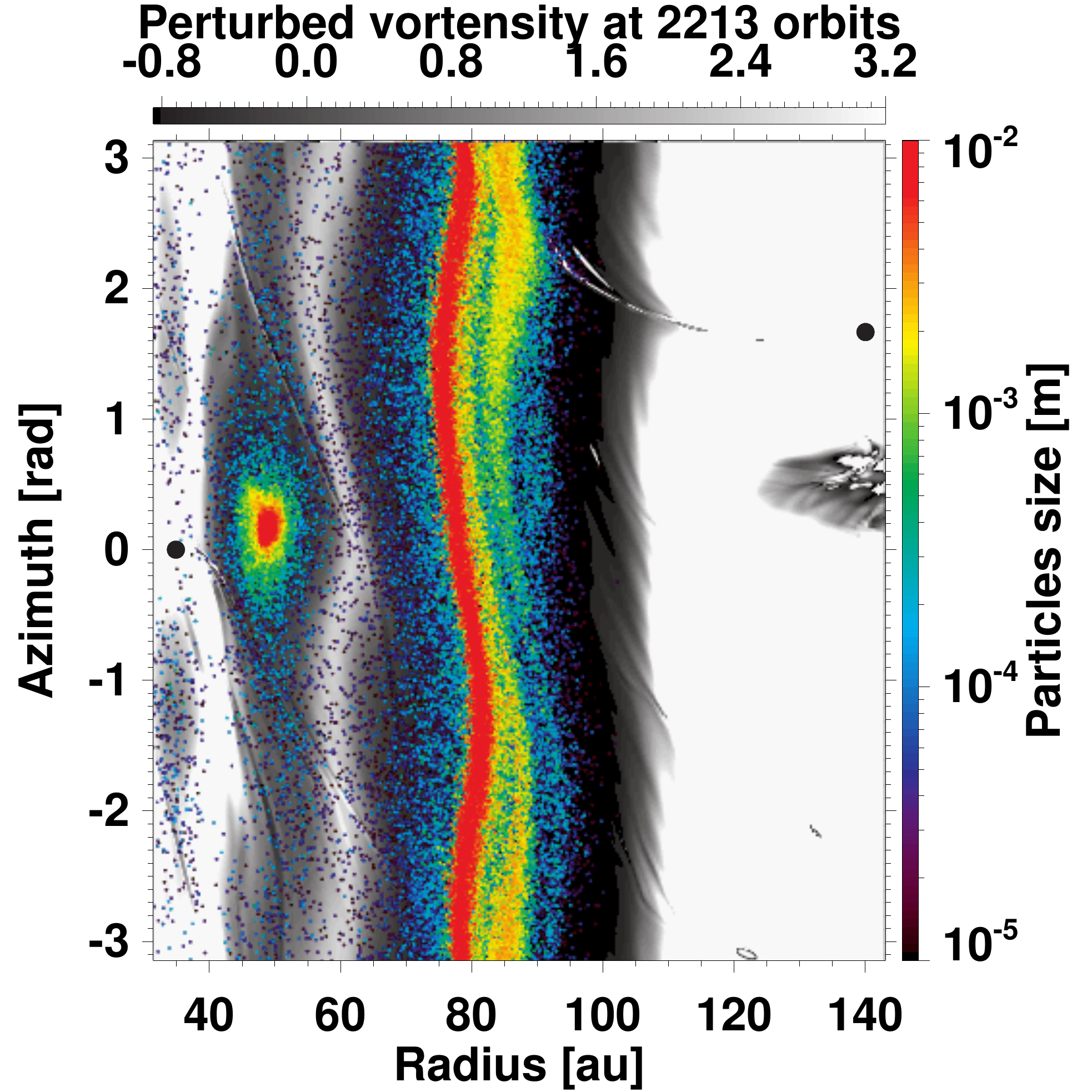}
}
\caption{Gas structure and dust spatial distribution in the disc
  region between the planets. Results are shown in polar cylindrical
  coordinates at 611, 1277 and 2213 orbits from top to
  bottom. \textbf{Left:} perturbed gas surface density relative to its
  initial profile, $(\Sigma - \Sigma_0)/\Sigma_0$.  \textbf{Middle:}
  perturbed gas vortensity relative to its initial profile,
  ($\omega-\omega_0)/\omega_0$.  The color scale has been adjusted to
  highlight the large-scale azimuthal minima of vortensity, which
  trace the gas vortices. \textbf{Right:} same as the middle panels,
  but with a sequential colormap for the contours of perturbed
  vortensity, and with the location of the dust particles overlaid by
  coloured dots. The color bar on the right-hand side shows dust size
  in metres. In each panel, the filled circles at 35 au and 140 au
  mark the position of the planets.}
\label{fig:hydro1}
\end{figure*}

\subsubsection{Gaps, spirals and vortices in the gas}
\label{sec:hydro_gas}
The planets in our disc model progressively carve a gap in the gas
around their orbit. This is illustrated in the left panels of
Fig.~\ref{fig:hydro1}, which display the perturbed gas surface density
relative to its initial radial profile, $(\Sigma -
\Sigma_0)/\Sigma_0$, at 611, 1277 and 2213 orbits after the beginning
of the simulation (which is about 0.10, 0.22 and 0.38 Myr after the
planets have reached their final mass).  Results are shown in polar
coordinates with the radius range ($x-$axis) narrowed to highlight the
disc structure between the planets. The white circles spot the
position of the planets.  At 2213 orbits (near the end of the
simulation), the azimuthally-averaged surface density of the gas has
decreased by about $70\%$ of its initial value at the bottom of the
inner planet's gap, and by about $95\%$ at the bottom of the outer
planet's gap. The inner edge of the gap carved by the outer planet in
the disc gas, which is at $\sim$90 au, is consistent with the
truncation radius of the C$^{18}$O ring in \citet{Boehler2018}'s ALMA
band 7 data (see their Figure 4). The left panels of
Fig.~\ref{fig:hydro1} also show multiple spiral arms in the disc gas,
with some spirals more prominent than others. It is not
straightforward to tell from these panels where each spiral
originates, since spirals are triggered by the planets and the gas
vortices that form in the disc. It is actually not straightforward
either to tell where the vortices are, since the gas density
perturbation in the spirals is as large, if not larger, than that in
the vortices located between the planets.

The presence of gas vortices is more easily seen in the middle panels
of Fig.~\ref{fig:hydro1}, which display the perturbation of the gas
vortensity\textsuperscript{\ref{def_vortensity}} relative to its
initial profile (the quantity $\{\omega - \omega_0\} /
\omega_0$). Anticyclonic vortices show up as local minima in the gas
perturbed vortensity in both the radial and azimuthal directions. A
large-scale vortex is clearly visible at the inner edge of the outer
planet's gap (near 85 au) at 611 and 1277 orbits. It is, however, no
longer active at 2213 orbits, which can be seen by the fact that the
vortensity minimum at this location has turned axisymmetric (based on
the gas vortensity, we estimate the lifetime of this vortex to be
about 1800 orbits). The fact that the outer planet forms a vortex at
the inner edge of its gap via the RWI is the consequence of the rather
large planet's mass, which initially allows the persistence of a
radial pressure bump (or vortensity minimum) at this location
notwithstanding the local turbulent viscosity \citep{Bae2016}. We have
checked with a dedicated simulation that a very similar vortex
structure forms in the absence of the inner planet.

The vortensity panels in Fig.~\ref{fig:hydro1} also show a vortex
around the L5 Lagrange point located behind the outer planet in
azimuth (at 140 au).  Although not shown in Fig.~\ref{fig:hydro1},
another vortex is situated at the outer edge of the outer planet's gap
near 230 au. There is no indication for vortices at both these
locations in the (sub)millimetre observations of MWC 758, but it could
be that the amount of dust trapped there is too small to have a
measurable effect on the continuum emission.

A vortex also forms at the outer edge of the inner planet's gap, near
50 au. This vortex has a rather unusual time evolution: it is active
during the first 1000 orbits, it then decays and becomes inactive over
the next 500 orbits, and finally builds up again and remains active
for at least 700 more orbits (until the end of the simulation).  This
can be seen in the second column of panels in Fig.~\ref{fig:hydro1} by
the presence at 611 and 2213 orbits of an azimuthal minimum in the gas
perturbed vortensity at $\sim$50 au that is absent at 1277 orbits. The
vortex decay coincides with a moderate increase in the eccentricity of
the disc gas between the planets.  This can be noticed by looking at
the contours of perturbed vortensity.  For instance, the vortensity
maximum that is located around 60 au at 611 orbits is associated with
gas on non-circular trajectories at 1277 orbits, with an eccentricity
close to 0.1. This point will be further emphasized when describing
the dust's spatial distribution in Section~\ref{sec:hydro_dust}.

The gas between the planets retains some level of eccentricity until
the end of the simulation, yet the vortex forms again near 50 au from
$\sim$1500 orbits. A possible explanation is the progressive increase
in the vortensity maximum around 60 au, due to the eccentric motion of
the gas and repeated interactions with the shock wakes of the planets.
This evolution makes the vortensity minimum at the outer edge of the
inner planet's gap increasingly pronounced, which ultimately allows
the RWI to set in again and form a vortex at this location. It does
not affect the vortensity near the inner edge of the outer planet's
gap, however, and the vortex initially formed at this location
progressively decays on account of the gas turbulent viscosity. Vortex
decay will be further discussed in Section~\ref{sec:decay}.

The reason why the disc gas between the planets becomes moderately
eccentric is not clear. The growth of a global eccentric mode with
azimuthal wavenumber $m=1$ has been observed in some simulations of
massive self-gravitating discs \citep[e.g.,][]{PierensLin2018,
  SebaHD169}, however, our Toomre Q parameter seems too high to
support this idea (furthermore, the increase in the gas eccentricity
is confined to the disc parts between the planets, and is therefore
not global). Another possibility is that disc-planet interactions
could account for this local increase in the disc eccentricity. This
proposal deserves a specific study, which is beyond the scope of this
paper.

We finally discuss the aspect ratio $\chi$ of the vortices, which
measures their elongation.  A rough estimate of $\chi$ can be obtained
by using contours of constant perturbed vortensity in the second
column of panels in Fig.~\ref{fig:hydro1}. For instance, at 611
orbits, the aspect ratio of the vortex near 85 au can be estimated by
using the light blue contour which marks a relative perturbation of
vortensity of about -0.76. In doing so, the vortex can be approximated
as an ellipse that extends from about -1.5 to 2.5 rad in azimuth, and
from about 80 to 100 au in radius, which corresponds to $\chi \sim
18$. Interestingly, similar aspect ratios can be estimated for the
inner vortex at 611 orbits, and for the outer vortex at 1277
orbits. In the same vein, we have checked that similar values for the
aspect ratio of the outer vortex can be obtained based on the raw
fluxes of continuum emission at 0.9 and 9 mm at 1277 orbits (see also
Section~\ref{sec:results_submm_clump1}) as well as on the
reconstructed dust's surface density (by approximating the spatial
distribution of aforementioned quantities as ellipses, and measuring
their aspect ratio).

\subsubsection{Dust trapping in the vortices}
\label{sec:hydro_dust}
The third column of panels in Fig.~\ref{fig:hydro1} displays the
spatial distribution of the dust particles on top of the vortensity
perturbation at 611, 1277 and 2213 orbits. The dust particles have a
Stokes number ranging from about $10^{-5}$ to 0.1 in all the
panels. The first thing to notice is that nearly all particles are
confined between the planets, and more particularly along two
rings. This is because the particles are inserted between the planets
at 300 orbits after the beginning of the simulation, when the planets
have already started to carve their gap. Particles thus tend to drift
towards the inner edge of the outer planet's gap, or towards the outer
edge of the inner planet's gap, since both locations are pressure
maxima in the radial direction. We see that a few (mostly small)
particles can cross the inner planet's gap, which is due to the effect
of dust turbulent diffusion kicking particles away from (and inside
of) the gap's edge.

The panels further illustrate the dust's azimuthal trapping in the
vortices. The top-right panel makes it clear that dust trapping
correlates with minima in the gas vortensity rather than with maxima
in the gas surface density.  We also notice in this panel that the
large particles trapped in the outer vortex get deflected inwards upon
crossing the primary wake induced by the outer planet.  We will come
back to this effect in Section~\ref{sec:impact_sizedistribution}.

In the middle-right panel at 1277 orbits, we see that dust is still
trapped in the outer vortex, that it is slightly eccentric, and that
the azimuth at which these particles are trapped varies continuously
with increasing particles size.  By comparison, it is quite clear that
the dust particles at the outer edge of the inner planet's gap have
lost azimuthal trapping as the inner vortex has decayed.  They form a
narrow eccentric ring, and from the orbital radius of the largest,
cm-sized particles, which range from about 45 to 55 au, we infer a
local eccentricity $\approx$0.1, very similar to that of the
background gas (see Section~\ref{sec:hydro_gas}).  This eccentricity
is consistent with that of the inner ring passing by Clump 2 in the
ALMA observation of \citet{Dong18} (see also
Section~\ref{sec:results_submm}).  This dust ring is not axisymmetric,
which reflects the fact that particles progressively lose memory of
their former azimuthal trapping on different timescales depending on
their size, much like in the simulations carried out in
\citet{Fuente2017} to account for recent submillimetre observations of
the AB Aurigae disc. We also note the presence of a faint eccentric
ring of $\lesssim 1$ mm dust particles slightly interior to Clump 1,
between $\sim$70 and 80 au, which is reminiscent to the ring of
emission at $\sim$0\farcs43 seen in the ALMA observation of
\citet{Dong18}. Comparison with Fig.~\ref{fig:sizepc}
(Section~\ref{sec:impact_sizedistribution}) suggests that this ring
could be due in part to deflections by the inner wake of the outer
planet.

In the bottom-right panel of Fig.~\ref{fig:hydro1}, at 2213 orbits,
the inner vortex is formed again and it efficiently traps the dust at
the outer edge of the inner planet's gap. The outer vortex is now
decayed, and the dust that was trapped at the outer vortex
progressively acquires a near axisymmetric spatial distribution. This
redistribution is not similar to that of the dust in the inner vortex
when the latter had decayed. The dust leaving the outer vortex shifts
radially inwards at a rate that depends on dust size. This shift,
which arises because of repeated deflections upon crossing the inner
wake of the outer planet, will be further described in
Section~\ref{sec:impact_sizedistribution}.

Anticipating the results of the radiative transfer calculations in
Section~\ref{sec:results_submm}, we find that the faint emission of
the inner ring passing by Clump 2 in the VLA observation of
\citet{Casassus2019} cannot be accounted for by a compact dust
distribution such as the one obtained in the early and late stages of
our hydrodynamical simulation when the inner vortex is active. When
the inner vortex is active, we find that the predicted peak intensity
at 9 mm is indeed 1.5-2 times larger at Clump 2 than at Clump 1, while
the observed ratio is $\sim$0.4 (see
Section~\ref{sec:results_submm_peaks}). We thus need the inner vortex
to have decayed for the dust trapped at the outer edge of the inner
planet's gap to be distributed along an eccentric ring, and yet have a
non-axisymmetric distribution. This is precisely the kind of
distribution that the dust has in our simulations between $\sim$1000
and 1500 orbits, and which we have illustrated at 1277 orbits in the
middle row of panels in Fig.~\ref{fig:hydro1}.

We actually found a fair number of outputs in our simulation between
1000 and 1500 orbits that could account for the main observational
features of the continuum maps at 0.9 mm (eccentric inner ring passing
by Clump 2 with Clump 2 near pericentre, compact emission at Clump 1,
azimuthal shift of about 130$^{\circ}$ between Clumps 1 and 2) and at
9 mm (compact emission at Clump 1, faint emission at Clump 2). Out of
these outputs, the one that best reproduces the observed maps is that
at 1277 orbits.

\begin{figure}
\centering
\resizebox{\hsize}{!}
{
\includegraphics[width=0.99\hsize]{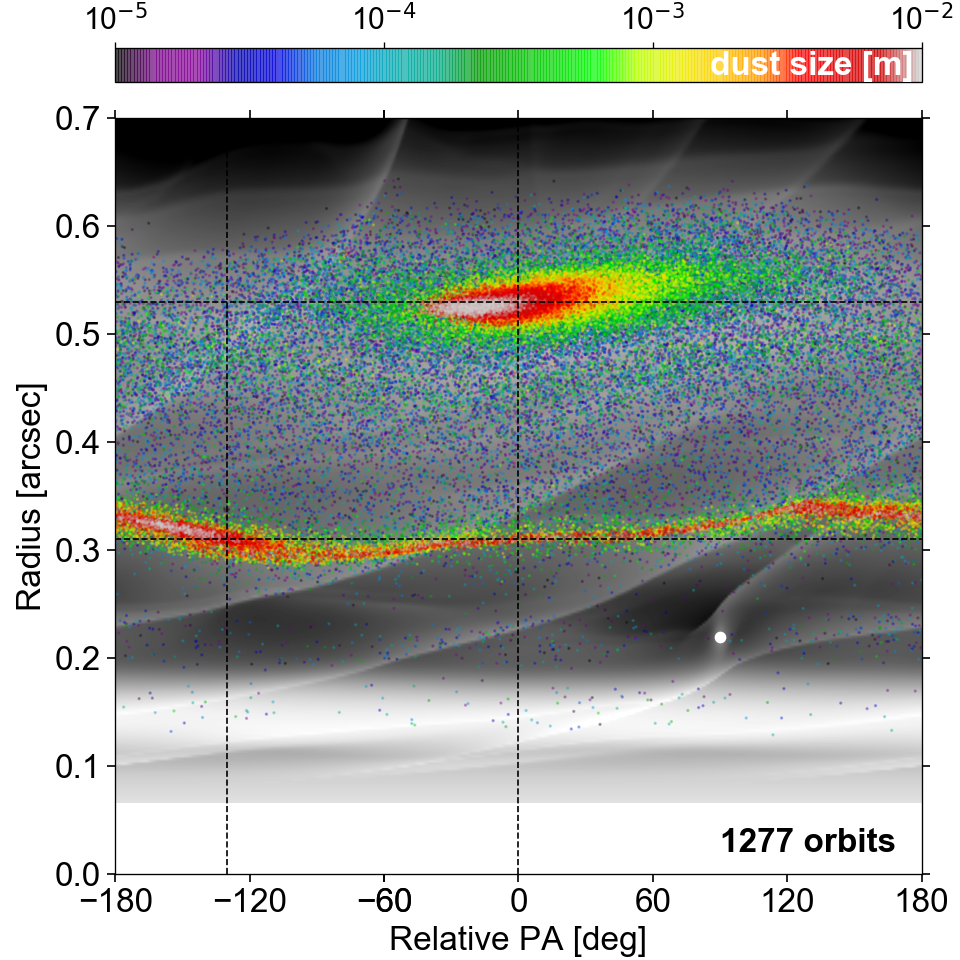}
}
\caption{Gas surface density and dust spatial distribution in the
  hydrodynamical simulation at 1277 orbits ($\sim$0.2 Myr). The gas
  surface density is shown by black and white contours with a
  logarithmic colour scale (density increases from 0.1 to 10 g
  cm$^{-2}$ from black to white). The location of the dust particles
  is marked by coloured dots (colour varies with dust size, see colour
  bar on top of the image), and that of the inner planet by a white
  circle. Results are displayed in the disc plane, with the $y$-axis
  indicating distance from the central star in arcseconds (assuming a
  disc distance of 160 pc), and the $x$-axis the position angle (or
  azimuth) relative to the approximate location of the largest
  particles in the outer vortex. The image has been flipped
  horizontally to reflect that the disc is rotating clockwise in the
  observations, but counter-clockwise in the simulation. The dashed
  lines show the radius and position angle of the two clumps of
  emission in the deprojected ALMA continuum image of
  Fig.~\ref{fig:mapsALMA}.}
\label{fig:hydro2}
\end{figure}

We finish this section by describing Fig.~\ref{fig:hydro2}, which
depicts again the dust's spatial distribution at 1277 orbits, but now
on top of the gas surface density in log scale. Results are displayed
in polar coordinates, with the orbital radius in arcseconds in
$y$-axis, and the azimuth (or position angle) relative to the
approximate location of the largest particles in the outer vortex in
$x$-axis. The inner planet is visible at $x=90^{\circ}$, $y\sim$
0\farcs22. The figure is meant to be compared with the deprojected
synthetic images of the (sub)millimetre continuum emission and of the
near-infrared scattered light shown in
Figs.~\ref{fig:mapsALMA},~\ref{fig:mapsVLA} and~\ref{fig:mapsPI}. It
helps to understand how the dust particles in the simulation
contribute to the (sub)millimetre continuum synthetic maps, which are
most sensitive to the largest (mm- to cm-sized) particles that lie
near the midplane. It also helps to see how the spiral density waves
in the gas contribute to the polarised intensity image.

\subsection{Continuum emission in the (sub)millimetre}
\label{sec:results_submm}

\begin{figure*}
\begin{center}
\includegraphics[width=0.49\textwidth]{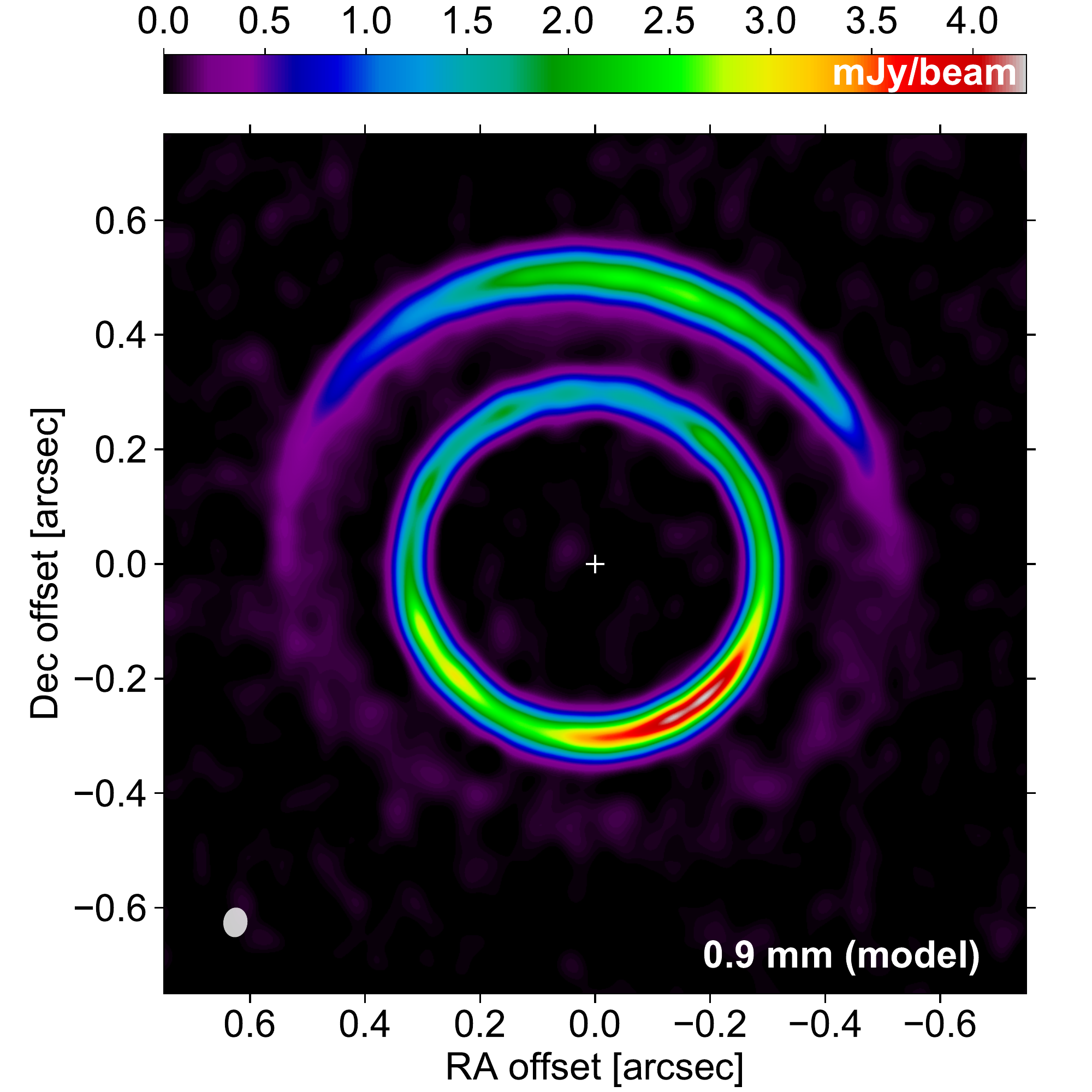}
\includegraphics[width=0.49\textwidth]{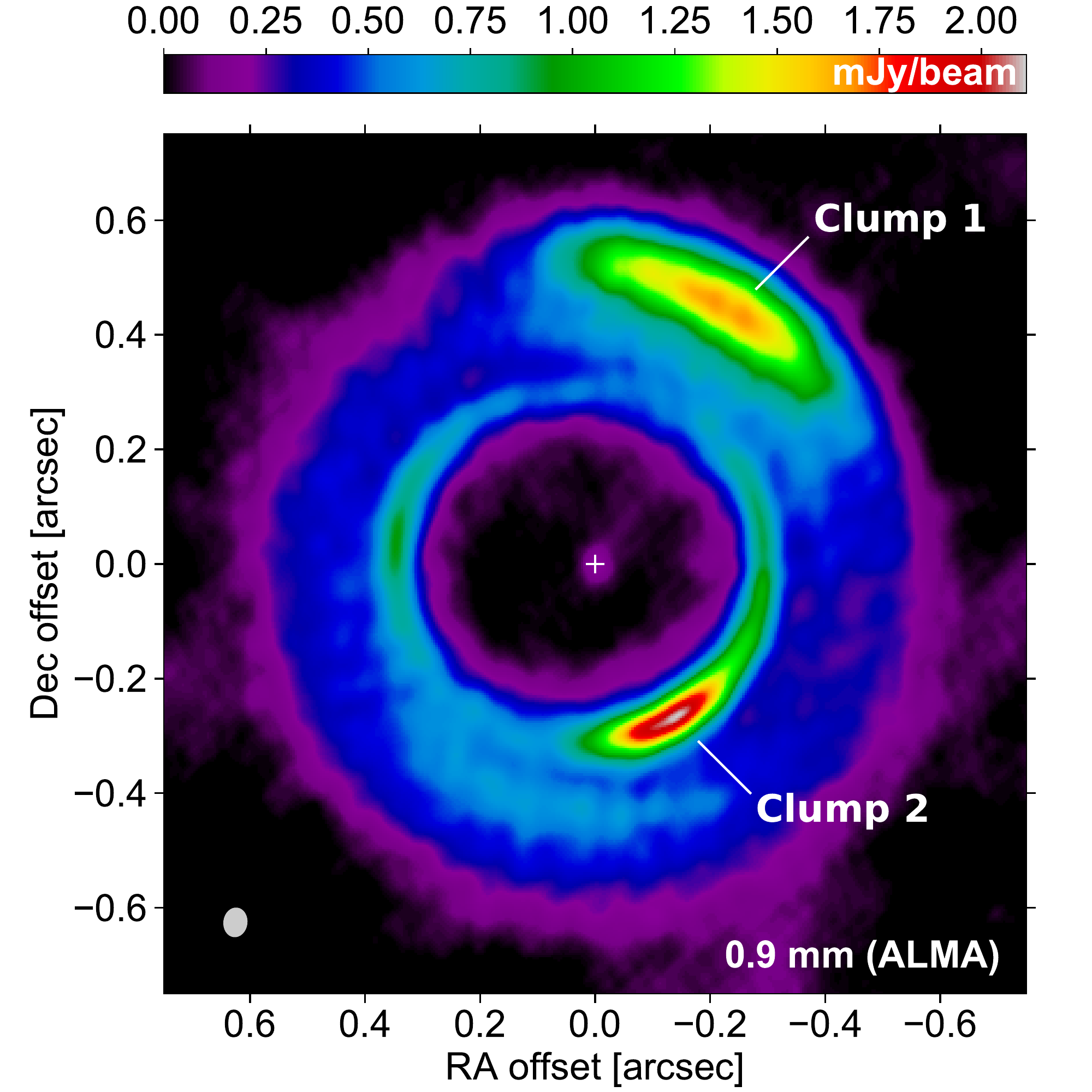}
\includegraphics[width=0.49\textwidth]{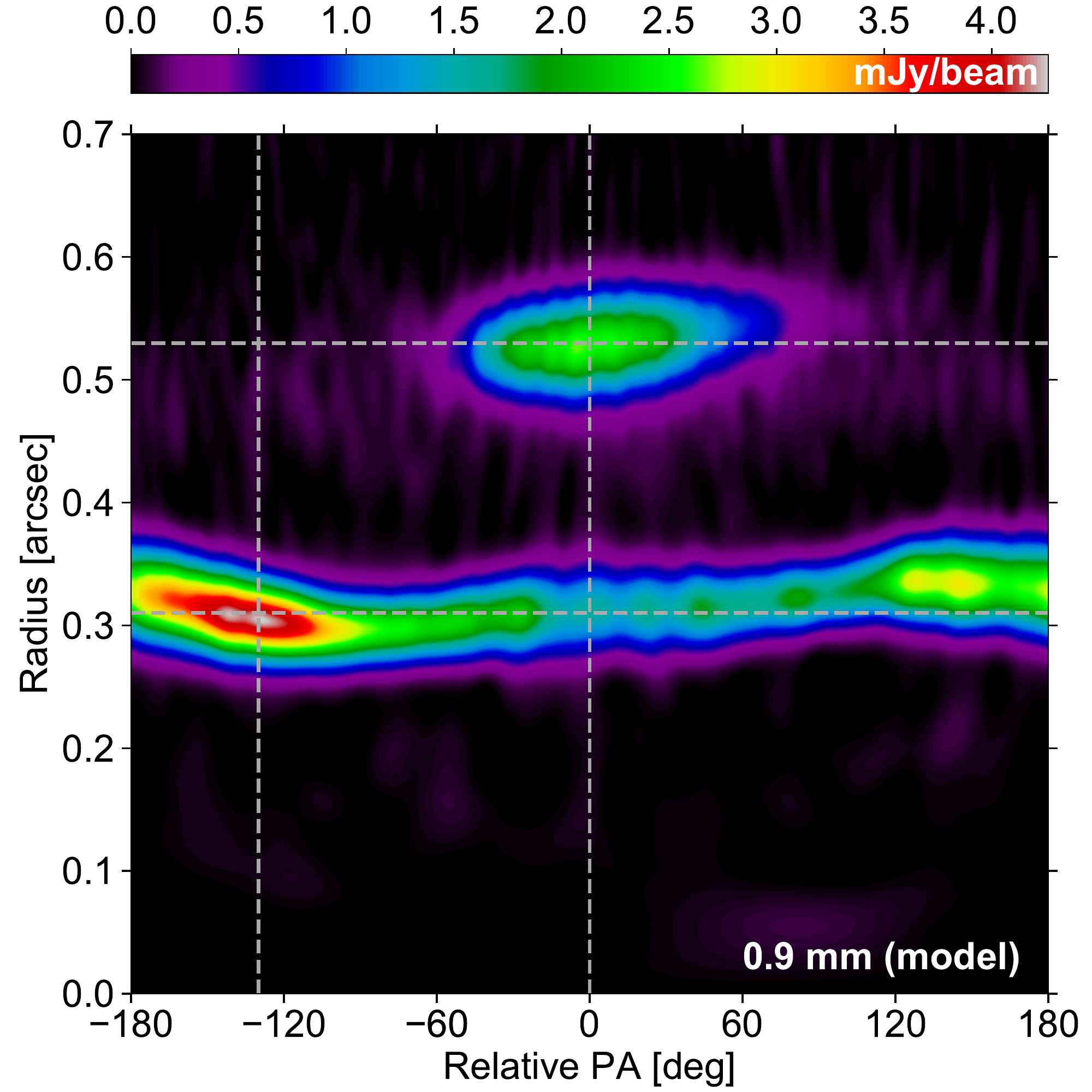}
\includegraphics[width=0.49\textwidth]{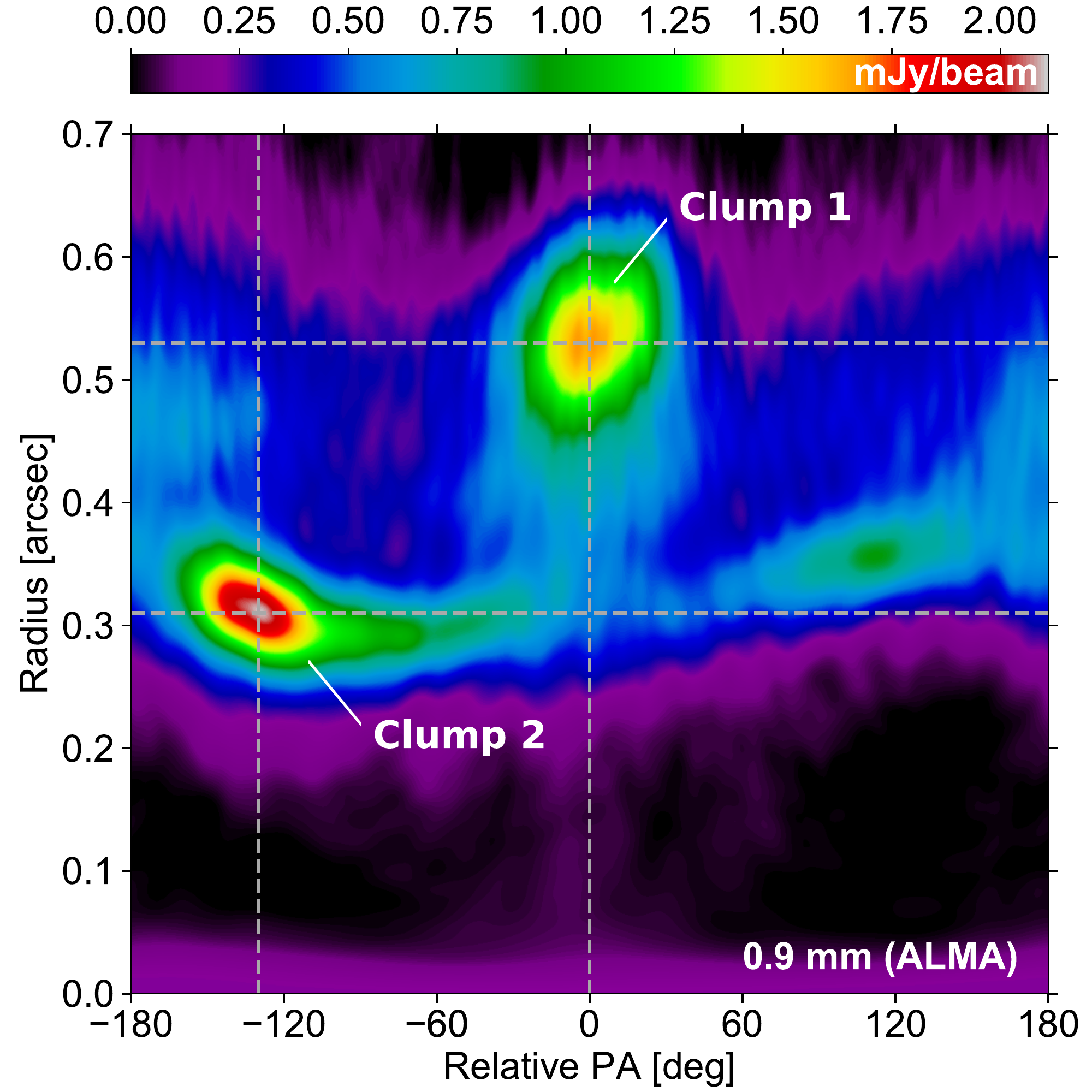}
\end{center}
\caption{Predicted continuum emission at 0.9 mm with white noise (left
  panels) compared with the ALMA band 7 observation of \citet[][right
  panels, images in natural weights]{Dong18}. {\bf Upper panels}:
  projected maps. The $x$- and $y$-axes indicate the offset from the
  stellar position in the right ascension (RA) and declination (Dec)
  in arcseconds, i.e., north is up and east is to the left. The beam
  ($0\farcs052\times0\farcs042$ PA -7.1$^\circ$) is shown by the
  ellipse in the bottom-left corner in each panel, and the star by a
  plus symbol.  {\bf Lower panels}: deprojected maps (assuming a disc
  inclination of $21^{\circ}$ and a position angle of
  $62^{\circ}$). The $y$-axis shows orbital radius in arcseconds, the
  $x$-axis the position angle in degrees relative to that of Clump
  1. The dashed curves mark the position angle (relative to Clump 1)
  and orbital radius of each clump in the ALMA image.}
\label{fig:mapsALMA}
\end{figure*}

\begin{figure*}
\begin{center}
\includegraphics[width=0.329\textwidth]{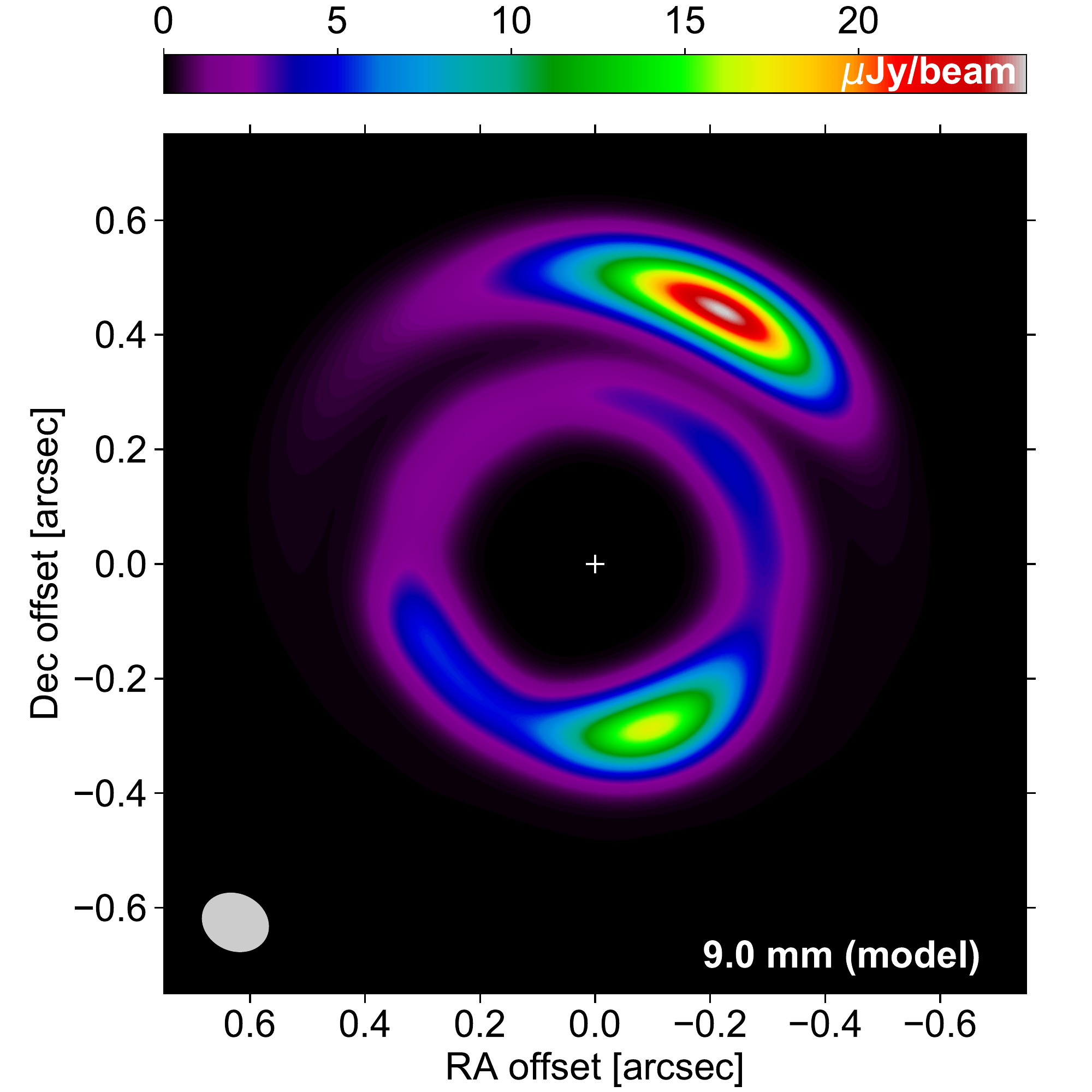}
\includegraphics[width=0.329\textwidth]{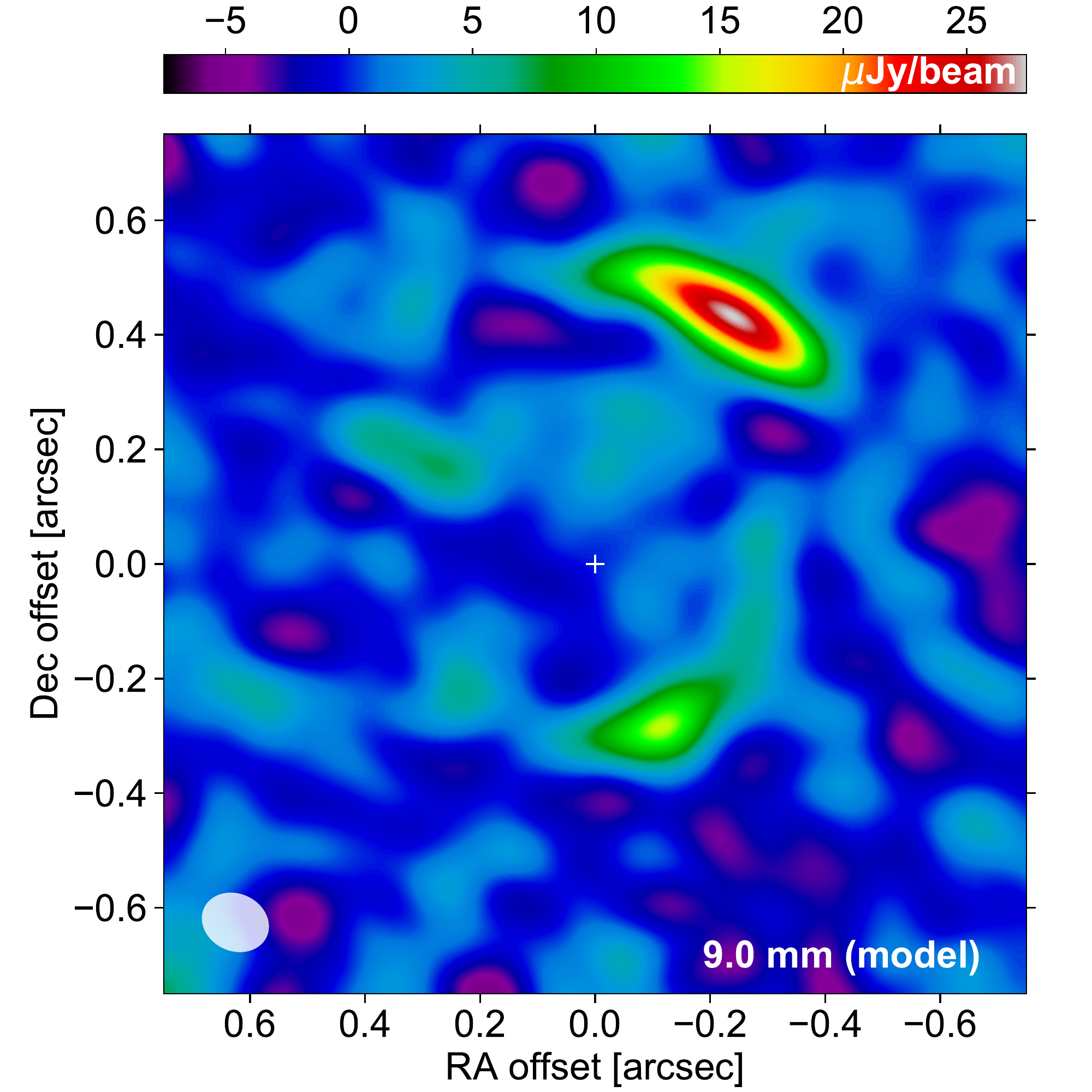}
\includegraphics[width=0.329\textwidth]{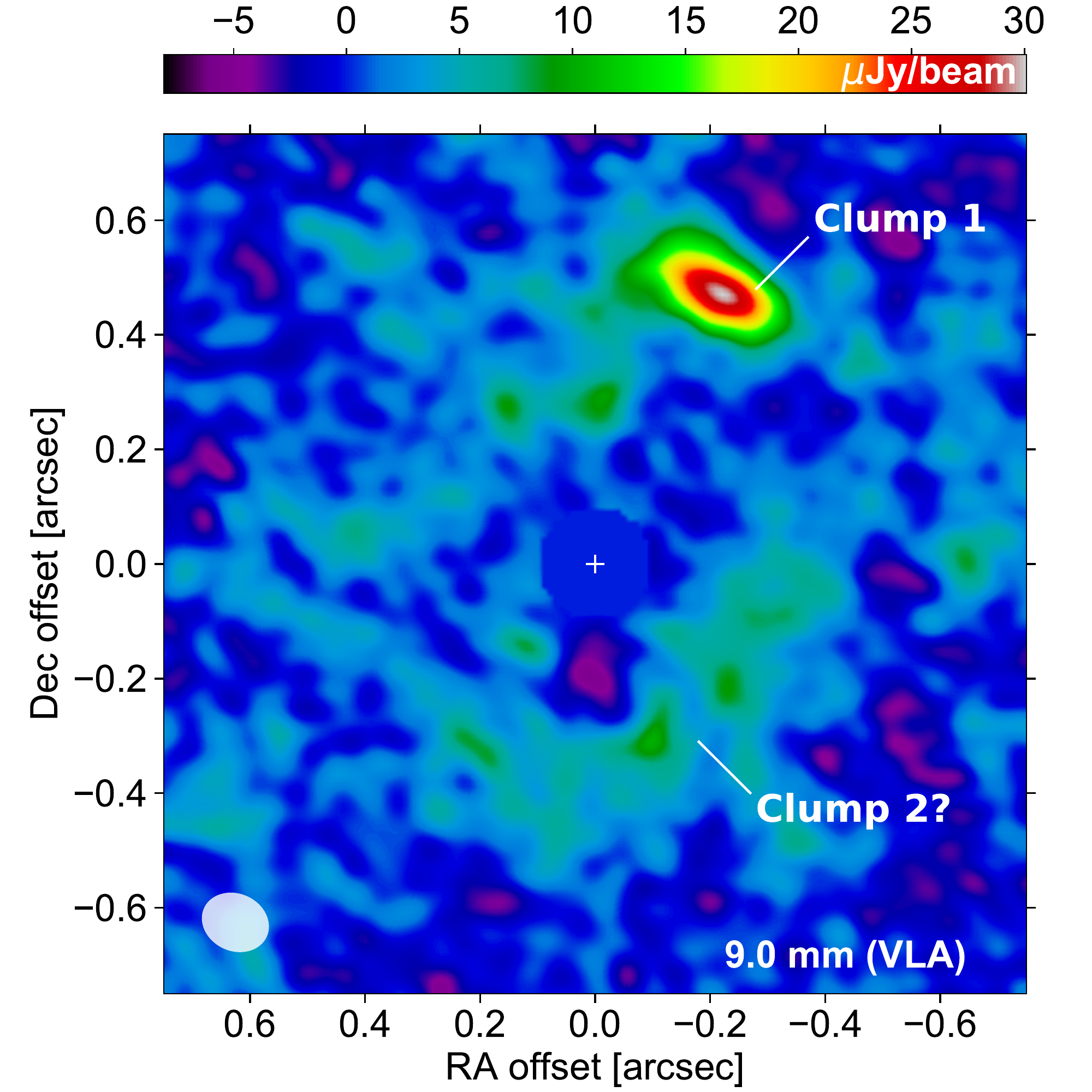}
\includegraphics[width=0.329\textwidth]{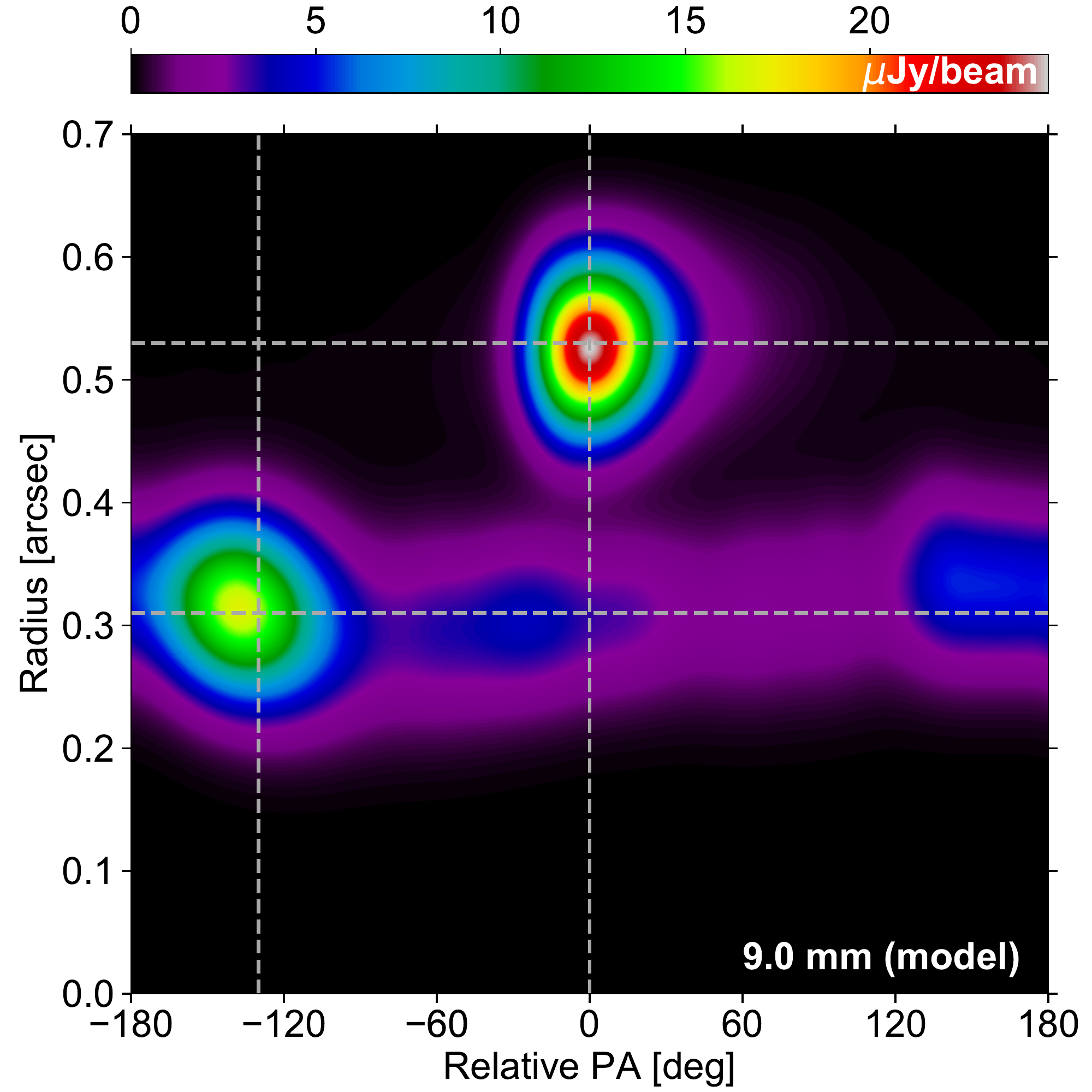}
\includegraphics[width=0.329\textwidth]{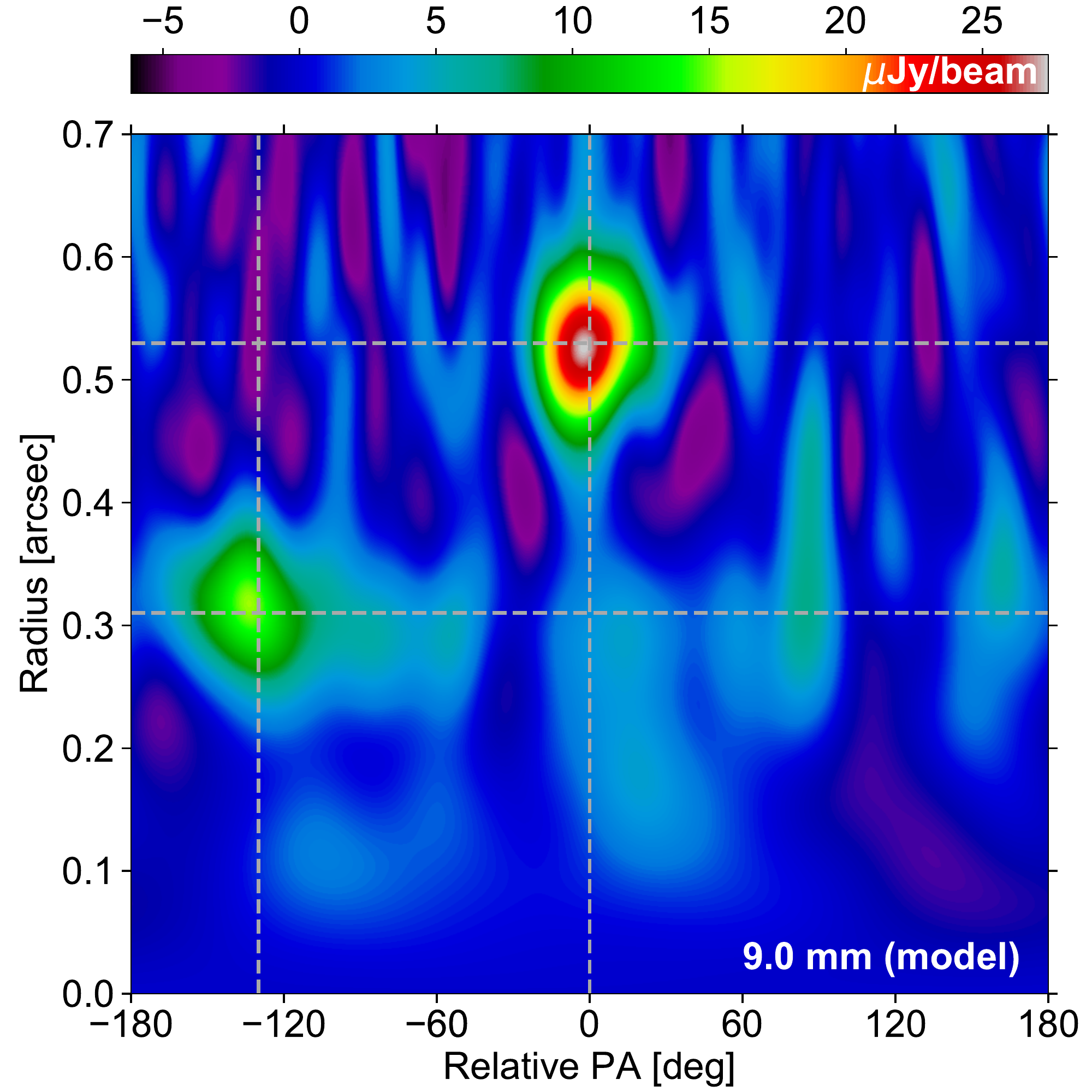}
\includegraphics[width=0.329\textwidth]{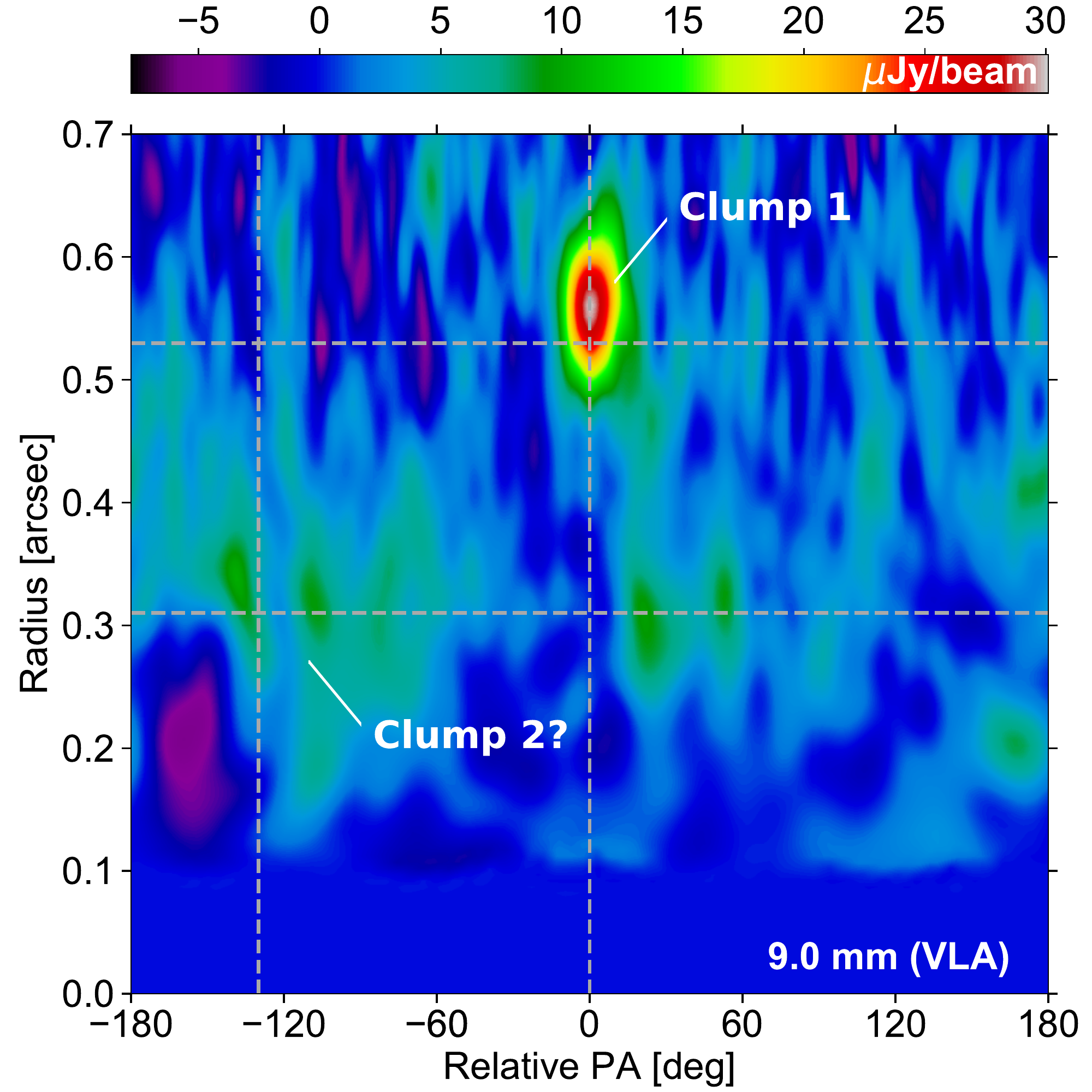}
\end{center}
\caption{Predicted continuum emission at 9 mm without and with white
  noise (left and middle panels, respectively) compared with the VLA
  observation of \citet[][right panels]{Casassus2019}. The star has
  been subtracted in the VLA images. {\bf Upper panels}: projected
  maps (as in the upper panels of Fig.~\ref{fig:mapsALMA}). The beam
  ($0\farcs12\times0\farcs10$ PA 65$^\circ$) is shown by the ellipse
  in the bottom-left corner in each panel, and the star by a plus
  symbol. {\bf Lower panels}: deprojected maps (as in the lower panels
  of Fig.~\ref{fig:mapsALMA}). The dashed curves mark the position
  angle (relative to Clump 1) and orbital radius of each clump in the
  ALMA image shown in Fig.~\ref{fig:mapsALMA}.}
\label{fig:mapsVLA}
\end{figure*}

A side-by-side comparison of our synthetic maps of continuum emission
with the ALMA image of \citet[][$\lambda\approx0.9$ mm]{Dong18} and
the VLA image of \citet[][$\lambda\approx9$ mm]{Casassus2019} is
displayed in Figs.~\ref{fig:mapsALMA} and~\ref{fig:mapsVLA},
respectively.  In each figure, the projected maps (flux maps in the
sky plane) are in the upper panels, and the deprojected maps (flux
maps in the disc plane) are in the lower panels.
Fig.~\ref{fig:mapsALMA} shows only the synthetic map at 0.9 mm with
white noise since its amplitude is very small (the predicted peak
intensity at Clump 2 is $\sim$200 times larger than the rms noise
level). Fig.~\ref{fig:mapsVLA}, however, displays the synthetic map at
9 mm without and with white noise since the predicted peak intensity
at Clump 1 is only $\sim$12 times the rms noise level. For comparison
with our synthetic maps at 9 mm, the star has been subtracted in the
VLA images shown in the right column of panels in
Fig.~\ref{fig:mapsVLA}.  We point out that the noise in the VLA
observation includes smaller scales than our synthetic map with white
noise. This is most likely due to the core of the VLA dirty beam not
being perfectly represented by an elliptical Gaussian.

Overall, for the assumed dust properties, our two-vortex model
captures the main features of the two clumps of emission observed in
the ALMA and VLA images. The dust-trapping vortex at the inner edge of
the outer planet's gap can reproduce the compact emission associated
with Clump 1 (at $\sim$1 o'clock in both the ALMA and VLA images). In
particular, we recover the fact that Clump 1 is azimuthally broader at
0.9 mm than at 9 mm, as expected with azimuthal dust trapping.  The
decaying vortex at the outer edge of the inner planet's gap reproduces
well the eccentric inner ring passing by Clump 2, with Clump 2 near
pericentre (at $\sim$5 o'clock in the ALMA image). It also forms a
secondary clump of emission near apocentre which could account for the
emission seen to the east of the ring in the ALMA image (at $\sim$9
o'clock). And importantly, we recover the more diffuse and rather low
level of emission seen in the VLA image for Clump 2. By comparing the
synthetic deprojected maps in Figs.~\ref{fig:mapsALMA}
and~\ref{fig:mapsVLA} with Fig.~\ref{fig:hydro2}, it is easy to see
that the location of the maxima in the synthetic maps corresponds to
that of the largest dust particles in the simulation (those ranging
from a few mm to a cm in size).

Comparison between the synthetic and observed flux maps at 0.9 mm
further indicates that Clump 1 and Clump 2 lie on top of a fainter
ring of background emission. This background emission could trace a
rather massive population of small dust well coupled to the gas. This
idea will be presented in Section~\ref{sec:impact_bin0} and
illustrated in Fig.~\ref{fig:mapsbin0}. We also note the presence in
our synthetic map at 0.9 mm of a faint ring of emission slightly
interior to Clump 1, at a level of emission ($\sim$0.1 mJy/beam) that
is close to the rms noise level adopted in the synthetic map. This
faint ring of emission traces the dust ring situated between 0\farcs44
and 0\farcs50 in Fig.~\ref{fig:hydro2} (or, equivalently, between
$\sim$70 and 80 au in the middle-right panel of
Fig.~\ref{fig:hydro1}). As already stated in
Section~\ref{sec:hydro_dust}, this ring is reminiscent to the ring of
emission around 0\farcs43 in the ALMA observation of \citet{Dong18},
and which peaks at about 0.6 mJy/beam in the right panels of
Fig.~\ref{fig:mapsALMA}.

We also briefly comment that the inclusion of anisotropic scattering
in the radiative transfer calculations has a minor impact on our
(sub)millimetre continuum synthetic maps. At 0.9 mm, anisotropic
scattering increases the overall level of flux by only a few percent
compared to a radiative transfer calculation that only includes
thermal absorption. At 9 mm, anisotropic scattering increases the
overall level of flux by about 20\%.

In the following, we provide a more quantitative comparison between
our predictions and the observations, in terms of the peak
intensities, and the widths and aspect ratio of Clump 1. We also quote
the peak absorption optical depths in the synthetic maps.

\subsubsection{Peak intensities}
\label{sec:results_submm_peaks}
At Clump 1, the predicted peak intensity at 0.9 mm is $\sim$2.6
mJy/beam, which is $\sim$50\% larger than the observed value, while
the predicted peak intensity at 9 mm is $\sim$24.8 $\mu$Jy/beam
(without noise), which is close to the observed value of $29.1 \pm
2.0$ $\mu$Jy/beam. At Clump 2, the predicted peak intensity at 0.9 mm
is $\sim$4.2 mJy/beam (about twice the observed value), that at 9 mm
is $\sim$17.0 $\mu$Jy/beam (without noise), which is $\sim$1.6 times
larger than the observed value of $10.8 \pm 2.0$ $\mu$Jy/beam. Our model
thus tends to produce a little too much flux at Clump 2, although a
direct comparison is not trivial because of different shapes and
emitting areas for Clump 2 in the predictions and the observations.

The predicted peak intensity ratio between Clump 1 and Clump 2 is
about 0.6 at 0.9 mm and 1.4 at 9 mm, while the observed values are
about 0.8 at 0.9 mm and 2.7 $\pm$ 0.3 at 9 mm. We interpret the larger
peak intensity ratio at 9 mm as a consequence of the vortex decay that
leads to Clump 2 in our model. We have checked this by computing
synthetic flux maps at 611 orbits, when the inner vortex is still
active (see top panels in Fig.~\ref{fig:hydro1}), which show that the
peak intensity ratio takes very similar values at both wavelengths (it
is about 0.5 at 0.9 mm, and 0.6 at 9 mm).

We point out that in the ALMA band 7 observations of
\citet{Boehler2018}, for which the angular resolution is about twice
as large as in the ALMA band 7 observations of \citet{Dong18}, Clump 1
has a larger peak intensity than Clump 2. The peak intensity ratio
between Clump 1 and Clump 2 is about 1.6 in \citet{Boehler2018}, while
it is about 0.8 in \citet{Dong18}. By convolving our raw flux maps
with the same beam as \citet{Boehler2018}'s observations
($0\farcs11\times0\farcs08$ PA 38$^\circ$), we find a peak intensity
ratio between Clump 1 and Clump 2 of $\sim$0.85, thus larger than the
value of $\sim$0.6 that we get with the same resolution as in
\citet{Dong18}, but still smaller than 1. Our model cannot reproduce
the reversal in the peak intensity ratio between both aforementioned
angular resolutions.

We finally note that Clump 1 is offset by about 0\farcs03 in the ALMA
and VLA images, which is not reproduced in our models. On the
contrary, there is no significant azimuthal shift between the
positions of Clump 1 in the ALMA and VLA images, whereas the synthetic
images show an azimuthal shift by about 10$^{\circ}$, with the peak at
9 mm shifted clockwise (in the direction opposite to that of
rotation). A very similar azimuthal shift is actually predicted for
Clump 2. We stress that this azimuthal shift is not a systematic
prediction of our models: at some other outputs in the simulation, the
azimuthal shift is in the opposite direction (shift by
$\sim$-10$^{\circ}$), while at some other outputs we find no azimuthal
shift. The predicted azimuthal shift is likely related to Clump 1's
eccentricity.

\subsubsection{Widths and aspect ratio of Clump 1}
\label{sec:results_submm_clump1}
From the polar maps displayed in Fig.~\ref{fig:mapsALMA}
and~\ref{fig:mapsVLA}, we see that Clump 1 tends to have a slightly
larger azimuthal width and smaller radial width in the synthetic maps
than in the observations. In the azimuthal direction, the predicted
FWHM of the convolved intensity across the peak is about 45$^{\circ}$
and 19$^{\circ}$ at 0.9 mm and 9 mm, respectively, while the observed
values are about 28$^{\circ}$ and 12$^{\circ}$. Similarly, in the
radial direction, the predicted FWHM of the convolved intensity across
the peak is about 0\farcs04 and 0\farcs05 at 0.9 mm and 9 mm,
respectively, while the observed values are about 0\farcs08 and
0\farcs06.  A similar analysis can be done for Clump 2, with the same
conclusion that, in our model, Clump 2 tends to be a little narrower
in radius and somewhat broader in azimuth than in the ALMA observation
of \citet{Dong18}.  Fine tuning of the disc gas parameters (surface
density, temperature, alpha viscosity) and of the dust parameters
(size distribution, mass, internal density) could potentially
reconcile our predictions and the data. A full parameter search is
beyond the scope of this work.

Moreover, from the radial and azimuthal widths of Clump 1 in the
emission maps, and the radial location of the peak intensity, we can
estimate again the aspect ratio $\chi$ for Clump 1. We find that the
predicted $\chi$ values amount to $\sim$10.2 and $\sim$3.5 at 0.9 mm
and 9 mm, respectively, while the observed values are $\sim$3.2 and
$\sim$1.8 (with about 10\% relative uncertainties). These values are
much smaller than the one obtained via the gas perturbed vortensity in
the vortex from which Clump 1 originates (see last paragraph in
Section~\ref{sec:hydro_gas}).  We have checked that this discrepancy
comes about because of the beam convolution. Upon calculating the
synthetic maps of raw intensity (intensity of continuum emission prior
to beam convolution), we obtain $\chi \approx 16 \pm 2$ at both
wavelengths, which agrees well with the vortex's aspect ratio measured
via the gas vortensity, and also with the value reported in
\citet{Casassus2019} where steady-state dust trapping predictions
based on \citet{LyraLin2013} are used to model the raw intensity for
Clump 1 at 0.9 and 9 mm.

\subsubsection{Peak optical depths in absorption}
\label{sec:results_submm_tau}
An interesting information that is accessible with our synthetic maps
of continuum emission is the absorption optical depth that our model
predicts at the location of the peak intensities. At 0.9 mm, we find
maxima in the absorption optical depth of about 30 and 15 for Clump 1
and Clump 2, respectively. The continuum emission at 0.9 mm is
therefore very optically thick near the centre of both clumps. At 9
mm, the absorption optical depth peaks at about 0.9 and 0.4 for Clump
1 and Clump 2, respectively, which are perhaps surprisingly large
values at this wavelength.  The peak absorption optical depth at 9 mm
near the centre of Clump 1 is a bit larger than the value of $\sim$0.3
predicted by the steady-state dust trapping predictions used in
\citet{Casassus2019}, but note that the gas vortex and dust
distribution assumed to give rise to Clump 1 in \citet{Casassus2019}
have different physical parameters than in our simulation (different
alpha turbulent viscosity, temperature, dust's internal density and
opacity; see Section~3.3.2 in \citealp{Casassus2019} for comparison).

\subsection{Polarised scattered light in $Y$-band}
\label{sec:results_Yband}
\begin{figure*}
\begin{center}
\includegraphics[width=0.99\hsize]{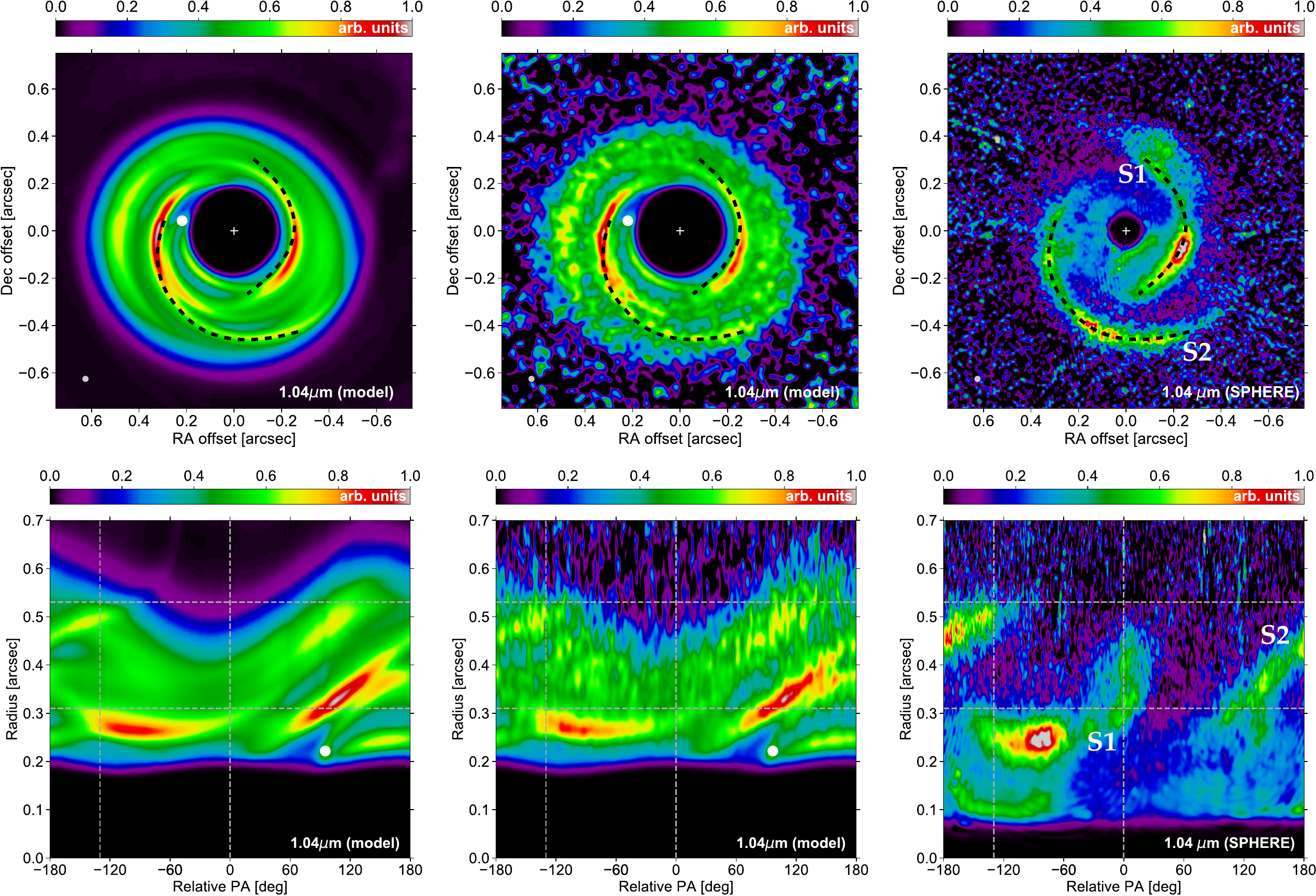}
\end{center}
\caption{Synthetic scattered light image at 1.04 $\mu$m without and
  with noise (left and middle panels, respectively) compared with the
  $Y$-band SPHERE image of \citet[][right panels]{Benisty2015}. In
  each image, the polarised intensity is scaled by the square of the
  deprojected distance from the central star, and normalised such that
  the intensity of the strongest pixel is 1. To highlight the spirals
  structure, a mask of 0\farcs2 radius is applied to the synthetic
  images. {\bf Upper panels}: projected maps (as in the upper panels
  of Fig.~\ref{fig:mapsALMA}). The beam ($0\farcs026\times0\farcs026$)
  is shown by the ellipse in the bottom-left corner in each panel, the
  star by a plus symbol, and the inner planet by a white circle. The
  outer planet is outside the image domain (at 0\farcs87, PA
  -60$^{\circ}$).  The dashed curves in all three panels are used to
  compare the synthetic and observed asymmetries, but they are not
  fits to the observed spiral traces. {\bf Lower panels}: deprojected
  maps (as in the lower panels of Fig.~\ref{fig:mapsALMA}). The
  position of the inner planet is marked by a white circle in the
  synthetic images. The dashed curves mark the position angle
  (relative to Clump 1) and orbital radius of each clump in the ALMA
  image shown in Fig.~\ref{fig:mapsALMA}.}
\label{fig:mapsPI}
\end{figure*}

A side-by-side comparison between our synthetic map of polarised
scattered light with the $Y$-band ($\lambda = 1.04$ $\mu$m) scattered
light image of \citet{Benisty2015} is displayed in
Fig.~\ref{fig:mapsPI}.  The projected maps are in the upper panels,
and the deprojected maps in the lower panels. The synthetic maps in
the first column of panels do not include noise, those in the second
column include white noise with the method described in
Section~\ref{sec:PIsetup}.  All images are multiplied by the square of
the deprojected distance from the central star, and normalised such
that the intensity of the strongest pixel is 1.  To highlight the
asymmetric structures in our synthetic map, a mask of $0\farcs2$
radius is applied. Recall that the dust density within 0\farcs15 has
been truncated, and that beyond 0\farcs35 has been reduced (see
Section~\ref{sec:PIsetup}). The same dashed curves are superimposed in
the upper panels to compare the predicted and observed asymmetries;
however, we stress that they are not fits to the observed spiral-like
features -- they are just drawn to guide the eye.

Overall, we see that our model predicts several spiral arms, but two
are more prominent. Just like the spirals in the gas surface density
of the simulation (see Figs.~\ref{fig:hydro1} and~\ref{fig:hydro2}),
the spirals in the synthetic scattered light image do not all have a
clear origin, as they trace spiral density waves due to either the
planets, the vortices, or a combination thereof. Some insight into the
origin of the spirals and/or asymmetries in the synthetic image can be
gained by comparing the deprojected maps with the gas surface density
contours in Fig.~\ref{fig:hydro2}.  Interestingly, we see that our
synthetic map can reproduce both the location and the winding of the
observed spiral arm to the south-east very well (see the dashed curve
denoted by S2 in Fig.~\ref{fig:mapsPI}).  According to our model, this
spiral arm would be the (primary) inner wake of the outer planet,
which propagates from the outer planet's position. Our synthetic image
shows a second prominent spiral to the west, which corresponds to a
secondary spiral density wave induced by the outer planet. This spiral
nearly coincides with the bright concentric arc to the west of the
SPHERE image (see the lower part of the dashed curve denoted by S1 in
Fig.~\ref{fig:mapsPI}). The same spiral is (mainly) responsible for
the emission to the east of the synthetic map (near 9 o'clock beyond
0\farcs4), but this emission is not observed in the SPHERE image. The
arc-shaped peak slightly inside of Clump 1 does not have a clear
counterpart in our synthetic map.

The synthetic $Y$-band polarized light images in \citet{Dong2015} and
\citet{Dong18}, which are obtained from 3D hydrodynamical simulations,
had previously shown that the two inner wakes of a planet several
times the mass of Jupiter could qualitatively account for the two
spiral arms in the SPHERE image of MWC 758. Note, however, that the
planet that they considered was located at $\sim$0\farcs6 (100 au),
which is quite close to Clump 1 in the (sub)millimetre images
($\sim$85 au), and it seems difficult for such a massive planet to
form a dust-trapping vortex at this short separation.

We finally stress that a realistic energy equation and
three-dimensional effects, which are not taken into account in our
hydrodynamical simulations, could well affect the way spirals would
look like in polarised intensity images, in particular the appearance
of the secondary wake induced by the outer planet \citep{ZhuDong2015,
  Fung2015, DongFung2017}.  For this reason we will not press the
comparison between our synthetic polarised intensity image and the
observed image too far. It could well be that the reproduction of the
S2 spiral is actually coincidental.

\section{Discussion and summary}
\label{sec:conclusion}

\subsection{Importance of gas self-gravity}
\label{sec:impact_sg}
The simulations carried out in this study include gas self-gravity,
which might seem unnecessary since the (azimuthally-averaged) Toomre
$Q$-parameter remains much larger than unity in our disc model. $Q$
reaches a local minimum of about 30 at the inner edge of the outer
planet's gap, near 85 au (Clump 1's location).  However, as shown in
previous studies
\citep{LP11a,Lin2012planetvortex,lovelace2013,Zhu2016,Regaly2017}, gas
self-gravity significantly weakens large-scale RWI vortices (with
azimuthal wavenumber $m=1$) when the product of $Q$ and $h$ becomes
smaller than $\sim$$\pi/2$ at the vortex location. Given our uniform
aspect ratio $h = 0.088$, vortices become significantly impacted by
self-gravity for $Q \lesssim 18$. Had we taken a larger initial
surface density for the gas in our simulations, the vortex formed at
$\sim$85 au would have been weaker and would have therefore decayed
earlier. Said differently, with gas self-gravity included, a rather
low surface density for the gas at the vortex location is necessary
for the dust trap to match the high concentration of dust grains
inside the vortex core, and consequently the compactness of Clump 1
seen in the ALMA and VLA observations.  This rather low surface
density is consistent with that estimated by \citet{Boehler2018} based
on ALMA band 7 observations of $^{13}$CO and C$^{18}$O.

\subsection{Vortex decay}
\label{sec:decay}
In this study, we present a scenario where the asymmetric eccentric
ring and the compact crescent seen in the ALMA band 7 continuum data
of the MWC 758 disc are due to planet-induced vortices. To explain the
low and diffuse signal obtained with the VLA at the location of the
ring, we propose that this ring results from a decaying vortex. In our
simulations, vortex decay is mediated by viscous diffusion ($\alpha =
10^{-4}$), and is found to coincide with a moderate increase in the
gas eccentricity between the planets (see
Section.~\ref{sec:hydro_gas}).  Notwithstanding this moderate
eccentricity of order 0.1, the lifetime of the vortex that gives rise
to the asymmetric ring, about a thousand orbits of the inner planet,
is consistent with previous 2D simulations of planet-induced vortices
for similar disc parameters and planet masses
\citep[e.g.,][]{Fu2014}. A smaller turbulent viscosity would increase
the lifetime of the vortices in our simulations. However, as we have
checked with preliminary simulations, it would also cause a higher
concentration of the large dust particles in the vortices, and would
therefore increase the compactness of the (sub)millimetre emission at
both dust traps in a way that is not consistent with neither the ALMA
nor the VLA observations. On the other hand, a larger turbulent
viscosity would shorten the lifetime of the vortices, thereby making
the dust's azimuthal trapping scenario for Clump 1 less likely.
Furthermore, several factors other than the disc's turbulent viscosity
affect the growth and decay timescales of planet-induced vortices in
2D viscous disc models, including the timescale for planetary growth
\citep{Hammer17,Hammer19}, which is taken to be the same for both
planets in our model, a non-isothermal energy equation
\citep{LesLin15}, gas self-gravity \citep{LP11a}, or dust feedback on
the gas \citep{Fu2014feedback}.

About dust feedback, we stress that for the total mass and size
distribution of the dust adopted in our radiative transfer
calculations, the dust-to-gas mass ratio in the vortices that
correspond to Clump 1 and Clump 2 does not exceed $\sim$0.1. This
rather low value implies that dust feedback should have a
small-to-moderate impact on the vortices if it were included in our
hydrodynamical simulations \citep[see, e.g.,][who showed that vortices
could be destroyed for dust-to-gas mass ratios $\geq$ 0.3-0.5 within
the vortices]{CrnkovicZhuStone}.

The level of viscosity adopted in our simulations is meant to model
the effects of MHD turbulence in the outer regions of a protoplanetary
disc, where non-ideal MHD effects, in particular ambipolar diffusion,
should play an important role.  Modelling turbulence as a diffusion
process is uncertain when the typical length scale of interest (here,
that of the gas vortex) is of the order of the pressure scale
height. \citet{ZhuStone2014} have investigated growth of and dust
trapping in planet-induced vortices via global 3D MHD simulations. One
of their simulations including ambipolar diffusion shows that the
vortex formed at the outer edge of the gap carved by a 9 Jupiter-mass
planet decays in about 1000 orbital timescales, with a similar
lifetime found in a 2D hydrodynamical simulation using an equivalent
turbulent viscosity. This result gives credit to the vortex's decay
timescales obtained in 2D viscous disc simulations.

MHD turbulence put aside, the growth and persistence of vortices in 3D
disc models have been intensively examined over the last decade or
so. We mention here a few results that are relevant to our
study. \citet{LesurPapa09} have found that anti-cyclonic vortices
could be unstable against the elliptic instability, a parametric
instability that mainly occurs at small scales, and which is stronger
for vortices with small aspect ratios ($\chi \lesssim 4$). The global
3D simulations of \citet{Meheut2012} have shown that a vortex formed
by the Rossby-Wave Instability developed meridional circulation and
could survive over hundreds of dynamical timescales before decaying,
most probably because of the elliptic instability. Regarding
planet-induced vortices in 3D, the impact of gas self-gravity has been
examined by \citet{Lin2012planetvortex}, essentially recovering the
predictions of 2D simulations, and the impact of a layered disc
structure (through a viscosity increasing with height from the
midplane) has been tackled by \citet{Lin2014}, who found that a high
($\alpha \sim 10^{-2}$) viscosity in the disc's upper layers could
largely decrease the vortex's lifetime, despite a modest viscosity
($\alpha \sim 10^{-4}$) in the disc midplane. About gas self-gravity,
we also highlight the recent work of \citet{LinPierens18}, who have
shown with 3D shearing box simulations that gas self-gravity could
prevent the decay of 3D vortices against elliptic instabilities, which
would otherwise destroy them in non self-gravitating discs.

To our knowledge, the impact of the elliptic instability on the growth
and survival of planet-induced vortices has not been investigated
yet. This would probably require 3D global, vertically-stratified,
high-resolution simulations of planet-disc interactions. It would also
be of interest to examine how the elliptic instability behaves in the
presence of other sources of turbulence, such as non-ideal MHD
turbulence.

We finally discuss to what extent the proposed vortex decay scenario
is associated with a transient phenomenon. To interpret the ALMA and
VLA observations of MWC 758, the inner vortex is required to have
decayed significantly (but with the dust retaining some degree of
non-axisymmetry), while the outer vortex has not yet significantly
decayed. Given the planets mass and disc parameters in our model, it
takes $\sim$0.17 Myr after the planets have reached their final mass
for the inner vortex to decay, and $\sim$0.30 Myr for the outer
vortex. Both timescales look short in comparison to the estimated age
of the system (5 to 10\% of the age), and one may argue that we are
catching the system at a special time. Interestingly, this ratio is
actually consistent with the fraction of discs with (sub)millimetre
continuum annular substructures that have high-contrast azimuthal
asymmetries like in the MWC 758 disc (7 out of 43, or $\sim$16\%; see
\citealp{DSHARP2}.)  However, one should bear in mind that it takes
time to grow the planets in the first place, and it is unclear how
long it should take to form a 1.5 Jupiter-mass planet at 35 au and a 5
Jupiter-mass planet at 140 around an A5-type star.  The lifetime of
the vortices in our model is also uncertain, and it would be
interesting to investigate how slow planetary growth, over a
significant fraction of the disc age, would change our scenario, since
it has been shown that slow growth tends to produce weaker vortices
\citep{Hammer17, Hammer19}. Similarly, the planets in the simulation
have been kept on circular orbits, and it is therefore relevant to
explore how migration would affect the growth and survival of the
vortices, and the eccentricity of the disc gas between the planets.

\subsection{Impact of dust parameters}
We discuss here how the size distribution, the total mass, the
internal density and the opacities of the dust particles affect the
results of our simulations and of our (sub)millimetre continuum
synthetic maps (Sections~\ref{sec:impact_sizedistribution}
to~\ref{sec:impact_opacities}). We then present in
Section~\ref{sec:impact_bin0} (sub)millimetre continuum synthetic maps
which include a population of small dust between the planets.

\subsubsection{Dust's size distribution}
\label{sec:impact_sizedistribution}
Throughout this work, we assume that the size of the dust particles
follows a power-law distribution, with thus three parameters: the
minimum and maximum sizes, and the power-law exponent. We recall that
the results presented in Section~\ref{sec:results_submm} are for a
size distribution $n(s) \propto s^{-3}$ for particles sizes $s$
between 10 $\mu$m and 1 cm. Decreasing the minimum size is found to
have no impact on our synthetic maps in the (sub)millimetre, as
particles smaller than 10 $\mu$m have a very small contribution to the
continuum emission at both 0.9 mm and 9 mm. A notable exception will
be presented in Section~\ref{sec:impact_bin0}, where we relax the
assumption of a power-law size distribution by adding a rather massive
population of small dust.

\begin{figure}
\begin{center}
\includegraphics[width=0.99\hsize]{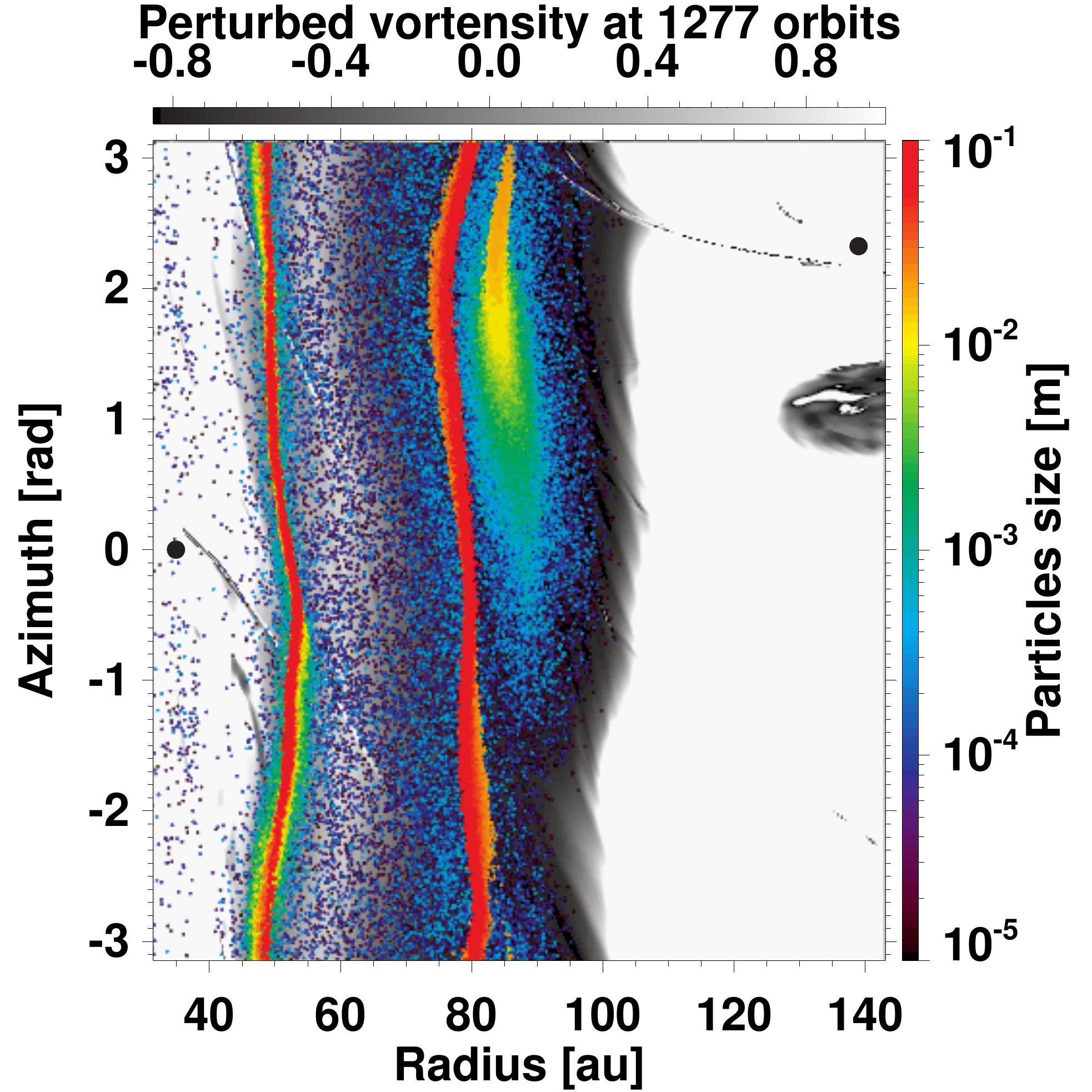}
\end{center}
\caption{Same as middle-right panel of Fig.~\ref{fig:hydro1}, but
  including dust particles between 1 cm and 10 cm in size.}
\label{fig:sizepc}
\end{figure}

The maximum size assumed for the dust particles has, however, a much
higher impact on our results. To see why this is the case, we display
in Fig.~\ref{fig:sizepc} the dust's spatial distribution at 1277
orbits overlaid on the gas perturbed vortensity, just like in the
middle-right panel of Fig.~\ref{fig:hydro1}, except that we now show
the location of dust particles up to 10 cm in size (instead of 1
cm). We see that particles larger than about 1.5 cm are not trapped in
the vortex at $\sim$85 au, but are shifted radially inwards by about 5
to 10 au at this time in the simulation. This radial shift stems from
the interaction between the large dust particles and the inner wake of
the outer planet. Since this wake carries negative fluxes of energy
and angular momentum, it pushes dust particles inwards each time they
cross the wake. Fig.~\ref{fig:sizepc} shows that, at the vortex's
orbital radius, the radial deflection caused by the planet wake is
larger than the radial drift back towards the vortex for dust
particles larger than about 1 cm, which have a Stokes number $\gtrsim$
0.1.  These particles thus find an equilibrium location interior to
the vortex, where they form a nearly axisymmetric ring. Investigation
of dust-wake interactions and their observational implications for the
continuum emission in the (sub)millimetre will also be detailed in a
forthcoming paper (Baruteau et al., in prep.). For the present work
dedicated to MWC 758, we have checked that this ring of $\gtrsim$
cm-sized dust particles would make the (sub)millimetre emission near
Clump 1 much more extended in the azimuthal direction than predicted
in Figs.~\ref{fig:mapsALMA} and~\ref{fig:mapsVLA} for the same total
dust mass, which would be incompatible with the compactness of Clump 1
as observed by ALMA and especially the VLA. This could have
interesting implications for dust growth in the MWC 758 disc, as our
model suggests inefficient growth beyond cm sizes around the location
of Clump 1, which we speculate might be due to bouncing and/or
fragmentation barriers.

Lastly, we have checked that the slope of the dust's size distribution
has, overall, a mild impact on the predicted (sub)millimetre emission
for Clump 1 and Clump 2. Qualitatively, we find that the shallower the
size distribution, the more compact the emission is for Clump 1 at 0.9
mm as well as at 9 mm.

\subsubsection{Dust's total mass}
\label{sec:impact_totalmass}
For the dust mass adopted in Section~\ref{sec:results} ($1.6\times
10^{-4} M_{\star}$ between $\sim$40 au and $\sim$100 au), the peak
absorption optical depth at 0.9 mm is $\sim$30 for Clump 1 and
$\sim$15 for Clump 2 (see Section~\ref{sec:results_submm_tau}). We
have checked that increasing the total dust mass would result in even
larger absorption optical depths at both clumps, and thus more
extended emissions in the azimuthal direction, which would be
incompatible with the observed compactness of both clumps at 0.9 mm.
Increasing the total dust mass too much would also conflict with the
assumption that dust feedback has been discarded in the hydrodynamical
simulation.

Decreasing the total dust mass would reduce the peak intensity of
Clump 2 faster than for Clump 1, thereby increasing the peak intensity
ratio between Clump 1 and Clump 2 at both wavelengths. For instance,
by reducing the total dust mass used in Section~\ref{sec:results} by a
factor 2, the peak intensity ratio between Clump 1 and Clump 2
increases from about 0.6 to 0.7 at 0.9 mm (observed value is
$\sim$0.8), and from about 1.4 to 1.7 at 9 mm (observed value is
$\sim$2.7 $\pm$ 0.3). However, since the emission at 9 mm is only
marginally optically thick at both clumps, decreasing the dust mass
would also decrease the overall flux level, which would reduce the
level of agreement between the predicted and observed peak intensities
for Clump 1 at 9 mm.  In the above example, halving the total dust
mass reduces the predicted peak intensity for Clump 1 at 9 mm from
$\sim$24.8 to $\sim$14.5 mJy/beam (observed value is 29.1 $\pm$ 2
mJy/beam).

\begin{figure*}
\begin{center}
\includegraphics[width=0.329\textwidth]{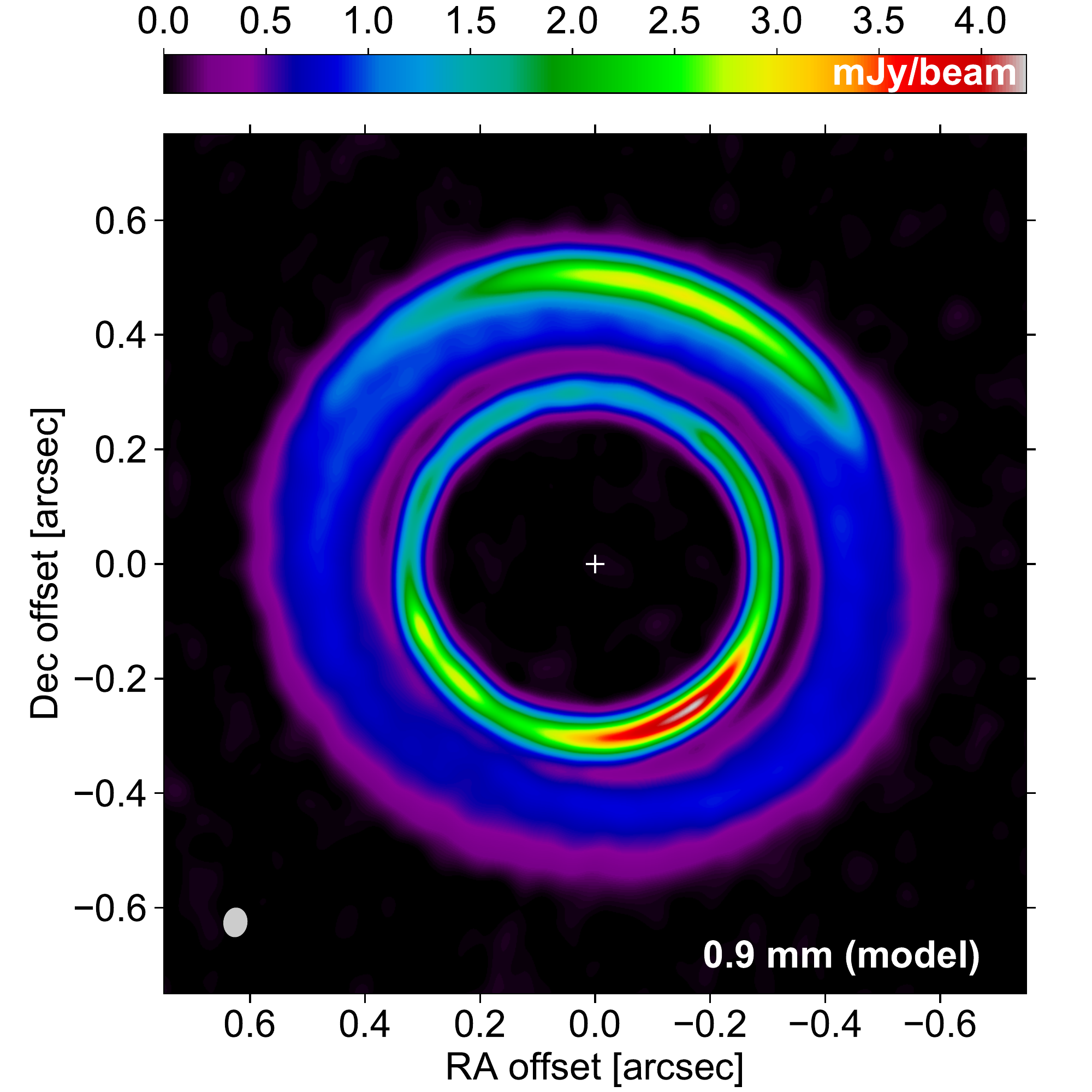}
\includegraphics[width=0.329\textwidth]{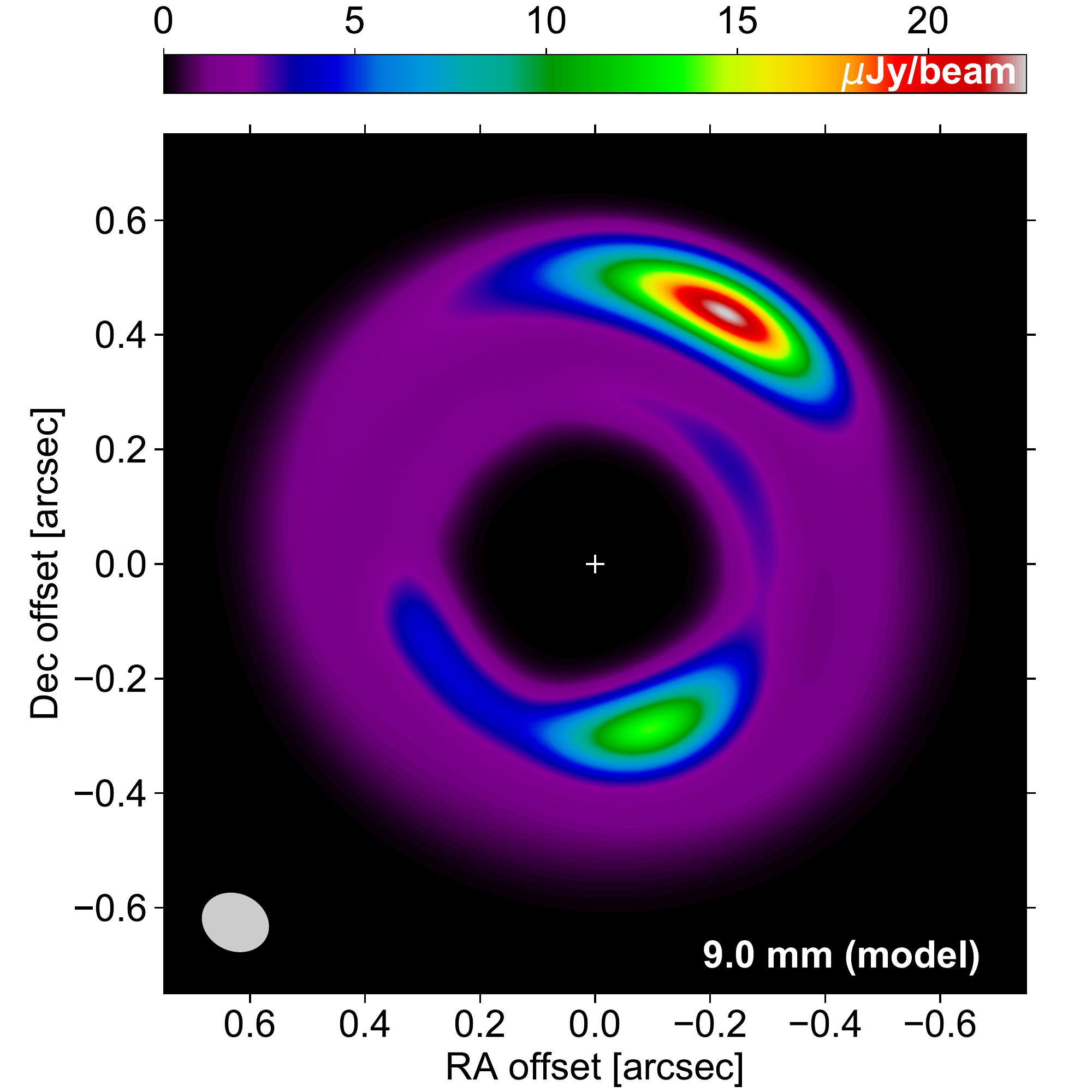}
\includegraphics[width=0.329\textwidth]{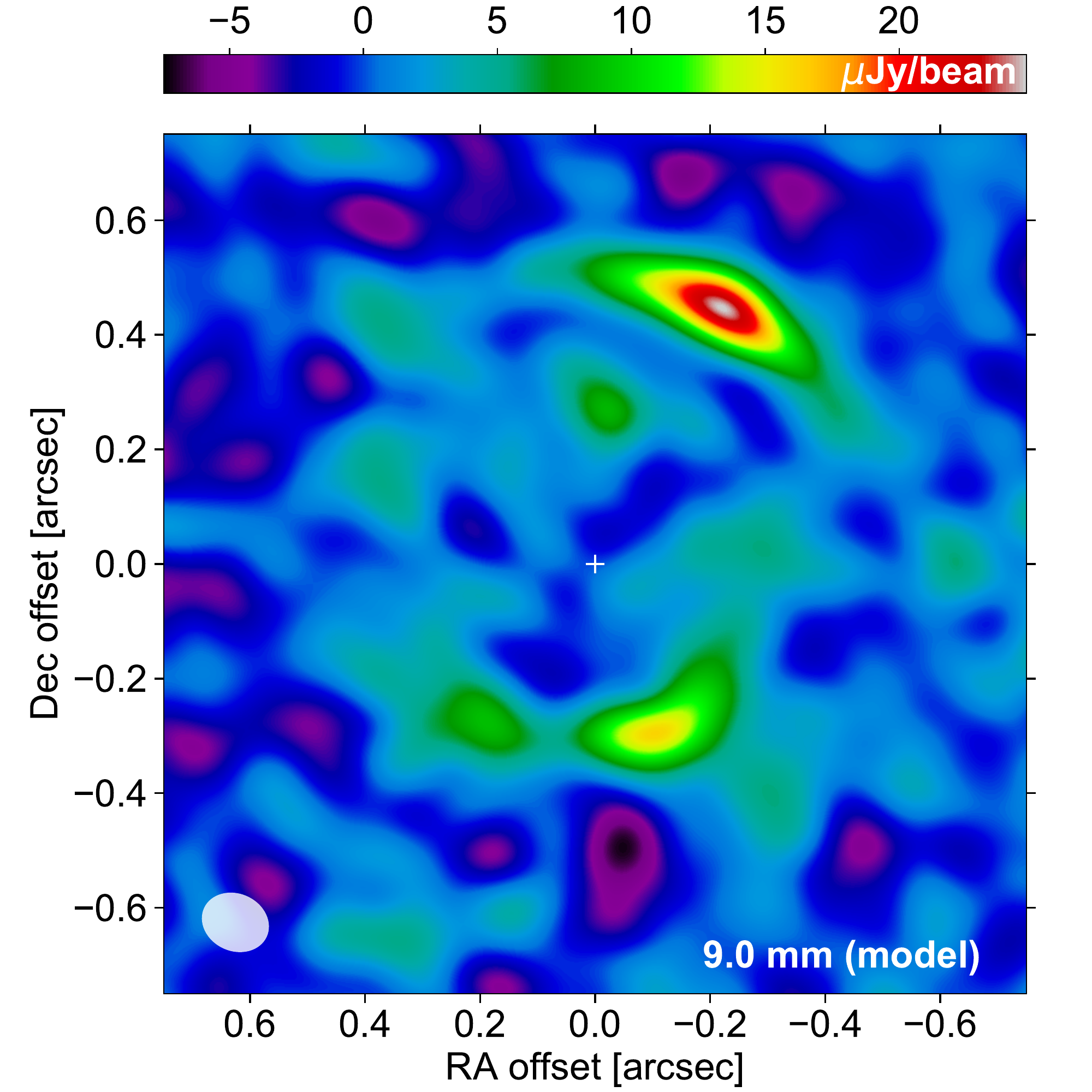}
\end{center}
\caption{Synthetic maps of continuum emission at 0.9 mm (left panel)
  and at 9 mm (without and with white noise, middle and right panels)
  obtained by adding a rather massive ($\sim$24 $M_{\oplus}$) population
  of small dust well coupled to the gas between Clumps 1 and 2.}
\label{fig:mapsbin0}
\end{figure*}

\subsubsection{Dust's internal density}
\label{sec:impact_internal_density}

The particles internal density adopted throughout this study,
$\rho_{\rm int} = 0.1$ g cm$^{-3}$, may seem a little surprising as
simulations and/or radiative transfer calculations usually model dust
as compact grains with an internal density of a few g
cm$^{-3}$. During the course of this project, we carried out a
simulation with a more conventional internal density of 1 g cm$^{-3}$,
and obtained synthetic maps in the (sub)millimetre that have nearly
the same morphology than for $\rho_{\rm int} = 0.1$ g cm$^{-3}$ {\it
  if} the maximum particle size is set to 1 mm instead of 1 cm. This
reduced maximum size comes about because the particles dynamics is
primarily\footnote{The particles dynamics is primarily set by the
  Stokes number because gas self-gravity also affects the dynamics,
  especially when the Stokes number exceeds about 0.1
  \citep{Baruteau2016}.} set by the Stokes number, which scales with
the product of particles size and internal density.  For $\rho_{\rm
  int} = 1$ g cm$^{-3}$, particles larger than about 1 mm are shifted
interior to the vortex located at $\sim$85 au because of the inner
wake of the outer planet, just like the particles larger than about 1
cm in our simulation with $\rho_{\rm int} = 0.1$ g cm$^{-3}$, as shown
in Section~\ref{sec:impact_sizedistribution}.

The main difference between both internal densities is that, in the
continuum emission at 0.9 mm, the peak intensity as well as the
azimuthal width are a little larger at $\rho_{\rm int} = 1$ g
cm$^{-3}$ for both clumps, but probably a similar level of agreement
could be attained by fine tuning the other parameters for the
dust. Our results therefore cannot rule out that the dust in the MWC
758 disc could be compact, given the degeneracy in the predictions,
and the large uncertainties in the dust opacities (at either internal
density, actually).  It would be interesting to explore the
possibility of even smaller internal densities \citep[see,
e.g.,][]{Kataoka13b}.

\subsubsection{Dust opacities}
\label{sec:impact_opacities}
The dust composition is another free parameter that we have tested for
the (sub)millimetre continuum synthetic maps. The dust composition
only impacts the calculation of the dust absorption and scattering
opacities. Recall that all the results in Section~\ref{sec:results}
are obtained for porous dust particles for which the solid component
comprises 70\% water ices and 30\% silicates. By testing a different
mixture, namely 70\% silicates and 30\% amorphous carbons, we find
that the overall level of flux at 0.9 mm is reduced by only
$\sim$15\%, and that at 9 mm by a factor $\sim$3. Again, given the
degeneracy of the predictions, we would find it difficult to constrain
the dust composition from the comparison between the synthetic maps of
continuum emission and the observations.

\subsubsection{Inclusion of small dust}
\label{sec:impact_bin0}
The synthetic and observed flux maps at 0.9 mm shown in
Fig.~\ref{fig:mapsALMA} indicate that Clump 1 and Clump 2 lie on top
of a fainter ring of background emission. As alluded to in
Section~\ref{sec:results_submm}, this emission might come about
because of small dust forming as a result of dust fragmentation
between the planets. The spiral shocks and the (moderately) eccentric
gas between the planets in our disc model
(Section~\ref{sec:hydro_gas}) {\it a priori} offer an environment ripe
for dust fragmentation. Since dust growth and fragmentation are not
implemented in our hydrodynamical code, we have tested this idea by
simply adding an extra bin of small dust in our radiative transfer
calculations. Since the absorption and scattering opacities of our
porous dust is largely independent of dust sizes below 10 $\mu$m, we
need not specify what size range this extra bin of small dust
precisely corresponds to. We will just assume these dust particles
remain larger than 0.3 $\mu$m, to distinguish them from the compact
dust particles that contribute to the near-infrared polarised
scattered light (Section~\ref{sec:PIsetup}).

The extra bin of small dust that is added to our synthetic maps of
continuum emission is assumed to have a spatial distribution that
follows that of the gas between 0\farcs4 and 0\farcs6, and a total
mass $\sim$0.6\% the mass of the gas (or 24 $M_{\oplus}$; the total
dust-to-gas mass ratio is therefore increased from 2\% to
$\sim$2.6\%). The results of this numerical experiment are displayed
in Fig.~\ref{fig:mapsbin0}, where we see a level of background
emission at 0.9 mm between Clumps 1 and 2 that resembles that in the
ALMA image of \citet{Dong18}. Interestingly, part of this emission has
a spiral shape towards the south-west, which is reminiscent of the
spiral- or arc-shaped emission seen to the south of Clump 2 in the
ALMA image of \citet{Dong18} (around 6 o'clock in the top-right panel
of Fig.~\ref{fig:mapsALMA}). In our disc model, this spiral, which
also has an observational counterpart in polarised intensity (see
spiral trace S2 in Fig.~\ref{fig:mapsPI}), corresponds to the primary
inner wake of the outer planet.  The inclusion of small dust has a
minor impact on the synthetic map at 9 mm (the peak intensities at
Clump 1 and Clump 2 are decreased by about 10\%).

\subsection{On the location and mass of the possible planets in the
  MWC 758 disc}
\label{sec:planets_mwc758}
\citet{Reggiani2018} have recently reported the detection of a
point-like source at about 20 au from MWC 758. This source, however,
has not been detected in H$_{\alpha}$ line observations with
SPHERE/ZIMPOL \citep{Huelamo18, Cugno19}. Assuming this source is a
companion candidate, we have investigated under which circumstances it
could produce the asymmetric ring at 0.9 mm passing by Clump 2 at
$\sim$50 au. With our disc model, we find that the mass of the
companion candidate should be close to $30 M_{\rm Jup}$ so that the
pressure maximum located at the outer edge of its gap matches the
orbital radius of Clump 2. While such a massive companion could
account for the large submillimetre cavity in the disc, it would most
likely open a very large gap or even a cavity in the gas which would
be devoid of the smallest dust grains. But such a large gap or cavity
is not seen in scattered light images of MWC 758, making this scenario
less likely. If the companion candidate at about 20 au were present in
addition to the planet at 35 au that we think is responsible for Clump
2, then both companions would be close to their 2:1 mean-motion
resonance and likely carving a common gap in the gas, which, again,
lacks observational support.

With a moderately massive inner planet in the submillimetre cavity,
the outer planet is required in our simulations to explain the
multiple spiral arms which are observed in polarised scattered light
(see Section~\ref{sec:results_Yband}). By adopting for the outer
planet's mass the upper mass limit obtained by \citet{Reggiani2018},
we find that the outer planet should be located at about 140 au to
produce the dust trap at the orbital distance of Clump 1 ($\sim$85
au). However, preliminary simulations have shown that a similar dust
trap could be achieved by adopting a more massive companion at a
larger orbital distance like, for instance, a $0.1 M_{\odot}$
substellar companion at $\sim$300 au from the central star.

Despite the fact that the massive outer planet in our scenario has not
been detected in the recent Keck $L'$-band observations of
\citet{Reggiani2018}, the detection limit was computed assuming a hot
start evolutionary model. Assuming a cold start model can dramatically
increase the mass detection limit for the same contrast
\citep{SpiegelBurrows2012}. Also, if the planet has a circumplanetary
disc or envelope, the material surrounding the planet could affect gas
accretion on to the planet and consequently the planet's luminosity
\citep{Szulagyi2014}, potentially hindering its direct observation. A
direct detection will be possible with future high spatial resolution
ALMA observations of molecular lines, where the strong impact that a
massive planet could have on the observed kinematics of the gas can be
identified \citep{Perez2015,Perez2018}. Kinematic evidence for a
massive planet in the HD 163296 disc based on $^{12}$CO ALMA
observations has been recently reported in \citet{Pinte2018}.

Based on multi-epoch near-infrared observations of the MWC 758 disc,
\citet{Ren2018} have estimated the rotation (or pattern) speed of the
two most prominent spirals. Assuming that these spirals are the inner
wakes of a massive companion, they find, based on the spirals pattern
speed, that the companion's best fit orbital distance would be at
about 95 au (0\farcs59). This is very close to the location of Clump 1
in the (sub)millimetre images ($\sim$85 au), and it seems unlikely
that a massive companion at about 95 au could form a dust-trapping
vortex at the location of Clump 1. However, we stress that the
uncertainty in the estimation of the pattern speed is large, as
highlighted in \citet{Ren2018}, and that there is actually no
constraint on the upper distance of the putative companion. The
$3\sigma$ lower limit on the pattern speed goes to 0 indeed
\citep[that is, no rotation;][]{Ren2018}, which would formally
correspond to a companion at infinite distance. The orbital distance
that we assume for the outermost planet in our model (140 au) is
therefore entirely consistent with the current measurement of the
spirals pattern speed.

\subsection{Summary}
We carried out 2D global hydrodynamical simulations including both gas
and dust of a self-gravitating circumstellar disc under the influence
of two giant planets. The results of our simulations, which were
post-processed with 3D dust radiative transfer calculations, support a
scenario where the asymmetries observed in the (sub)millimetre and
near-infrared scattered light of the transition disc MWC 758 could be
due to the presence of two massive planets molding the global
structure of the disc.

In this scenario, our model suggests the presence of a 5~$M_{\rm Jup}$
planet at 140~au and a $1.5 M_{\rm Jup}$ planet at 35~au. The outer
more massive planet triggers several spiral arms, of which two can
account for the brightest spirals or arcs seen in $Y$-band polarised
scattered light. It also forms a vortex at the inner edge of its gap
(at $\sim$85 au), where the dust concentration reproduces quite well
the compact crescent-shaped structure seen at $\sim$0\farcs53 in the
ALMA and VLA observations (Clump 1), if assuming moderately porous
dust particles, with an internal density of 0.1 g cm$^{-3}$, up to a
centimetre in size. Because it is less massive, the inner planet
produces dim spirals in scattered light, and it forms a vortex at the
outer edge of its gap (at $\sim$50 au) which decays due to the disc's
turbulent viscosity, as the gas between the planets become moderately
eccentric. This decay can explain why the eccentric and asymmetric
emission ring seen with ALMA at $\sim$0\farcs32 has a weak counterpart
in the VLA observations of \citet{Casassus2019}. This scenario of a
decaying vortex has been recently proposed by \citet{Fuente2017} to
explain multi-wavelength NOEMA observations of the lopsided emission
ring in the AB Aurigae transition disc.

We finally point out the striking similarities between the transition
discs around MWC 758, HD 135344B \citep{Marel16HD135} and V1247
Orionis \citep{Kraus2017}. Similarly to MWC 758, HD 135344B and V1247
Ori have a moderately asymmetric emission ring surrounded by a
crescent-shaped structure in ALMA band 7 continuum observations.  Both
discs also have at least one spiral arm seen in scattered light (see
\citealp{Garufi13HD135} for HD 135344B and \citealp{OhtaV1247PI} for
V1247 Ori). It is therefore tempting to suggest that the asymmetries
in the discs around HD 135344B and V1247 Ori could also result from
the presence of two massive planets, just like in the proposed
scenario for the MWC 758 disc.

\section*{Acknowledgments}
Numerical simulations were performed on the CALMIP Supercomputing
Centre of the University of Toulouse. M.B., S.P. and S.C.  acknowledge
support from the Millennium Science Initiative (Chilean Ministry of
Economy), through grant ``Nucleus P10-022-F''. M.B. acknowledges
CONICYT-PFCHA/Mag\'isterNacional/2017-22171601 and the support from
Departamento de Postgrado y Post\'itulo de la Vicerrector\'ia de
Asuntos Acad\'emicos, Universidad de Chile. This research was
partially supported by the supercomputing infrastructure of the NLHPC
(ECM-02).  S.C. and S.P. acknowledge financial support from FONDECYT
grants 1130949, 3140601 and 3150643, respectively. FD thanks the
European Research Council (ERC) for funding under the H2020 research
\& innovation programme (grant agreement \#740651 NewWorlds).  We
thank Jérémie Lasue, Romane Le Gal and Gaylor Wafflard-Fernandez for
helpful discussions, as well as our referee for constructive comments.

\bibliographystyle{mnras}

\begin{thebibliography}{}
\makeatletter
\relax
\def\mn@urlcharsother{\let\do\@makeother \do\$\do\&\do\#\do\^\do\_\do\%\do\~}
\def\mn@doi{\begingroup\mn@urlcharsother \@ifnextchar [ {\mn@doi@}
  {\mn@doi@[]}}
\def\mn@doi@[#1]#2{\def\@tempa{#1}\ifx\@tempa\@empty \href
  {http://dx.doi.org/#2} {doi:#2}\else \href {http://dx.doi.org/#2} {#1}\fi
  \endgroup}
\def\mn@eprint#1#2{\mn@eprint@#1:#2::\@nil}
\def\mn@eprint@arXiv#1{\href {http://arxiv.org/abs/#1} {{\tt arXiv:#1}}}
\def\mn@eprint@dblp#1{\href {http://dblp.uni-trier.de/rec/bibtex/#1.xml}
  {dblp:#1}}
\def\mn@eprint@#1:#2:#3:#4\@nil{\def\@tempa {#1}\def\@tempb {#2}\def\@tempc
  {#3}\ifx \@tempc \@empty \let \@tempc \@tempb \let \@tempb \@tempa \fi \ifx
  \@tempb \@empty \def\@tempb {arXiv}\fi \@ifundefined
  {mn@eprint@\@tempb}{\@tempb:\@tempc}{\expandafter \expandafter \csname
  mn@eprint@\@tempb\endcsname \expandafter{\@tempc}}}

\bibitem[\protect\citeauthoryear{{Allard}}{{Allard}}{2014}]{Allard14}
{Allard} F.,  2014, in {Booth} M.,  {Matthews} B.~C.,   {Graham} J.~R.,  eds,
  IAU Symposium Vol. 299, Exploring the Formation and Evolution of Planetary
  Systems. pp 271--272, \mn@doi{10.1017/S1743921313008545}

\bibitem[\protect\citeauthoryear{{Andrews}, {Wilner}, {Espaillat}, {Hughes},
  {Dullemond}, {McClure}, {Qi}  \& {Brown}}{{Andrews}
  et~al.}{2011}]{Andrews2011}
{Andrews} S.~M.,  {Wilner} D.~J.,  {Espaillat} C.,  {Hughes} A.~M.,
  {Dullemond} C.~P.,  {McClure} M.~K.,  {Qi} C.,   {Brown} J.~M.,  2011,
  \mn@doi [\apj] {10.1088/0004-637X/732/1/42}, \href
  {http://adsabs.harvard.edu/abs/2011ApJ...732...42A} {732, 42}

\bibitem[\protect\citeauthoryear{{Ataiee}, {Baruteau}, {Alibert}  \&
  {Benz}}{{Ataiee} et~al.}{2018}]{Ataiee18}
{Ataiee} S.,  {Baruteau} C.,  {Alibert} Y.,   {Benz} W.,  2018, \mn@doi [\aap]
  {10.1051/0004-6361/201732026}, \href
  {http://adsabs.harvard.edu/abs/2018A%26A...615A.110A} {615, A110}

\bibitem[\protect\citeauthoryear{{Avenhaus} et~al.,}{{Avenhaus}
  et~al.}{2017}]{Avenhaus2017}
{Avenhaus} H.,  et~al., 2017, \mn@doi [\aj] {10.3847/1538-3881/aa7560}, \href
  {http://cdsads.u-strasbg.fr/abs/2017AJ....154...33A} {154, 33}

\bibitem[\protect\citeauthoryear{{Bae}, {Zhu}  \& {Hartmann}}{{Bae}
  et~al.}{2016}]{Bae2016}
{Bae} J.,  {Zhu} Z.,   {Hartmann} L.,  2016, \mn@doi [\apj]
  {10.3847/0004-637X/819/2/134}, \href
  {http://adsabs.harvard.edu/abs/2016ApJ...819..134B} {819, 134}

\bibitem[\protect\citeauthoryear{{Baruteau} \& {Masset}}{{Baruteau} \&
  {Masset}}{2008a}]{BaruteauMasset2008a}
{Baruteau} C.,  {Masset} F.,  2008a, \mn@doi [\apj] {10.1086/523667}, \href
  {http://adsabs.harvard.edu/abs/2008ApJ...672.1054B} {672, 1054}

\bibitem[\protect\citeauthoryear{{Baruteau} \& {Masset}}{{Baruteau} \&
  {Masset}}{2008b}]{BaruteauMasset2008b}
{Baruteau} C.,  {Masset} F.,  2008b, \mn@doi [\apj] {10.1086/529487}, \href
  {http://adsabs.harvard.edu/abs/2008ApJ...678..483B} {678, 483}

\bibitem[\protect\citeauthoryear{{Baruteau} \& {Zhu}}{{Baruteau} \&
  {Zhu}}{2016}]{Baruteau2016}
{Baruteau} C.,  {Zhu} Z.,  2016, \mn@doi [\mnras] {10.1093/mnras/stv2527},
  \href {http://adsabs.harvard.edu/abs/2016MNRAS.458.3927B} {458, 3927}

\bibitem[\protect\citeauthoryear{{Benisty} et~al.,}{{Benisty}
  et~al.}{2015}]{Benisty2015}
{Benisty} M.,  et~al., 2015, \mn@doi [\aap] {10.1051/0004-6361/201526011},
  \href {http://adsabs.harvard.edu/abs/2015A\%26A...578L...6B} {578, L6}

\bibitem[\protect\citeauthoryear{{Bentley} et~al.,}{{Bentley}
  et~al.}{2016}]{Bentley16_67P}
{Bentley} M.~S.,  et~al., 2016, \mn@doi [\nat] {10.1038/nature19091}, \href
  {http://cdsads.u-strasbg.fr/abs/2016Natur.537...73B} {537, 73}

\bibitem[\protect\citeauthoryear{{Boehler} et~al.,}{{Boehler}
  et~al.}{2018}]{Boehler2018}
{Boehler} Y.,  et~al., 2018, \mn@doi [\apj] {10.3847/1538-4357/aaa19c}, \href
  {http://adsabs.harvard.edu/abs/2018ApJ...853..162B} {853, 162}

\bibitem[\protect\citeauthoryear{{Bohren} \& {Huffman}}{{Bohren} \&
  {Huffman}}{1983}]{Bohren1983}
{Bohren} C.~F.,  {Huffman} D.~R.,  1983, {Absorption and scattering of light by
  small particles}

\bibitem[\protect\citeauthoryear{{Casassus} et~al.,}{{Casassus}
  et~al.}{2019}]{Casassus2019}
{Casassus} S.,  et~al., 2019, \mn@doi [\mnras] {10.1093/mnras/sty3269}, \href
  {http://adsabs.harvard.edu/abs/2019MNRAS.483.3278C} {483, 3278}

\bibitem[\protect\citeauthoryear{{Charnoz}, {Fouchet}, {Aleon}  \&
  {Moreira}}{{Charnoz} et~al.}{2011}]{Charnoz2011}
{Charnoz} S.,  {Fouchet} L.,  {Aleon} J.,   {Moreira} M.,  2011, \mn@doi [\apj]
  {10.1088/0004-637X/737/1/33}, \href
  {http://adsabs.harvard.edu/abs/2011ApJ...737...33C} {737, 33}

\bibitem[\protect\citeauthoryear{{Chavanis}}{{Chavanis}}{2000}]{Chavanis2000}
{Chavanis} P.~H.,  2000, \aap, \href
  {http://cdsads.u-strasbg.fr/abs/2000A%26A...356.1089C} {356, 1089}

\bibitem[\protect\citeauthoryear{{Crnkovic-Rubsamen}, {Zhu}  \&
  {Stone}}{{Crnkovic-Rubsamen} et~al.}{2015}]{CrnkovicZhuStone}
{Crnkovic-Rubsamen} I.,  {Zhu} Z.,   {Stone} J.~M.,  2015, \mn@doi [\mnras]
  {10.1093/mnras/stv828}, \href
  {http://cdsads.u-strasbg.fr/abs/2015MNRAS.450.4285C} {450, 4285}

\bibitem[\protect\citeauthoryear{{Cugno} et~al.,}{{Cugno}
  et~al.}{2019}]{Cugno19}
{Cugno} G.,  et~al., 2019, \mn@doi [\aap] {10.1051/0004-6361/201834170}, \href
  {http://cdsads.u-strasbg.fr/abs/2019A%26A...622A.156C} {622, A156}

\bibitem[\protect\citeauthoryear{{Dong} \& {Fung}}{{Dong} \&
  {Fung}}{2017}]{DongFung2017}
{Dong} R.,  {Fung} J.,  2017, \mn@doi [\apj] {10.3847/1538-4357/835/1/38},
  \href {http://cdsads.u-strasbg.fr/abs/2017ApJ...835...38D} {835, 38}

\bibitem[\protect\citeauthoryear{{Dong}, {Zhu}, {Rafikov}  \& {Stone}}{{Dong}
  et~al.}{2015}]{Dong2015}
{Dong} R.,  {Zhu} Z.,  {Rafikov} R.~R.,   {Stone} J.~M.,  2015, \mn@doi [\apjl]
  {10.1088/2041-8205/809/1/L5}, \href
  {http://adsabs.harvard.edu/abs/2015ApJ...809L...5D} {809, L5}

\bibitem[\protect\citeauthoryear{{Dong} et~al.,}{{Dong} et~al.}{2018}]{Dong18}
{Dong} R.,  et~al., 2018, \mn@doi [\apj] {10.3847/1538-4357/aac6cb}, \href
  {http://cdsads.u-strasbg.fr/abs/2018ApJ...860..124D} {860, 124}

\bibitem[\protect\citeauthoryear{{Draine} \& {Lee}}{{Draine} \&
  {Lee}}{1984}]{Draine1984}
{Draine} B.~T.,  {Lee} H.~M.,  1984, \mn@doi [\apj] {10.1086/162480}, \href
  {http://adsabs.harvard.edu/abs/1984ApJ...285...89D} {285, 89}

\bibitem[\protect\citeauthoryear{{Dullemond}, {Juhasz}, {Pohl}, {Sereshti},
  {Shetty}, {Commercon}  \& {Flock}}{{Dullemond} et~al.}{2015}]{Dullemond2015}
{Dullemond} C.,  {Juhasz} A.,  {Pohl} A.,  {Sereshti} F.,  {Shetty}
  R.and~{Peters} T.,  {Commercon} B.,   {Flock} M.,  2015, {RADMC3D},
  \url{http://www.ita.uni-heidelberg.de/\~dullemond/software/radmc-3d/}

\bibitem[\protect\citeauthoryear{{Fu}, {Li}, {Lubow}  \& {Li}}{{Fu}
  et~al.}{2014a}]{Fu2014}
{Fu} W.,  {Li} H.,  {Lubow} S.,   {Li} S.,  2014a, \mn@doi [\apjl]
  {10.1088/2041-8205/788/2/L41}, \href
  {http://adsabs.harvard.edu/abs/2014ApJ...788L..41F} {788, L41}

\bibitem[\protect\citeauthoryear{{Fu}, {Li}, {Lubow}, {Li}  \& {Liang}}{{Fu}
  et~al.}{2014b}]{Fu2014feedback}
{Fu} W.,  {Li} H.,  {Lubow} S.,  {Li} S.,   {Liang} E.,  2014b, \mn@doi [\apjl]
  {10.1088/2041-8205/795/2/L39}, \href
  {http://adsabs.harvard.edu/abs/2014ApJ...795L..39F} {795, L39}

\bibitem[\protect\citeauthoryear{{Fuente} et~al.,}{{Fuente}
  et~al.}{2017}]{Fuente2017}
{Fuente} A.,  et~al., 2017, \mn@doi [\apjl] {10.3847/2041-8213/aa8558}, \href
  {http://adsabs.harvard.edu/abs/2017ApJ...846L...3F} {846, L3}

\bibitem[\protect\citeauthoryear{{Fung} \& {Dong}}{{Fung} \&
  {Dong}}{2015}]{Fung2015}
{Fung} J.,  {Dong} R.,  2015, \mn@doi [\apjl] {10.1088/2041-8205/815/2/L21},
  \href {http://adsabs.harvard.edu/abs/2015ApJ...815L..21F} {815, L21}

\bibitem[\protect\citeauthoryear{{Gaia Collaboration} et~al.,}{{Gaia
  Collaboration} et~al.}{2018}]{Gaia2018}
{Gaia Collaboration} et~al., 2018, \mn@doi [\aap]
  {10.1051/0004-6361/201833051}, \href
  {http://cdsads.u-strasbg.fr/abs/2018A%26A...616A...1G} {616, A1}

\bibitem[\protect\citeauthoryear{{Garufi} et~al.,}{{Garufi}
  et~al.}{2013}]{Garufi13HD135}
{Garufi} A.,  et~al., 2013, \mn@doi [\aap] {10.1051/0004-6361/201322429}, \href
  {http://cdsads.u-strasbg.fr/abs/2013A%26A...560A.105G} {560, A105}

\bibitem[\protect\citeauthoryear{{Grady} et~al.,}{{Grady}
  et~al.}{2013}]{Grady2013}
{Grady} C.~A.,  et~al., 2013, \mn@doi [\apj] {10.1088/0004-637X/762/1/48},
  \href {http://adsabs.harvard.edu/abs/2013ApJ...762...48G} {762, 48}

\bibitem[\protect\citeauthoryear{{G{\"u}ttler} et~al.,}{{G{\"u}ttler}
  et~al.}{2019}]{GuttlerReviewRosettaDust}
{G{\"u}ttler} C.,  et~al., 2019, arXiv e-prints, \href
  {http://cdsads.u-strasbg.fr/abs/2019arXiv190210634G} {}

\bibitem[\protect\citeauthoryear{{Hammer}, {Kratter}  \& {Lin}}{{Hammer}
  et~al.}{2017}]{Hammer17}
{Hammer} M.,  {Kratter} K.~M.,   {Lin} M.-K.,  2017, \mn@doi [\mnras]
  {10.1093/mnras/stw3000}, \href
  {http://cdsads.u-strasbg.fr/abs/2017MNRAS.466.3533H} {466, 3533}

\bibitem[\protect\citeauthoryear{{Hammer}, {Pinilla}, {Kratter}  \&
  {Lin}}{{Hammer} et~al.}{2019}]{Hammer19}
{Hammer} M.,  {Pinilla} P.,  {Kratter} K.~M.,   {Lin} M.-K.,  2019, \mn@doi
  [\mnras] {10.1093/mnras/sty2946}, \href
  {http://cdsads.u-strasbg.fr/abs/2019MNRAS.482.3609H} {482, 3609}

\bibitem[\protect\citeauthoryear{{Huang} et~al.,}{{Huang}
  et~al.}{2018}]{DSHARP2}
{Huang} J.,  et~al., 2018, \mn@doi [\apjl] {10.3847/2041-8213/aaf740}, \href
  {http://cdsads.u-strasbg.fr/abs/2018ApJ...869L..42H} {869, L42}

\bibitem[\protect\citeauthoryear{{Hu{\'e}lamo} et~al.,}{{Hu{\'e}lamo}
  et~al.}{2018}]{Huelamo18}
{Hu{\'e}lamo} N.,  et~al., 2018, \mn@doi [\aap] {10.1051/0004-6361/201832874},
  \href {http://cdsads.u-strasbg.fr/abs/2018A%26A...613L...5H} {613, L5}

\bibitem[\protect\citeauthoryear{{Isella}, {Natta}, {Wilner}, {Carpenter}  \&
  {Testi}}{{Isella} et~al.}{2010}]{Isella2010}
{Isella} A.,  {Natta} A.,  {Wilner} D.,  {Carpenter} J.~M.,   {Testi} L.,
  2010, \mn@doi [\apj] {10.1088/0004-637X/725/2/1735}, \href
  {http://adsabs.harvard.edu/abs/2010ApJ...725.1735I} {725, 1735}

\bibitem[\protect\citeauthoryear{{Kataoka}, {Tanaka}, {Okuzumi}  \&
  {Wada}}{{Kataoka} et~al.}{2013}]{Kataoka13b}
{Kataoka} A.,  {Tanaka} H.,  {Okuzumi} S.,   {Wada} K.,  2013, \mn@doi [\aap]
  {10.1051/0004-6361/201322151}, \href
  {http://cdsads.u-strasbg.fr/abs/2013A%26A...557L...4K} {557, L4}

\bibitem[\protect\citeauthoryear{{Kataoka}, {Okuzumi}, {Tanaka}  \&
  {Nomura}}{{Kataoka} et~al.}{2014}]{Kataoka14}
{Kataoka} A.,  {Okuzumi} S.,  {Tanaka} H.,   {Nomura} H.,  2014, \mn@doi [\aap]
  {10.1051/0004-6361/201323199}, \href
  {http://cdsads.u-strasbg.fr/abs/2014A%26A...568A..42K} {568, A42}

\bibitem[\protect\citeauthoryear{{Kraus} et~al.,}{{Kraus}
  et~al.}{2017}]{Kraus2017}
{Kraus} S.,  et~al., 2017, \mn@doi [\apjl] {10.3847/2041-8213/aa8edc}, \href
  {http://cdsads.u-strasbg.fr/abs/2017ApJ...848L..11K} {848, L11}

\bibitem[\protect\citeauthoryear{{Langevin} et~al.,}{{Langevin}
  et~al.}{2016}]{Langevin16_67P}
{Langevin} Y.,  et~al., 2016, \mn@doi [\icarus] {10.1016/j.icarus.2016.01.027},
  \href {http://cdsads.u-strasbg.fr/abs/2016Icar..271...76L} {271, 76}

\bibitem[\protect\citeauthoryear{{Les} \& {Lin}}{{Les} \&
  {Lin}}{2015}]{LesLin15}
{Les} R.,  {Lin} M.-K.,  2015, \mn@doi [\mnras] {10.1093/mnras/stv712}, \href
  {http://adsabs.harvard.edu/abs/2015MNRAS.450.1503L} {450, 1503}

\bibitem[\protect\citeauthoryear{{Lesur} \& {Papaloizou}}{{Lesur} \&
  {Papaloizou}}{2009}]{LesurPapa09}
{Lesur} G.,  {Papaloizou} J.~C.~B.,  2009, \mn@doi [\aap]
  {10.1051/0004-6361/200811577}, \href
  {http://cdsads.u-strasbg.fr/abs/2009A%26A...498....1L} {498, 1}

\bibitem[\protect\citeauthoryear{{Li} \& {Greenberg}}{{Li} \&
  {Greenberg}}{1997}]{Li1997}
{Li} A.,  {Greenberg} J.~M.,  1997, \aap, \href
  {http://adsabs.harvard.edu/abs/1997A\%26A...323..566L} {323, 566}

\bibitem[\protect\citeauthoryear{{Li}, {Finn}, {Lovelace}  \& {Colgate}}{{Li}
  et~al.}{2000}]{Li2000}
{Li} H.,  {Finn} J.~M.,  {Lovelace} R.~V.~E.,   {Colgate} S.~A.,  2000, \mn@doi
  [\apj] {10.1086/308693}, \href
  {https://ui.adsabs.harvard.edu/#abs/2000ApJ...533.1023L} {533, 1023}

\bibitem[\protect\citeauthoryear{{Li}, {Colgate}, {Wendroff}  \& {Liska}}{{Li}
  et~al.}{2001}]{Li2001}
{Li} H.,  {Colgate} S.~A.,  {Wendroff} B.,   {Liska} R.,  2001, \mn@doi [\apj]
  {10.1086/320241}, \href
  {https://ui.adsabs.harvard.edu/#abs/2001ApJ...551..874L} {551, 874}

\bibitem[\protect\citeauthoryear{{Lin}}{{Lin}}{2012}]{Lin2012planetvortex}
{Lin} M.-K.,  2012, \mn@doi [\mnras] {10.1111/j.1365-2966.2012.21955.x}, \href
  {http://cdsads.u-strasbg.fr/abs/2012MNRAS.426.3211L} {426, 3211}

\bibitem[\protect\citeauthoryear{{Lin}}{{Lin}}{2014}]{Lin2014}
{Lin} M.-K.,  2014, \mn@doi [\mnras] {10.1093/mnras/stt1909}, \href
  {http://cdsads.u-strasbg.fr/abs/2014MNRAS.437..575L} {437, 575}

\bibitem[\protect\citeauthoryear{{Lin} \& {Papaloizou}}{{Lin} \&
  {Papaloizou}}{2011}]{LP11a}
{Lin} M.-K.,  {Papaloizou} J.~C.~B.,  2011, \mn@doi [\mnras]
  {10.1111/j.1365-2966.2011.18798.x}, \href
  {http://cdsads.u-strasbg.fr/abs/2011MNRAS.415.1426L} {415, 1426}

\bibitem[\protect\citeauthoryear{{Lin} \& {Pierens}}{{Lin} \&
  {Pierens}}{2018}]{LinPierens18}
{Lin} M.-K.,  {Pierens} A.,  2018, \mn@doi [\mnras] {10.1093/mnras/sty947},
  \href {http://cdsads.u-strasbg.fr/abs/2018MNRAS.478..575L} {478, 575}

\bibitem[\protect\citeauthoryear{{Lovelace} \& {Hohlfeld}}{{Lovelace} \&
  {Hohlfeld}}{2013}]{lovelace2013}
{Lovelace} R.~V.~E.,  {Hohlfeld} R.~G.,  2013, \mn@doi [\mnras]
  {10.1093/mnras/sts361}, \href
  {http://adsabs.harvard.edu/abs/2013MNRAS.429..529L} {429, 529}

\bibitem[\protect\citeauthoryear{{Lovelace}, {Li}, {Colgate}  \&
  {Nelson}}{{Lovelace} et~al.}{1999}]{Lovelace1999}
{Lovelace} R.~V.~E.,  {Li} H.,  {Colgate} S.~A.,   {Nelson} A.~F.,  1999,
  \mn@doi [\apj] {10.1086/306900}, \href
  {https://ui.adsabs.harvard.edu/#abs/1999ApJ...513..805L} {513, 805}

\bibitem[\protect\citeauthoryear{{Lyra} \& {Lin}}{{Lyra} \&
  {Lin}}{2013}]{LyraLin2013}
{Lyra} W.,  {Lin} M.-K.,  2013, \mn@doi [\apj] {10.1088/0004-637X/775/1/17},
  \href {http://adsabs.harvard.edu/abs/2013ApJ...775...17L} {775, 17}

\bibitem[\protect\citeauthoryear{{Lyra}, {Johansen}, {Klahr}  \&
  {Piskunov}}{{Lyra} et~al.}{2009}]{Lyra2009}
{Lyra} W.,  {Johansen} A.,  {Klahr} H.,   {Piskunov} N.,  2009, \mn@doi [\aap]
  {10.1051/0004-6361:200810797}, \href
  {http://adsabs.harvard.edu/abs/2009A\%26A...493.1125L} {493, 1125}

\bibitem[\protect\citeauthoryear{{Lyra}, {Turner}  \& {McNally}}{{Lyra}
  et~al.}{2015}]{Lyra15}
{Lyra} W.,  {Turner} N.~J.,   {McNally} C.~P.,  2015, \mn@doi [\aap]
  {10.1051/0004-6361/201424919}, \href
  {http://adsabs.harvard.edu/abs/2015A%26A...574A..10L} {574, A10}

\bibitem[\protect\citeauthoryear{{Marino}, {Casassus}, {Perez}, {Lyra},
  {Roman}, {Avenhaus}, {Wright}  \& {Maddison}}{{Marino}
  et~al.}{2015}]{Marino2015mwc}
{Marino} S.,  {Casassus} S.,  {Perez} S.,  {Lyra} W.,  {Roman} P.~E.,
  {Avenhaus} H.,  {Wright} C.~M.,   {Maddison} S.~T.,  2015, \mn@doi [\apj]
  {10.1088/0004-637X/813/1/76}, \href
  {http://adsabs.harvard.edu/abs/2015ApJ...813...76M} {813, 76}

\bibitem[\protect\citeauthoryear{{Masset}}{{Masset}}{2000}]{Masset2000}
{Masset} F.,  2000, \mn@doi [\aaps] {10.1051/aas:2000116}, \href
  {http://adsabs.harvard.edu/abs/2000A\%26AS..141..165M} {141, 165}

\bibitem[\protect\citeauthoryear{{Meeus} et~al.,}{{Meeus}
  et~al.}{2012}]{Meeus12}
{Meeus} G.,  et~al., 2012, \mn@doi [\aap] {10.1051/0004-6361/201219225}, \href
  {http://adsabs.harvard.edu/abs/2012A%26A...544A..78M} {544, A78}

\bibitem[\protect\citeauthoryear{{Meheut}, {Keppens}, {Casse}  \&
  {Benz}}{{Meheut} et~al.}{2012}]{Meheut2012}
{Meheut} H.,  {Keppens} R.,  {Casse} F.,   {Benz} W.,  2012, \mn@doi [\aap]
  {10.1051/0004-6361/201118500}, \href
  {http://cdsads.u-strasbg.fr/abs/2012A%26A...542A...9M} {542, A9}

\bibitem[\protect\citeauthoryear{{Ohta} et~al.,}{{Ohta}
  et~al.}{2016}]{OhtaV1247PI}
{Ohta} Y.,  et~al., 2016, \mn@doi [\pasj] {10.1093/pasj/psw051}, \href
  {http://adsabs.harvard.edu/abs/2016PASJ...68...53O} {68, 53}

\bibitem[\protect\citeauthoryear{{Perez}, {Dunhill}, {Casassus}, {Roman},
  {Szul{\'a}gyi}, {Flores}, {Marino}  \& {Montesinos}}{{Perez}
  et~al.}{2015}]{Perez2015}
{Perez} S.,  {Dunhill} A.,  {Casassus} S.,  {Roman} P.,  {Szul{\'a}gyi} J.,
  {Flores} C.,  {Marino} S.,   {Montesinos} M.,  2015, \mn@doi [\apjl]
  {10.1088/2041-8205/811/1/L5}, \href
  {http://adsabs.harvard.edu/abs/2015ApJ...811L...5P} {811, L5}

\bibitem[\protect\citeauthoryear{{P{\'e}rez}, {Casassus}  \&
  {Ben{\'{\i}}tez-Llambay}}{{P{\'e}rez} et~al.}{2018}]{Perez2018}
{P{\'e}rez} S.,  {Casassus} S.,   {Ben{\'{\i}}tez-Llambay} P.,  2018, \mn@doi
  [\mnras] {10.1093/mnrasl/sly109}, \href
  {http://cdsads.u-strasbg.fr/abs/2018MNRAS.tmpL.113P} {}

\bibitem[\protect\citeauthoryear{{P{\'e}rez}, {Casassus}, {Baruteau}, {Dong},
  {Hales}  \& {Cieza}}{{P{\'e}rez} et~al.}{2019}]{SebaHD169}
{P{\'e}rez} S.,  {Casassus} S.,  {Baruteau} C.,  {Dong} R.,  {Hales} A.,
  {Cieza} L.,  2019, arXiv e-prints, \href
  {http://cdsads.u-strasbg.fr/abs/2019arXiv190205143P} {}

\bibitem[\protect\citeauthoryear{{Pierens} \& {Lin}}{{Pierens} \&
  {Lin}}{2018}]{PierensLin2018}
{Pierens} A.,  {Lin} M.-K.,  2018, \mn@doi [\mnras] {10.1093/mnras/sty1314},
  \href {http://cdsads.u-strasbg.fr/abs/2018MNRAS.479.4878P} {479, 4878}

\bibitem[\protect\citeauthoryear{{Pinte}, {Dent}, {M{\'e}nard}, {Hales},
  {Hill}, {Cortes}  \& {de Gregorio-Monsalvo}}{{Pinte} et~al.}{2016}]{Pinte16}
{Pinte} C.,  {Dent} W.~R.~F.,  {M{\'e}nard} F.,  {Hales} A.,  {Hill} T.,
  {Cortes} P.,   {de Gregorio-Monsalvo} I.,  2016, \mn@doi [\apj]
  {10.3847/0004-637X/816/1/25}, \href
  {http://cdsads.u-strasbg.fr/abs/2016ApJ...816...25P} {816, 25}

\bibitem[\protect\citeauthoryear{{Pinte} et~al.,}{{Pinte}
  et~al.}{2018}]{Pinte2018}
{Pinte} C.,  et~al., 2018, \mn@doi [\apjl] {10.3847/2041-8213/aac6dc}, \href
  {http://cdsads.u-strasbg.fr/abs/2018ApJ...860L..13P} {860, L13}

\bibitem[\protect\citeauthoryear{{Reg{\'a}ly} \& {Vorobyov}}{{Reg{\'a}ly} \&
  {Vorobyov}}{2017}]{Regaly2017}
{Reg{\'a}ly} Z.,  {Vorobyov} E.,  2017, \mn@doi [\mnras]
  {10.1093/mnras/stx1801}, \href
  {http://adsabs.harvard.edu/abs/2017MNRAS.471.2204R} {471, 2204}

\bibitem[\protect\citeauthoryear{{Reg{\'a}ly}, {Juh{\'a}sz}, {S{\'a}ndor}  \&
  {Dullemond}}{{Reg{\'a}ly} et~al.}{2012}]{Regaly12}
{Reg{\'a}ly} Z.,  {Juh{\'a}sz} A.,  {S{\'a}ndor} Z.,   {Dullemond} C.~P.,
  2012, \mn@doi [\mnras] {10.1111/j.1365-2966.2011.19834.x}, \href
  {http://adsabs.harvard.edu/abs/2012MNRAS.419.1701R} {419, 1701}

\bibitem[\protect\citeauthoryear{{Reggiani} et~al.,}{{Reggiani}
  et~al.}{2018}]{Reggiani2018}
{Reggiani} M.,  et~al., 2018, \mn@doi [\aap] {10.1051/0004-6361/201732016},
  \href {http://adsabs.harvard.edu/abs/2018A\%26A...611A..74R} {611, A74}

\bibitem[\protect\citeauthoryear{{Ren} et~al.,}{{Ren} et~al.}{2018}]{Ren2018}
{Ren} B.,  et~al., 2018, \mn@doi [\apjl] {10.3847/2041-8213/aab7f5}, \href
  {http://cdsads.u-strasbg.fr/abs/2018ApJ...857L...9R} {857, L9}

\bibitem[\protect\citeauthoryear{{Riols} \& {Lesur}}{{Riols} \&
  {Lesur}}{2018}]{Riols18}
{Riols} A.,  {Lesur} G.,  2018, \mn@doi [\aap] {10.1051/0004-6361/201833212},
  \href {http://adsabs.harvard.edu/abs/2018A%26A...617A.117R} {617, A117}

\bibitem[\protect\citeauthoryear{{S{\'a}ndor}, {Lyra}  \&
  {Dullemond}}{{S{\'a}ndor} et~al.}{2011}]{Sandor2011}
{S{\'a}ndor} Z.,  {Lyra} W.,   {Dullemond} C.~P.,  2011, \mn@doi [\apjl]
  {10.1088/2041-8205/728/1/L9}, \href
  {http://adsabs.harvard.edu/abs/2011ApJ...728L...9S} {728, L9}

\bibitem[\protect\citeauthoryear{{Simon}, {Lesur}, {Kunz}  \&
  {Armitage}}{{Simon} et~al.}{2015}]{Simon15}
{Simon} J.~B.,  {Lesur} G.,  {Kunz} M.~W.,   {Armitage} P.~J.,  2015, \mn@doi
  [\mnras] {10.1093/mnras/stv2070}, \href
  {http://cdsads.u-strasbg.fr/abs/2015MNRAS.454.1117S} {454, 1117}

\bibitem[\protect\citeauthoryear{{Simon}, {Bai}, {Flaherty}  \&
  {Hughes}}{{Simon} et~al.}{2018}]{Simon18}
{Simon} J.~B.,  {Bai} X.-N.,  {Flaherty} K.~M.,   {Hughes} A.~M.,  2018,
  \mn@doi [\apj] {10.3847/1538-4357/aad86d}, \href
  {http://cdsads.u-strasbg.fr/abs/2018ApJ...865...10S} {865, 10}

\bibitem[\protect\citeauthoryear{{Spiegel} \& {Burrows}}{{Spiegel} \&
  {Burrows}}{2012}]{SpiegelBurrows2012}
{Spiegel} D.~S.,  {Burrows} A.,  2012, \mn@doi [\apj]
  {10.1088/0004-637X/745/2/174}, \href
  {http://adsabs.harvard.edu/abs/2012ApJ...745..174S} {745, 174}

\bibitem[\protect\citeauthoryear{{Szul{\'a}gyi}, {Morbidelli}, {Crida}  \&
  {Masset}}{{Szul{\'a}gyi} et~al.}{2014}]{Szulagyi2014}
{Szul{\'a}gyi} J.,  {Morbidelli} A.,  {Crida} A.,   {Masset} F.,  2014, \mn@doi
  [\apj] {10.1088/0004-637X/782/2/65}, \href
  {http://adsabs.harvard.edu/abs/2014ApJ...782...65S} {782, 65}

\bibitem[\protect\citeauthoryear{{Varni{\`e}re} \& {Tagger}}{{Varni{\`e}re} \&
  {Tagger}}{2006}]{Varniere06}
{Varni{\`e}re} P.,  {Tagger} M.,  2006, \mn@doi [\aap]
  {10.1051/0004-6361:200500226}, \href
  {http://adsabs.harvard.edu/abs/2006A%26A...446L..13V} {446, L13}

\bibitem[\protect\citeauthoryear{{Yang}, {Mac Low}  \& {Johansen}}{{Yang}
  et~al.}{2018}]{YMLJ18}
{Yang} C.-C.,  {Mac Low} M.-M.,   {Johansen} A.,  2018, \mn@doi [\apj]
  {10.3847/1538-4357/aae7d4}, \href
  {http://adsabs.harvard.edu/abs/2018ApJ...868...27Y} {868, 27}

\bibitem[\protect\citeauthoryear{{Youdin}}{{Youdin}}{2010}]{Youdin2010}
{Youdin} A.~N.,  2010, in {Montmerle} T.,  {Ehrenreich} D.,   {Lagrange} A.-M.,
   eds,  EAS Publications Series Vol. 41, EAS Publications Series. pp 187--207
  (\mn@eprint {arXiv} {0807.1114}), \mn@doi{10.1051/eas/1041016}

\bibitem[\protect\citeauthoryear{{Zhu} \& {Baruteau}}{{Zhu} \&
  {Baruteau}}{2016}]{Zhu2016}
{Zhu} Z.,  {Baruteau} C.,  2016, \mn@doi [\mnras] {10.1093/mnras/stw202}, \href
  {http://adsabs.harvard.edu/abs/2016MNRAS.458.3918Z} {458, 3918}

\bibitem[\protect\citeauthoryear{{Zhu} \& {Stone}}{{Zhu} \&
  {Stone}}{2014}]{ZhuStone2014}
{Zhu} Z.,  {Stone} J.~M.,  2014, \mn@doi [\apj] {10.1088/0004-637X/795/1/53},
  \href {http://cdsads.u-strasbg.fr/abs/2014ApJ...795...53Z} {795, 53}

\bibitem[\protect\citeauthoryear{{Zhu}, {Dong}, {Stone}  \& {Rafikov}}{{Zhu}
  et~al.}{2015}]{ZhuDong2015}
{Zhu} Z.,  {Dong} R.,  {Stone} J.~M.,   {Rafikov} R.~R.,  2015, \mn@doi [\apj]
  {10.1088/0004-637X/813/2/88}, \href
  {http://adsabs.harvard.edu/abs/2015ApJ...813...88Z} {813, 88}

\bibitem[\protect\citeauthoryear{{van der Marel}, {Cazzoletti}, {Pinilla}  \&
  {Garufi}}{{van der Marel} et~al.}{2016}]{Marel16HD135}
{van der Marel} N.,  {Cazzoletti} P.,  {Pinilla} P.,   {Garufi} A.,  2016,
  \mn@doi [\apj] {10.3847/0004-637X/832/2/178}, \href
  {http://cdsads.u-strasbg.fr/abs/2016ApJ...832..178V} {832, 178}

\makeatother
\end{thebibliography}

\bsp	
\label{lastpage}
\end{document}